\documentclass[nofootinbib,preprint,preprintnumbers,a4paper,10pt]{revtex4}
\usepackage{multirow}
\usepackage{textcomp}
\usepackage{amsmath,graphicx,color,epsfig}

\usepackage{footnote}
\usepackage{ulem}
\usepackage{booktabs}
\usepackage{array}
\usepackage{amssymb}
\usepackage{subfigure}
\usepackage{longtable}
\usepackage{verbatim}
\usepackage{amsfonts}
\usepackage{hyperref}
\usepackage{cancel}

\begin{document}

\title{Isospin sum rules for bottom-baryon weak decays}

\author{Wei-Chen Fu$^{1}$}
\author{Si-Jia Wen$^{1}$}
\author{Di Wang$^{1}$}\email{wangdi@hunnu.edu.cn}

\address{%
$^1$Department of Physics, Hunan Normal University, Changsha 410081, China
}

\begin{abstract}
Isospin symmetry, as the most precise flavor symmetry, can be used to extract information about hadronic dynamics.
The effective Hamiltonian operators of bottom quark weak decays are zero under a series of isospin lowering operators $I_-^n$, which permits us to generate isospin sum rules without the Wigner-Eckart invariants.
In this work, we derive hundreds of isospin sum rules for the two- and three-body non-leptonic decays of bottom baryons.
They provide hints for new decay modes and the isospin partners of pentaquark states.

\end{abstract}

\maketitle

\section{Introduction}
Bottom baryon decays provide ideal laboratories for studying strong and weak interactions in heavy-to-light baryonic transitions.
In recent years, a number of bottom-baryon decay modes have been observed at the Large Hadron Collider (LHC)\cite{PDG}, allowing us to extract information about the dynamics of bottom-baryon decays.
Flavor symmetry is a powerful tool for analyzing weak decays of heavy hadrons.
Flavor symmetry analysis has been applied to bottom baryon decays in the literature \cite{Huang:2022zsy,Han:2023oac,Roy:2019cky,He:2015fsa,He:2015fwa,Wang:2024rwf,Roy:2020nyx,He:2018joe,Gavrilova:2022hbx}.
Flavor symmetry results in certain relations between several decay modes, known as flavor sum rules.
Isospin symmetry is the most precise flavor symmetry.
Isospin breaking is naively expected to be $\delta_I \simeq (m_u-m_d)/\Lambda_{\rm QCD}\sim 1\%$.
Isospin sum rules may provide hints for the isospin partners of exotic hadrons through bottom baryon decays.

In Refs.~\cite{Wang:2023pnb,Luo:2023vbx,Wang:2024tnx}, we propose a simple approach to generate isospin sum rules for heavy hadron decays, in which the Wigner-Eckart invariants \cite{Eckart30,Wigner59} are not needed.
The effective Hamiltonian operators of heavy quark weak decays are zero under isospin lowering operators $I_-^n$.
This fact allows us to generate isospin sum rules by acting with $I_-^n$ on the initial and final states of heavy hadron weak decays.
In this work, we apply this approach to bottom baryon decays, and we derive master formulas for generating isospin sum rules for two- and three-body bottom baryon decays.
Many new isospin sum rules are obtained from these master formulas.

The rest of this paper is structured as follows.
The theoretical framework for generating isospin sum rules for $b$-baryon decays is presented in Sect. \ref{th}, and phenomenological analysis of the isospin sum rules is discussed in Sect. \ref{PA}.
Section \ref{SUM} provides a brief summary.
The coefficient matrices generated by $I_-$ operating on hadron states and decay modes are listed in
Appendix \ref{cmsda}.
The isospin sum rules for two- and three-body $b$-baryon decays are listed in Appendices \ref{I2} and \ref{I3}, respectively.

\section{Theoretical Framework}\label{th}

Taking the $\mathcal{B}_b\to D\mathcal{B}_8$ decays (where $D$ is a charm meson and $\mathcal{B}_8$ is a light octet baryon) as examples, we demonstrate the basic idea of generating isospin sum rules by $I_-^n$ as follows.
In the $SU(3)$ picture, the effective Hamiltonian of $b\to c\overline u q$ transition can be written as
\begin{equation}\label{h}
 \mathcal H_{\rm eff}=\sum_{i,j=1}^3 H^{i}_{j}O^{i}_{j},
\end{equation}
where $O^{i}_{j}$ denotes the four-quark operator and $H$ is the $3\times 3$ coefficient matrix.
The initial and final states of the weak decay, such as the light octet baryon, can be written as
\begin{align}
  |\mathcal{B}_8^\alpha\rangle = (\mathcal{B}_8^\alpha)^{i}_{j}|[\mathcal{B}_8]^{i}_{j} \rangle,
\end{align}
where $|[\mathcal{B}_8]^{i}_{j} \rangle$ is the quark composition of the meson state and $(\mathcal{B}_8^\alpha)$ is the coefficient matrix.
The decay amplitude for the $\mathcal{B}_b^\gamma\to D^\alpha \mathcal{B}_8^\beta$ mode is constructed as
\begin{align}\label{amp}
\mathcal{A}(\mathcal{B}_b^\gamma\to D^\alpha \mathcal{B}_8^\beta)& = \langle D^\alpha \mathcal{B}_8^\beta |\mathcal{H}_{\rm eff}| \mathcal{B}_b^\gamma\rangle\nonumber\\&~
=\sum_{\omega}\,(D^\alpha)^n\langle D^n|(\mathcal{B}_8^\beta)^l_m\langle [\mathcal{B}_8]^l_m||H^{j}_kO^{j}_k||(\mathcal{ B}_b^\gamma)_i|[\mathcal{B}_b]_i\rangle\nonumber\\& ~~=\sum_\omega\,\langle D^{n} [\mathcal{B}_8]^l_m |O^{j}_{k} |[\mathcal{B}_b]_{i}\rangle \times (D^\alpha)^{n}(\mathcal{B}_8^\beta)_{m}^{l} H^{j}_{k}(\mathcal{B}_b^\gamma)_i\nonumber\\& ~~~= \sum_\omega X_{\omega}(C_\omega)^{\alpha\beta\gamma},
\end{align}
where $\sum_{\omega}$ represents summing over all full contractions of tensor $\langle D^{n} [\mathcal{B}_8]^l_m |O^{j}_{k} |[\mathcal{B}_b]_{i}\rangle$, i.e., the  invariant tensor such as  $\langle D^{i} [\mathcal{B}_8]^k_j |O^{j}_{k} |[\mathcal{B}_b]_{i}\rangle$, etc.
According to the Wigner-Eckart theorem \cite{Eckart30,Wigner59}, the  invariant tensor $X_\omega$ is independent of decay channels, i.e., indices $\alpha$, $\beta$ and $\gamma$.
All the information for initial/final states is absorbed into the Clebsch-Gordan coefficient $(C_\omega)^{\alpha\beta\gamma}$.

The isospin lowering operator $I_-$ is
\begin{eqnarray}
 I_-=  \left( \begin{array}{ccc}
   0   & 0  & 0 \\
     1 &   0  & 0 \\
    0 & 0 & 0 \\
  \end{array}\right).
\end{eqnarray}
If the effective Hamiltonian \eqref{h} is zero under a series of operators $I_-^n$, i.e., $I_-^n\,H=I_-\{I_-\dots \{I_-\,H\}\dots\}=0$,
we have
\begin{equation}\label{rule}
  \langle D^\alpha \mathcal{B}_8^\beta |I_-^n\,\mathcal{H}_{\rm eff}| \mathcal{B}_b^\gamma\rangle
   = \sum_\omega\,\langle D^{n} [\mathcal{B}_8]^l_m |O^{j}_{k} |[\mathcal{B}_b]_{i}\rangle \times (D^\alpha)^{n}(\mathcal{B}_8^\beta)_{m}^{l} (I_-^n\,H)^{j}_{k}(\mathcal{B}_b^\gamma)_i = 0,
\end{equation}
since zero multiplied by any quantity is zero.
The times of operation $n$ is obtained through the following calculations.
The effective Hamiltonian of the $b\to c\overline u q$ transition is given by \cite{Buchalla:1995vs}
 \begin{align}\label{hsmb}
 \mathcal H_{\rm eff}={\frac{G_F}{\sqrt 2} }
 \sum_{q=d,s}V_{cb}V_{uq}^*\left[C_1(\mu)O_1(\mu)+C_2(\mu)O_2(\mu)\right]+h.c.,
 \end{align}
where $O_1$ and $O_2$ are
\begin{eqnarray}
O_1=(\bar{q}_{\alpha}u_{\beta})_{V-A}
(\bar{c}_{\beta}b_{\alpha})_{V-A},\qquad
O_2=(\bar{q}_{\alpha}u_{\alpha})_{V-A}
(\bar{c}_{\beta}b_{\beta})_{V-A}.
\end{eqnarray}
In the $SU(3)$ picture,
the non-zero coefficients of $O^{i}_j$ include
\begin{align}\label{ckm5}
 &H^{2}_1 = V_{cb}V_{ud}^*,  \qquad H_{1}^{3}=V_{cb}V_{us}^*.
\end{align}
$H^{2}_1$ and $H_{1}^{3}$ are responding to the $b\to c\overline u d$ and $b\to c\overline u s$ transitions, respectively.
Operator $O^{i}_j$ can be decomposed into irreducible representations as $3 \otimes  \overline 3 =  8\oplus 1$.
The coefficient matrices $ H(8)$ and $ H(1)$ are
\begin{eqnarray}
 H( 8)= \left( \begin{array}{ccc}
   0   & V_{cb}V_{ud}^*  & V_{cb}V_{us}^* \\
     0 & 0 &  0 \\
    0 & 0 & 0 \\
  \end{array}\right), \qquad \text{and}\qquad H(1) = 0.
\end{eqnarray}
Under $I_-^n$, $H(8)$ is transformed as
\begin{align}\label{I8x}
 I_-\, H(8) =I_-\cdot H(8)-H(8)\cdot I_- =
  \left( \begin{array}{ccc}
   -V_{cb}V_{ud}^*   & 0  & 0 \\
     0 & V_{cb}V_{ud}^* & V_{cb}V_{us}^* \\
    0 & 0 & 0 \\
  \end{array}\right),
\end{align}
\begin{align}\label{I8y}
 I_-^2 \,H(8) = I_-\{I_- \,H(8)\} =
  \left( \begin{array}{ccc}
   0  & 0  & 0 \\
    -2V_{cb}V_{ud}^* & 0 &  0 \\
    0 & 0 & 0 \\
  \end{array}\right),
\end{align}
\begin{align}\label{I8z}
 I_-^3\, H(8) = I_-\{I_-\{I_-\,H(8)\}\} =0.
\end{align}
Thus, the effective Hamiltonian of $b\to c\overline u d$ ($b\to c\overline u s$) transition is zero under $I_-^n$ with $n\geq 3$ ($n\geq 2$).

On the other hand, if we apply $I_-^n$ to the initial and final states, the LHS of Eq.~\eqref{rule} becomes
\begin{align}\label{rulea}
  \langle D^\alpha \mathcal{B}_8^\beta |I_-^n\,\mathcal{H}_{\rm eff}| \mathcal{B}_b^\gamma\rangle
   = &\sum_\omega\,\langle D^{n} [\mathcal{B}_8]^l_m |O^{j}_{k} |[\mathcal{B}_b]_{i}\rangle \times \nonumber\\ &~~\left[ (I_-^n\,(D^\alpha))^{n}(\mathcal{B}_8^\beta)_{m}^{l} H^{j}_{k}(\mathcal{B}_b^\gamma)_i + (D^\alpha)^{n}(I_-^n\,(\mathcal{B}_8^\beta))_{m}^{l} H^{j}_{k}(\mathcal{B}_b^\gamma)_i + (D^\alpha)^{n}(\mathcal{B}_8^\beta)_{m}^{l} H^{j}_{k}(I_-^n\,(\mathcal{B}_b^\gamma))_i\right],
\end{align}
and the RHS of Eq.~\eqref{rule} is still zero since $\langle D^\alpha \mathcal{B}_8^\beta |I_-^n\,\mathcal{H}_{\rm eff}| \mathcal{B}_b^\gamma\rangle$ is invariant.
One can expand the matrices $I_-^n\,(D^\alpha)$, $I_-^n\,(\mathcal{B}_8^\beta)$, and $I_-^n\,(\mathcal{B}_b^\gamma)$ by the coefficient matrices of the initial and final states.
Then Eq.~\eqref{rulea} becomes a sum of decay amplitudes with appropriate factors.
For example, the equation
\begin{align}
  I_-\,(\mathcal{B}_8^{\Sigma^+}) & = I_-\cdot (\mathcal{B}_8^{\Sigma^+}) - (\mathcal{B}_8^{\Sigma^+})\cdot I_-\nonumber\\&~~=
  \left( \begin{array}{ccc}
   0   & 0  & 0 \\
     1 &   0  & 0 \\
    0 & 0 & 0 \\
  \end{array}\right)\cdot
  \left( \begin{array}{ccc}
   0   & 1  & 0 \\
     0 &   0  & 0 \\
    0 & 0 & 0 \\
  \end{array}\right)-\left( \begin{array}{ccc}
   0   & 1  & 0 \\
     0 &   0  & 0 \\
    0 & 0 & 0 \\
  \end{array}\right)\cdot\left( \begin{array}{ccc}
   0   & 0  & 0 \\
     1 &   0  & 0 \\
    0 & 0 & 0 \\
  \end{array}\right)\nonumber\\&~~~~=\left( \begin{array}{ccc}
   -1   & 0  & 0 \\
     0 &   1  & 0 \\
    0 & 0 & 0 \\
  \end{array}\right) = -\sqrt{2}\left( \begin{array}{ccc}
   1/\sqrt{2}  & 0  & 0 \\
     0 &   -1/\sqrt{2}  & 0 \\
    0 & 0 & 0 \\
  \end{array}\right) =  -\sqrt{2}\,(\mathcal{B}_8^{\Sigma^0})
\end{align}
indicates that
\begin{align}\label{ruleb}
  \langle D^\alpha \Sigma^+ |I_-\,\mathcal{H}_{\rm eff}| \mathcal{B}_b^\gamma\rangle
   = &\sum_\omega\,\langle D^{n} [\mathcal{B}_8]^l_m |O^{j}_{k} |[\mathcal{B}_b]_{i}\rangle \times [... -\sqrt{2} (D^\alpha)^{n}(\mathcal{B}_8^{\Sigma^0})_{m}^{l} H^{j}_{k}(\mathcal{B}_b^\gamma)_i + ...]\nonumber\\& = ...\,\,-\sqrt{2}\,\mathcal{A}(\mathcal{B}_b^\gamma\to D^\alpha \Sigma^0)\,\,...\,\,.
\end{align}
Summing over all the contributions arising from $I_-\,(D^\alpha)$, $I_-\,(\mathcal{B}_8^\beta)$, and $I_-\,(\mathcal{B}_b^\gamma)$, the sum of decay amplitudes generated by $I_-$ for the $\mathcal{B}^\gamma_b\to D^\alpha\mathcal{B}_8^\beta$ mode is derived to be
\begin{align}\label{rulex1}
{SumI_-}\,[\gamma, \alpha,\beta]= \sum_\mu\left[-\{[I_-]_{D}\}_\alpha^\mu \mathcal{A}_{ \gamma \to \mu \beta} +  \{[I_-]_{\mathcal{B}_8}\}_\beta^\mu \mathcal{A}_{\gamma\to \alpha\mu } + \{[I_-]_{\mathcal{ B}_b}\}_\gamma^\mu \mathcal{A}_{\mu\to \alpha \beta }\right],
\end{align}
in which $[I_-]_{D}$, $[I_-]_{\mathcal{B}_8}$, and $[I_-]_{\mathcal{ B}_b}$ are the coefficient matrices given in Appendix \ref{cmsda}, and
$\mathcal{A}_{ \gamma \to \mu \beta}$, $\mathcal{A}_{\gamma\to \alpha\mu }$, and $\mathcal{A}_{\mu\to \alpha \beta }$ are the decay amplitudes of the $\mathcal{ B}_b^\gamma\to D^\mu \mathcal{B}_8^\beta$, $\mathcal{ B}_b^\gamma\to D^\alpha \mathcal{B}_8^\mu$, and $\mathcal{ B}_b^\mu\to D^\alpha \mathcal{B}_8^\beta$ modes, respectively.
The minus sign in the first term matches the minus sign in $I_-$ operating on the octet baryon, where $I_-\langle \mathcal{B}_8 | = I_- \cdot \langle \mathcal{B}_8 | - \langle \mathcal{B}_8 |\cdot I_-$.
One can apply Eq.~\eqref{rulex1} three times (two times) or more with appropriate $\alpha$, $\beta$, and $\gamma$ to obtain an isospin sum rule for $b\to c\overline u d$ ($b\to c\overline u s$) modes.

%%%%%%%%%%%%%%%%%%%%%%%%%%%%%%%%%%%%%%%%%%%%%%%%%%%%%%%
\begin{table*}[t!]
\caption{The values of $n$ for which the Hamiltonian operators of $b\to q_1\overline q_2 q_3$ transitions are zero under $I_-^n$. }\label{n}
\begin{tabular}{|c|c|c|c|c|c|c|c|c|}\hline
 ~~Mode~~ &   ~$b\to c\overline u d$~    & ~$b\to c\overline u s$~ &  ~$b\to c\overline cd$~ & ~$b\to c\overline cs$~ &  ~$b\to u\overline u d$~    & ~$b\to u\overline u s$~ &  ~$b\to u\overline cd$~ & ~$b\to u\overline cs$~ \\\hline
$n$ & $\geq3$ & $\geq2$ & $\geq2$ & $\geq1$ & $\geq3$ & $\geq2$ & $\geq2$ & $\geq1$ \\\hline
\end{tabular}
\end{table*}
%%%%%%%%%%%%%%%%%%%%%%%%%%%%%%%%%%%%%%%%%%

It is found in Refs.~\cite{Wang:2023pnb,Wang:2024tnx} that the effective Hamiltonian operators for the $b\to c\overline u d$, $b\to c\overline u s$, $b\to u\overline u d$, $b\to u\overline u s$, $b\to c\overline c d$, $b\to c\overline c s$, $b\to u\overline c d$, and $b\to u\overline c s$ transitions are zero under $I_-^n$ with $n\geq3$, $n\geq2$, $n\geq3$, $n\geq2$, $n\geq2$, $n\geq1$, $n\geq2$, and $n\geq1$, respectively.
The values of $n$ for which $I_-^n\,H=0$ holds for all types of transitions are listed in Table~\ref{n}.
Similarly to Eq.~\eqref{rulex1}, the summations of amplitudes for other $b$-baryon decay modes generated by $I_-$ can be obtained using the coefficient matrices given in Appendix \ref{cmsda}.
It is noted that minus signs should be introduced for the $D$ meson and the anti-triplet charmed baryon to match the minus sign of the commutator in meson and baryon octets.

The isospin sum rules for two- and three-body $b$-baryon decays are listed in Appendices \ref{I2} and \ref{I3}, respectively.
Many isospin sum rules are derived for the first time.
One can verify these isospin sum rules by writing the isospin amplitudes for each decay mode and substituting them into the isospin sum rules to check whether they are zero.
The isospin sum rules \eqref{r5}$\sim$\eqref{r4} were found in Refs. \cite{Roy:2019cky,Roy:2020nyx}, and \eqref{r3} was also found in Ref.~\cite{He:2018joe}.
The relative minus signs between Refs. \cite{Roy:2019cky,Roy:2020nyx} and this work arise from the different conventions of initial and final states.

It is noted that the order of final states in the isospin sum rules for three-body decays cannot be exchanged arbitrarily; otherwise the isospin relations of intermediate resonance strong decays are violated \cite{Luo:2023vbx}.
The isospin sum rules listed in Appendices \ref{I2} and \ref{I3} are valid for excited states.
For example, the pseudoscalar mesons $\pi$ and $K$ in the isospin sum rules can be replaced by the corresponding vector mesons $\rho$ and $K^*$.

The decay modes dominated by the $b\to c\overline cd/s$ ($b\to u\overline ud/s$) transitions also receive contributions from the $b\to u\overline ud/s$ ($b\to c\overline cd/s$) transitions.
According to Table~\ref{n}, the isospin sum rules derived from $I^n_- \,H_{u\overline ud/s}=0$ are not violated by the $b\to c\overline cd/s$ transitions, but the isospin sum rules derived from $I^n_- \,H_{c\overline cd/s}=0$ are violated by the $b\to u\overline ud/s$ contributions.
The breaking induced by $I^n_-\, H_{ u\overline us}\neq 0$ and $I^n_-\, H_{ u\overline ud}\neq 0$ is naively predicted to be $ (V_{ub}V_{us})/(V_{cb}V_{cs})\sim
\mathcal{O}(1\%)$ and $(V_{ub}V_{ud})/(V_{cb}V_{cd})\sim
\mathcal{O}(10\%)$, respectively.
Thus, we discarded the isospin sum rules derived from $I^n_- \,H_{c\overline cd}=0$ where $I^n_- \,H_{u\overline ud}\neq0$, while retaining the isospin sum rules derived from $I^n_- \,H_{c\overline cs}=0$ where $I^n_- \,H_{u\overline us}\neq0$ in Appendices \ref{I2cc} and \ref{I3cc}.

\section{Phenomenological analysis}\label{PA}
The isospin sum rules are useful tools in the phenomenological analysis of bottom baryon decays.
The decay amplitude for the $\mathcal{B}_{b}\to\mathcal{B}M$ mode is given by
\begin{eqnarray}\label{eq:A&B}
\mathcal{A}(\mathcal{B}_{b}\to\mathcal{B}M)=i\overline u_\mathcal{B}(A-B\gamma_5)u_{\mathcal{B}_b},
\end{eqnarray}
where $A$ and $B$ are the parity-violating $S$-wave and parity-conserving $P$-wave amplitudes with strong phases $\delta_S$ and $\delta_P$, respectively.
The decay width $\Gamma$ and Lee-Yang parameters $\alpha^\prime$, $\beta^\prime$, and $\gamma^\prime$ are computed as
\begin{align}
&\Gamma = \frac{p_c}{8\pi}\frac{(m_{\mathcal{B}_b}+m_\mathcal{B})^2-m_M^2}
{m_{\mathcal{B}_b}^2}\left(|A|^2
+ \kappa^2|B|^2\right),\nonumber\\
& \alpha^\prime=\frac{2\kappa |A^*B|\cos(\delta_P-\delta_S)}{|A|^2+\kappa^2 |B|^2},~~
\beta^\prime=\frac{2\kappa |A^*B|\sin(\delta_P-\delta_S)}{|A|^2+\kappa^2 |B|^2},~~
\gamma^\prime=\frac{|A|^2-\kappa^2 |B|^2}{|A|^2+\kappa^2 |B|^2},
\end{align}
where $p_c$ is the center of momentum (CM) three-momentum in the rest frame of the initial baryon, and $\kappa$ is defined as $\kappa=p_c/(E_\mathcal{B}+m_\mathcal{B})=\sqrt{(E_\mathcal{B}-m_\mathcal{B})
/(E_\mathcal{B}+m_\mathcal{B})}$.
The isospin sum rules work for all partial waves.
Therefore, if two decay channels form an isospin sum rule, their branching fractions are proportional, and their decay asymmetries $\alpha^\prime$, $\beta^\prime$, and $\gamma^\prime$ are identical.
One can use the isospin sum rules and experimental data to test isospin symmetry and predict the branching fractions and Lee-Yang parameters of unobserved decay modes.
If three decay channels form an isospin sum rule, their decay amplitudes form a triangle in the complex plane.
One can use the isospin triangle to extract the relative strong phases between different modes.
For three-body decays, the resonance states do not violate the isospin sum rules. The isospin sum rules can be used to study the isospin multiplets of exotic hadrons in bottom baryon decays.

According to the isospin sum rule
\begin{align}\label{rule1}
SumI_-[\Lambda_b^0,p,J/\psi,\overline{K}^0] &=\mathcal{A}(\Lambda_b^0\to nJ/\psi\overline{K}^0)-\mathcal{A}(\Lambda_b^0\to pJ/\psi K^-)=0,
\end{align}
and the branching fraction of $\Lambda_b^0\to pJ/\psi K^-$ decay \cite{PDG}, we predict that the branching fraction of the $\Lambda_b^0\to nJ/\psi\overline{K}^0$ decay is
\begin{align}
\mathcal{B}r(\Lambda_b^0\to nJ/\psi\overline{K}^0) \simeq \mathcal{B}r(\Lambda_b^0\to pJ/\psi K^-)=(3.2^{+0.6}_{-0.5})\times 10^{-4}.
\end{align}
The LHCb collaboration reported pentaquark states $P_c(4312)^+$, $P_c(4440)^+$, and $P_c(4457)^+$ in the $\Lambda_b^0\to P_c^+(\to pJ/\psi) K^-$ decay \cite{LHCb:2015yax,LHCb:2016ztz,LHCb:2019kea}.
If the isospin symmetry is exact, the neutral isospin partners of these pentaquark states will contribute to the $\Lambda_b^0\to nJ/\psi \overline{K}^0$ decay with $\Lambda_b^0\to P_c^0(\to nJ/\psi) \overline{K}^0$.
However, the isospin breaking is naively predicted to be $\mathcal{O}(1\%)$, which is comparable to the relative mass differences and decay widths of the three pentaquark states.
Therefore, it is an uncertain possibility that three neutral pentaquark states contributing to the $\Lambda_b^0\to nJ/\psi \overline{K}^0$ decay.

The recent LHCb measurement found that \cite{LHCb:2025lhk}
\begin{align}
\frac{\mathcal{B}r(\Lambda_b^0\to\Xi^-J/\psi K^+)}
{\mathcal{B}r(\Lambda_b^0\to\Lambda^0 J/\psi)}=(1.17\pm 0.14\pm 0.08)\times 10^{-2}.
\end{align}
According to the isospin sum rule
\begin{align}
SumI_-[\Lambda_b^0,\Xi^0,J/\psi,K^+] &=\mathcal{A}(\Lambda_b^0\to\Xi^0J/\psi K^0)-\mathcal{A}(\Lambda_b^0\to\Xi^-J/\psi K^+)=0,
\end{align}
the ratio $\mathcal{B}r(\Lambda_b^0\to\Xi^0J/\psi K^0) /\mathcal{B}r(\Lambda_b^0\to\Lambda^0 J/\psi)$ is predicted to be $(1.17\pm 0.16)\times 10^{-2}$.
According to the following isospin sum rules given in Appendix \ref{I3cc},
\begin{align}
SumI_-^2[\Xi_b^-,\Xi^0,J/\psi,\pi^+] =&\,2\big[\sqrt{2}\mathcal{A}(\Xi_b^0\to\Xi^0J/\psi\pi^0)+\mathcal{A}(\Xi_b^0\to\Xi^-J/\psi\pi^+) \notag \\ &-\mathcal{A}(\Xi_b^-\to\Xi^0J/\psi\pi^-)+\sqrt{2}\mathcal{A}(\Xi_b^-\to\Xi^-J/\psi\pi^0)\big]=0,
\end{align}
\begin{align}
SumI_-[\Xi_b^0,\Xi^0,J/\psi,\pi^+] &= -\sqrt{2}\mathcal{A}(\Xi_b^0\to\Xi^0J/\psi\pi^0)-\mathcal{A}(\Xi_b^0\to\Xi^-J/\psi\pi^+)=0,
\end{align}
\begin{align}
SumI_-[\Xi_b^-,\Xi^-,J/\psi,\pi^+] &=-\mathcal{A}(\Xi_b^0\to\Xi^-J/\psi\pi^+)-\sqrt{2}\mathcal{A}(\Xi_b^-\to\Xi^-J/\psi\pi^0)=0,
\end{align}
\begin{align}
SumI_-[\Xi_b^-,\Xi^0,J/\psi,\pi^0] =&-\mathcal{A}(\Xi_b^0\to\Xi^0J/\psi\pi^0)+\sqrt{2}\mathcal{A}(\Xi_b^-\to\Xi^0J/\psi\pi^-) \notag \\ &-\mathcal{A}(\Xi_b^-\to\Xi^-J/\psi\pi^0)=0,
\end{align}
we obtain
\begin{align}
&\mathcal{A}(\Xi_b^-\to\Xi^0J/\psi\pi^-) = \sqrt{2}\mathcal{A}(\Xi_b^-\to\Xi^-J/\psi\pi^0) \nonumber \\&~~~~~ = \sqrt{2}\mathcal{A}(\Xi_b^0\to\Xi^0J/\psi\pi^0) = -\mathcal{A}(\Xi_b^0\to\Xi^-J/\psi\pi^+).
\end{align}
The recent LHCb measurement showed that \cite{LHCb:2025lhk}
\begin{align}
\frac{\mathcal{B}r(\Xi_b^0\to\Xi^-J/\psi\pi^+)}
{\mathcal{B}r(\Xi_b^-\to\Xi^-J/\psi)}=(11.9\pm 1.4\pm 0.6)\times 10^{-2}.
\end{align}
The ratios between other $\Xi_b\to\Xi J/\psi\pi$ modes and $\Xi_b^-\to\Xi^-J/\psi$ are predicted to be
\begin{align}
\frac{\mathcal{B}r(\Xi_b^0\to\Xi^0J/\psi\pi^0)}
{\mathcal{B}r(\Xi_b^-\to\Xi^-J/\psi)}&=(6.0\pm 0.8)\times 10^{-2},\qquad \frac{\mathcal{B}r(\Xi_b^-\to\Xi^0J/\psi\pi^-)}
{\mathcal{B}r(\Xi_b^-\to\Xi^-J/\psi)}=(12.6\pm 1.6)\times 10^{-2},\nonumber\\
\frac{\mathcal{B}r(\Xi_b^-\to\Xi^-J/\psi\pi^0)}
{\mathcal{B}r(\Xi_b^-\to\Xi^-J/\psi)}&=(6.3\pm 0.8)\times 10^{-2}.
\end{align}

\section{Summary}\label{SUM}
Flavor symmetry is a powerful tool for analyzing the weak decays of heavy hadrons.
Isospin symmetry is the most precise flavor symmetry.
In this work, we derive isospin sum rules for the two- and three-body non-leptonic decays of bottom baryons using a systematic approach.
The isospin sum rules can be used to test isospin symmetry and provide hints about new decay modes and the isospin partners of exotic hadrons.

\begin{acknowledgements}

This work was supported in part by the National Natural Science Foundation of China under Grants No. 12105099.

\end{acknowledgements}

\begin{appendix}

%%%%%%%%%%%%%%%%%%%%%%%%%%%%%%%
\section{Coefficient matrices derived by $I_-$ acting on states}\label{cmsda}

The bottom-baryon anti-triplet is defined as
\begin{eqnarray}
 |\mathcal{B}_{b\overline 3}\rangle=  \left( \begin{array}{ccc}
   0   & \Lambda_b^0  & \Xi_b^0 \\
    -\Lambda_b^0 &   0   & \Xi_b^- \\
    -\Xi_b^0 & -\Xi_b^- & 0 \\
  \end{array}\right),
\end{eqnarray}
which can be expressed using the Levi-Civita tensor as
\begin{eqnarray}
|\mathcal{B}_{b\overline 3}\rangle_{ij}=\epsilon_{ijk}|\mathcal{B}_{b\overline 3}\rangle^{k}\qquad {\rm with}\qquad |\mathcal{B}_{b\overline 3}\rangle^{k}=\left( \begin{array}{ccc}
     \Xi_b^- \\
    -\Xi_b^0  \\
    \Lambda_b^0 \\
  \end{array}\right).
\end{eqnarray}
If we use the beauty baryon anti-triplet basis as
 $|[\mathcal{B}_{b\overline 3}]_\alpha\rangle = ( |\Xi^-_b\rangle,\,\, |\Xi^0_b\rangle ,\,\, |\Lambda^0_b\rangle )$,
 the coefficient matrix $[I_-]_{\mathcal{B}_{b\overline 3}}$ is
\begin{eqnarray}\label{mbb}
 [I_-]_{\mathcal{B}_{b\overline 3}}= \left( \begin{array}{ccc}
   0   & 0  & 0 \\
     -1 &  0  & 0 \\
    0 & 0 & 0 \\
  \end{array}\right).
\end{eqnarray}
If we use the charmed baryon anti-triplet basis as
 $|[\mathcal{B}_{c\overline 3}]_\beta\rangle = ( |\Xi^0_c\rangle,\,\, |\Xi^+_c\rangle ,\,\, |\Lambda^+_c\rangle )$,
the coefficient matrix $[I_-]_{\mathcal{B}_{c\overline 3}}$ is
a transposition of Eq.~\eqref{mbb}, $[I_-]_{\mathcal{B}_{c\overline 3}} = [I_-]^T_{\mathcal{B}_{b\overline 3}}$, since the anti-triplet in the final state can be regarded as triplet in the initial state.
The charm meson anti-triplet and triplet are $|D\rangle = ( |D^0\rangle,\,\, |D^+\rangle ,\,\, |D^+_s\rangle )$ and $|\overline D\rangle = ( |\overline D^0\rangle,\,\, |D^-\rangle ,\,\, |D^-_s\rangle )$, respectively.
The coefficient matrix $[I_-]_{ D}$ is derived as
\begin{eqnarray}\label{MD}
 [I_-]_{ D}= \left( \begin{array}{ccc}
   0   & 1  & 0 \\
     0 &  0  & 0 \\
    0 & 0 & 0 \\
  \end{array}\right),
\end{eqnarray}
and $[I_-]_{ \overline D} = [I_-]_{D}^T$.

The pseudoscalar meson octet is expressed as
\begin{eqnarray}\label{a1}
 |M_8\rangle =  \left( \begin{array}{ccc}
   \frac{1}{\sqrt 2} \pi^0 +  \frac{1}{\sqrt 6} \eta_8,    & \pi^+,  & K^+ \\
    \pi^-, &   - \frac{1}{\sqrt 2} \pi^0+ \frac{1}{\sqrt 6} \eta_8,   & K^0 \\
  K^- ,& \overline K^0, & -\sqrt{2/3}\eta_8 \\
  \end{array}\right).
\end{eqnarray}
If the basis of pseudoscalar meson octet is defined as
\begin{align}
 \langle [M_8]_\alpha| = ( \langle \pi^+|,\,\,\langle \pi^0|,\,\,\langle \pi^-|,\,\,\langle K^+|,\,\,\langle K^0|,\,\,\langle \overline K^0|,\,\,\langle K^-|,\,\,\langle \eta_8|    ),
\end{align}
the coefficient matrix $[I_-]_{M_8}$ is
\begin{eqnarray}\label{MP}
 [I_-]_{M_8}= \left( \begin{array}{cccccccc}
  0 & 0& 0& 0& 0& 0& 0& 0 \\
  -\sqrt{2}& 0& 0& 0& 0& 0& 0& 0 \\
 0& \sqrt{2}& 0& 0& 0& 0& 0& 0 \\
  0& 0& 0& 0& 0& 0& 0& 0 \\
  0& 0& 0& 1& 0& 0& 0& 0\\
 0& 0& 0& 0& 0& 0& 0& 0\\
 0& 0& 0& 0& 0& -1& 0& 0 \\
 0& 0& 0& 0& 0&0& 0& 0 \\
  \end{array}\right).
\end{eqnarray}
The light baryon octet is
\begin{eqnarray}
 |\mathcal{B}_8\rangle=  \left( \begin{array}{ccc}
 \frac{1}{\sqrt 2} \Sigma^0+  \frac{1}{\sqrt 6} \Lambda^0    & \Sigma^+  & p \\
 \Sigma^- &   - \frac{1}{\sqrt 2} \Sigma^0+ \frac{1}{\sqrt 6} \Lambda^0   & n \\
 \Xi^- & \Xi^0 & -\sqrt{2/3}\Lambda^0 \\
  \end{array}\right).
\end{eqnarray}
Matrix $ [I_-]_{\mathcal{B}_8}$ is the same as $[I_-]_{M_8}$ if the light baryon octet basis is defined as
\begin{align}
 \langle [\mathcal{B}_8]_\beta| = ( \langle \Sigma^+|,\,\,\langle \Sigma^0|,\,\,\langle \Sigma^-|,\,\,\langle p|,\,\,\langle n|,\,\,\langle \Xi^0|,\,\,\langle \Xi^-|,\,\,\langle \Lambda^0|).
\end{align}

The charmed baryon sextet is
\begin{eqnarray}
 |\mathcal{B}_{c6}\rangle=  \left( \begin{array}{ccc}
   \Sigma_c^{++}   &  \frac{1}{\sqrt{2}}\Sigma_c^{+}  & \frac{1}{\sqrt{2}}\Xi_c^{*+} \\
   \frac{1}{\sqrt{2}}\Sigma_c^{+} &   \Sigma_c^{0}   & \frac{1}{\sqrt{2}}\Xi_c^{*0} \\
    \frac{1}{\sqrt{2}}\Xi_c^{*+} & \frac{1}{\sqrt{2}}\Xi_c^{*0} & \Omega_c^0\\
  \end{array}\right).
\end{eqnarray}
If we define the charmed baryon sextet basis as
 $\langle[\mathcal{B}_{c6}]_\beta |= ( \langle\Sigma_{c}^{++}|,\,\, \langle\Sigma_{c}^{0}|,\,\, \langle\Omega_{c}^{0}|, \,\, \langle\Sigma_{c}^{+}|,\,\, \langle\Xi_{c}^{*+}|,\,\, \langle\Xi_{c}^{*0}| )$,
the coefficient matrix $[I_-]_{\mathcal{B}_{c6}}$ is derived as
\begin{eqnarray}
 [I_-]_{\mathcal{B}_{c6}}= \left( \begin{array}{cccccc}
  0 & 0& 0& 0& 0& 0 \\
  0 & 0& 0& \sqrt{2}& 0& 0 \\
  0 & 0& 0& 0& 0& 0 \\
  \sqrt{2} & 0& 0& 0& 0& 0 \\
  0 & 0& 0& 0& 0& 0 \\
  0 & 0& 0& 0& 1& 0 \\
  \end{array}\right).
\end{eqnarray}
The light baryon decuplet is given by
{\small \begin{align}\label{b10}
|\mathcal{B}_{10}\rangle = \left(\left( \begin{array}{ccc}
      \Delta^{++} &  \frac{1}{\sqrt{3}}\Delta^{+}  & \frac{1}{\sqrt{3}}\Sigma^{*+} \\
   \frac{1}{\sqrt{3}}\Delta^{+} &  \frac{1}{\sqrt{3}}\Delta^{0}  & \frac{1}{\sqrt{6}}\Sigma^{*0} \\
    \frac{1}{\sqrt{3}}\Sigma^{*+} & \frac{1}{\sqrt{6}}\Sigma^{*0} & \frac{1}{\sqrt{3}}\Xi^{*0} \\
  \end{array}\right)\left( \begin{array}{ccc}\frac{1}{\sqrt{3}}\Delta^{+} &  \frac{1}{\sqrt{3}}\Delta^{0}  & \frac{1}{\sqrt{6}}\Sigma^{*0} \\
   \frac{1}{\sqrt{3}}\Delta^{0} & \Delta^{-}  & \frac{1}{\sqrt{3}}\Sigma^{*-} \\
    \frac{1}{\sqrt{6}}\Sigma^{*0} & \frac{1}{\sqrt{3}}\Sigma^{*-} & \frac{1}{\sqrt{3}}\Xi^{*-}\\\end{array}\right)
    \left(\begin{array}{ccc}\frac{1}{\sqrt{3}}\Sigma^{*+} &  \frac{1}{\sqrt{6}}\Sigma^{*0}  & \frac{1}{\sqrt{3}}\Xi^{*0} \\
   \frac{1}{\sqrt{6}}\Sigma^{*0} & \frac{1}{\sqrt{3}}\Sigma^{*-}  & \frac{1}{\sqrt{3}}\Xi^{*-} \\
    \frac{1}{\sqrt{3}}\Xi^{*0} & \frac{1}{\sqrt{3}}\Xi^{*-} & \Omega^{-}\\\end{array}\right)\right).
\end{align}}
If the light baryon decuplet basis is defined as
\begin{align}
 \langle [\mathcal{B}_{10}]_\beta| = ( \langle \Delta^{++}|,\,\,\langle \Delta^+|,\,\,\langle \Delta^0|,\,\,\langle \Delta^-|,\,\,\langle \Sigma^{*+}|,\,\,\langle \Sigma^{*0}|,\,\,\langle \Sigma^{*-}|,\,\,\langle \Xi^{*0}|,\,\,\langle \Xi^{*-}|,\,\,\langle \Omega^{-}|),
\end{align}
the coefficient matrix $[I_-]_{\mathcal{B}_{10}}$ is derived as
\begin{eqnarray}
 [I_-]_{\mathcal{B}_{10}}= \left( \begin{array}{cccccccccc}
  0 & 0& 0& 0& 0& 0 &0 &0 &0&0\\
   \sqrt{3}& 0& 0& 0& 0& 0& 0&0&0&0\\
  0 & 2& 0&0& 0& 0& 0&0&0&0 \\
 0&0& \sqrt{3} & 0&0& 0& 0& 0& 0& 0 \\
 0& 0& 0& 0&0 & 0& 0& 0& 0& 0 \\
  0 & 0& 0& 0& \sqrt{2} & 0& 0&0& 0&0 \\
  0& 0& 0& 0&0 & \sqrt{2}&0&  0& 0&0\\0& 0& 0&0& 0& 0&0& 0& 0&0\\0& 0& 0&0& 0& 0& 0&1&0& 0\\ 0 & 0& 0& 0& 0& 0 &0 &0 &0&0
  \end{array}\right).
\end{eqnarray}
$J/\Psi$ is the isospin singlet and then $ [I_-]_{ J/\Psi} = 0$.

%%%%%%%%%%%%%%%%%%%%%%%%%%%%%%
\section{Isospin sum rules for two-body decays}\label{I2}

\subsection{$b\to c\overline c d/s$ modes}\label{I2cc}

\begin{align}
{ SumI_-}\,[\Xi^-_b, \Xi^0, J/\Psi]=-\big[\,\mathcal{A}( \Xi^0_b\to\Xi^0J/\Psi)+\mathcal{A}(\Xi^-_b\to\Xi^-J/\Psi)\big]=0,
\end{align}
\begin{align}
{ SumI_-}\,[\Xi^-_b, \Xi^{*0}, J/\Psi]=-\mathcal{A}( \Xi^0_b\to\Xi^{*0}J/\Psi)+\mathcal{A}(\Xi^-_b\to\Xi^{*-}J/\Psi)=0,
\end{align}
\begin{align}
{ SumI_-}\,[\Lambda^0_b, \Sigma^{*+}, J/\Psi]=\sqrt{2}\,\mathcal{A}( \Lambda^0_b\to\Sigma^{*0}J/\Psi)=0,
\end{align}
\begin{align}
{ SumI_-}\,[\Xi^-_b, \Xi^{+}_c, D^-_s]=-\mathcal{A}( \Xi^0_b\to\Xi^{+}_c D^-_s)+\mathcal{A}( \Xi^-_b\to\Xi^{0}_c D^-_s)=0,
\end{align}
\begin{align}
{ SumI_-}\,[\Lambda^0_b, \Xi^{+}_c, \overline D^0]=\mathcal{A}( \Lambda^0_b\to\Xi^{0}_c \overline D^0)+\mathcal{A}( \Lambda^0_b\to\Xi^{+}_c D^-)=0,
\end{align}
\begin{align}
{ SumI_-}\,[\Xi^-_b, \Omega^{0}_c, \overline D^0]=-\mathcal{A}( \Xi^0_b\to\Omega^{0}_c \overline D^0)+\mathcal{A}(\Xi^-_b\to\Omega^{0}_c  D^-)=0,
\end{align}
\begin{align}
{ SumI_-}\,[\Xi^-_b, \Xi^{*+}_c,  D^-_s]=-\mathcal{A}( \Xi^0_b\to\Xi^{*+}_c D^-_s)+\mathcal{A}(\Xi^-_b\to\Xi^{*0}_c D^-_s)=0,
\end{align}
\begin{align}
{ SumI_-}\,[\Lambda^0_b, \Sigma^{++}_c, D^-_s]=\sqrt{2}\,\mathcal{A}( \Lambda^0_b\to\Sigma^{+}_c D^-_s)=0,
\end{align}
\begin{align}
{ SumI_-}\,[\Lambda^0_b, \Xi^{*+}_c, \overline D^0]=\mathcal{A}( \Lambda^0_b\to\Xi^{*0}_c \overline D^0)+\mathcal{A}( \Lambda^0_b\to\Xi^{*+}_c D^-)=0.
\end{align}

\subsection{$b\to c\overline u d/s$ modes}

\begin{align}
{ SumI_-^3}\,[\Xi^-_b, \Sigma^{+},  D^+]=-6\,\big[\sqrt{2}\,\mathcal{A}( \Xi^0_b\to\Sigma^{0} D^0)-\mathcal{A}(\Xi^0_b\to\Sigma^{-} D^+)-\mathcal{A}(\Xi^-_b\to\Sigma^{-} D^0)\big]=0,
\end{align}
\begin{align}
{ SumI_-^2}\,[\Xi^-_b, \Xi^{0},  D^+]=2\,\big[\,\mathcal{A}( \Xi^0_b\to\Xi^{0} D^0)+\mathcal{A}(\Xi^0_b\to\Xi^{-} D^+)+\mathcal{A}(\Xi^-_b\to\Xi^{-} D^0)\big]=0,
\end{align}
\begin{align}
{ SumI_-^2}\,[\Lambda^0_b, \Sigma^{+},  D^+]=2\,\big[\sqrt{2}\,\mathcal{A}( \Lambda^0_b\to\Sigma^{0} D^0)-\mathcal{A}(\Lambda^0_b\to\Sigma^{-} D^+)\big]=0,
\end{align}
\begin{align}
{ SumI_-^3}\,[\Xi^-_b, \Sigma^{*+},  D^+]=6\,\big[\sqrt{2}\,\mathcal{A}( \Xi^0_b\to\Sigma^{*0} D^0)-\mathcal{A}(\Xi^0_b\to\Sigma^{*-} D^+)-\mathcal{A}(\Xi^-_b\to\Sigma^{*-} D^0)\big]=0,
\end{align}
\begin{align}
{ SumI_-^3}\,[\Lambda^0_b, \Delta^{++},  D^+]=6\,\big[-\sqrt{3}\,\mathcal{A}( \Lambda^0_b\to\Delta^{0} D^0)+\mathcal{A}(\Lambda^0_b\to\Delta^{-} D^+)\big]=0,
\end{align}
\begin{align}
{ SumI_-^2}\,[\Xi^-_b, \Xi^{*0},  D^+]=2\,\big[\,\mathcal{A}( \Xi^0_b\to\Xi^{*0} D^0)-\mathcal{A}(\Xi^0_b\to\Xi^{*-} D^+)-\mathcal{A}(\Xi^-_b\to\Xi^{*-} D^0)\big]=0,
\end{align}
\begin{align}
{ SumI_-^2}\,[\Lambda^0_b, \Sigma^{*+},  D^+]=-2\,\big[\sqrt{2}\,\mathcal{A}( \Lambda^0_b\to\Sigma^{*0} D^0)-\mathcal{A}(\Lambda^0_b\to\Sigma^{*-} D^+)\big]=0,
\end{align}
\begin{align}
{ SumI_-^3}\,[\Xi^-_b, \Xi^{+}_c,  \pi^+]=6\,\big[\sqrt{2}\,\mathcal{A}( \Xi^0_b\to\Xi^{0}_c \pi^0)+\mathcal{A}(\Xi^0_b\to\Xi^{+}_c \pi^-)-\mathcal{A}(\Xi^-_b\to\Xi^{0}_c \pi^-)\big]=0,
\end{align}
\begin{align}
{ SumI_-^2}\,[\Xi^-_b, \Xi^{+}_c,  \overline K^0]=2\,\big[\,\mathcal{A}( \Xi^0_b\to\Xi^{+}_c K^-)-\mathcal{A}(\Xi^0_b\to\Xi^{0}_c \overline K^0)-\mathcal{A}(\Xi^-_b\to\Xi^{0}_c K^-)\big]=0,
\end{align}
\begin{align}
{ SumI_-^2}\,[\Lambda^0_b, \Xi^{+}_c,  \pi^+]=-2\,\big[\sqrt{2}\,\mathcal{A}( \Lambda^0_b\to\Xi^{0}_c \pi^0)+\mathcal{A}(\Lambda^0_b\to\Xi^{+}_c \pi^-)\big]=0,
\end{align}
\begin{align}
{ SumI_-^3}\,[\Xi^-_b, \Sigma^{++}_c,  \overline K^0]=6\,\big[\sqrt{2}\,\mathcal{A}( \Xi^0_b\to\Sigma^{+}_c K^-)-\mathcal{A}(\Xi^0_b\to\Sigma^{0}_c \overline K^0)-\mathcal{A}(\Xi^-_b\to\Sigma^{0}_c K^-)\big]=0,
\end{align}
\begin{align}
{ SumI_-^3}\,[\Xi^-_b, \Xi^{*+}_c,  \pi^+]=6\,\big[\sqrt{2}\,\mathcal{A}( \Xi^0_b\to\Xi^{*0}_c \pi^0)+\mathcal{A}(\Xi^0_b\to\Xi^{*+}_c \pi^-)-\mathcal{A}(\Xi^-_b\to\Xi^{*0}_c \pi^-)\big]=0,
\end{align}
\begin{align}
{ SumI_-^3}\,[\Lambda^0_b, \Sigma^{++}_c,  \pi^+]=-6\sqrt{2}\,\big[\,\mathcal{A}( \Lambda^0_b\to\Sigma^{0}_c \pi^0)+\mathcal{A}(\Lambda^0_b\to\Sigma^{+}_c \pi^-)\big]=0,
\end{align}
\begin{align}
{ SumI_-^2}\,[\Xi^-_b, \Omega^{0}_c,  \pi^+]=2\,\big[\sqrt{2}\,\mathcal{A}( \Xi^0_b\to\Omega^{0}_c \pi^0)-\mathcal{A}(\Xi^-_b\to\Omega^{0}_c \pi^-)\big]=0,
\end{align}
\begin{align}
{ SumI_-^2}\,[\Xi^-_b, \Xi^{*+}_c,  \overline K^0]=-2\,\big[\,\mathcal{A}(\Xi^0_b\to\Xi^{*0}_c \overline K^0)-\mathcal{A}( \Xi^0_b\to\Xi^{*+}_c K^-)+\mathcal{A}(\Xi^-_b\to\Xi^{*0}_c K^-)\big]=0,
\end{align}
\begin{align}
{ SumI_-^2}\,[\Lambda^0_b, \Sigma^{++}_c,  \overline K^0]=-2\,\big[\sqrt{2}\,\mathcal{A}( \Lambda^0_b\to\Sigma^{+}_c K^-)-\mathcal{A}(\Lambda^0_b\to\Sigma^{0}_c \overline K^0)\big]=0,
\end{align}
\begin{align}
{ SumI_-^2}\,[\Lambda^0_b, \Xi^{*+}_c,  \pi^+]=-2\,\big[\sqrt{2}\,\mathcal{A}( \Lambda^0_b\to\Xi^{*0}_c \pi^0)+\mathcal{A}(\Lambda^0_b\to\Xi^{*+}_c \pi^-)\big]=0.
\end{align}

\subsection{$b\to u\overline u d/s$ modes}

\begin{align}
{ SumI_-^3}\,[\Xi^-_b, \Sigma^{+},  \pi^+]=&6\,\big[-2\,\mathcal{A}( \Xi^0_b\to\Sigma^{0} \pi^0)+\mathcal{A}(\Xi^0_b\to\Sigma^{-} \pi^+)+\mathcal{A}(\Xi^0_b\to\Sigma^{+} \pi^-)\nonumber\\&+\sqrt{2}\,\mathcal{A}(\Xi^-_b\to\Sigma^{0} \pi^-)+\sqrt{2}\,\mathcal{A}(\Xi^-_b\to\Sigma^{-} \pi^0)\big]=0,
\end{align}
\begin{align}\label{r5}
{ SumI_-^2}\,[\Xi^-_b, \Sigma^{+},  \overline K^0]=&2\,\big[\sqrt{2}\,\mathcal{A}( \Xi^0_b\to\Sigma^{0} \overline K^0)+\mathcal{A}(\Xi^0_b\to\Sigma^{+} K^-)\nonumber\\&+\sqrt{2}\,\mathcal{A}(\Xi^-_b\to\Sigma^{0} K^-)-\mathcal{A}(\Xi^-_b\to\Sigma^{-}\overline K^0)\big]=0,
\end{align}
\begin{align}\label{r6}
{ SumI_-^2}\,[\Xi^-_b, \Xi^{0},  \pi^+]=&2\,\big[\sqrt{2}\,\mathcal{A}( \Xi^0_b\to\Xi^{0} \pi^0)+\mathcal{A}(\Xi^0_b\to\Xi^{-} \pi^+)\nonumber\\&+\sqrt{2}\,\mathcal{A}(\Xi^-_b\to\Xi^{-} \pi^0)-\mathcal{A}(\Xi^-_b\to\Xi^{0} \pi^-)\big]=0,
\end{align}
\begin{align}\label{r1}
{ SumI_-^2}\,[\Lambda^0_b, \Sigma^{+},  \pi^+]=&2\,\big[2\,\mathcal{A}( \Lambda^0_b\to\Sigma^{0} \pi^0)-\mathcal{A}(\Lambda^0_b\to\Sigma^{-} \pi^+)-\mathcal{A}(\Lambda^0_b\to\Sigma^{+} \pi^-)\big]=0,
\end{align}
\begin{align}
{ SumI_-^3}\,[\Xi^-_b, \Delta^{++},  \overline K^0]=&6\,\big[-\sqrt{3}\,\mathcal{A}( \Xi^0_b\to\Delta^{0} \overline K^0)+\sqrt{3}\,\mathcal{A}(\Xi^0_b\to\Delta^{+} K^-)\nonumber\\&-\sqrt{3}\,\mathcal{A}(\Xi^-_b\to\Delta^{0} K^-)+\mathcal{A}(\Xi^-_b\to\Delta^{-}\overline K^0)\big]=0,
\end{align}
\begin{align}
{ SumI_-^3}\,[\Xi^-_b, \Sigma^{*+},  \pi^+]=&6\,\big[2\,\mathcal{A}( \Xi^0_b\to\Sigma^{*0} \pi^0)-\mathcal{A}(\Xi^0_b\to\Sigma^{*-} \pi^+)-\mathcal{A}(\Xi^0_b\to\Sigma^{*+} \pi^-)\nonumber\\&-\sqrt{2}\,\mathcal{A}(\Xi^-_b\to\Sigma^{*0} \pi^-)-\sqrt{2}\,\mathcal{A}(\Xi^-_b\to\Sigma^{*-} \pi^0)\big]=0,
\end{align}
\begin{align}\label{r2}
{ SumI_-^3}\,[\Lambda^0_b, \Delta^{++}, \pi^+]=&-6\,\big[\sqrt{6}\,\mathcal{A}(\Lambda^0_b\to\Delta^{0}  \pi^0)+\sqrt{3}\,\mathcal{A}(\Lambda^0_b\to\Delta^{+} \pi^-)-\mathcal{A}(\Lambda^0_b\to\Delta^{-} \pi^+)\big]=0,
\end{align}
\begin{align}
{ SumI_-^2}\,[\Xi^-_b, \Sigma^{*+},  \overline K^0]=&2\,\big[-\sqrt{2}\,\mathcal{A}( \Xi^0_b\to\Sigma^{*0} \overline K^0)+\mathcal{A}(\Xi^0_b\to\Sigma^{*+} K^-)\nonumber\\&-\sqrt{2}\,\mathcal{A}(\Xi^-_b\to\Sigma^{*0} K^-)+\mathcal{A}(\Xi^-_b\to\Sigma^{*-}\overline K^0)\big]=0,
\end{align}
\begin{align}
{ SumI_-^2}\,[\Xi^-_b, \Xi^{*0},  \pi^+]=&2\,\big[\sqrt{2}\,\mathcal{A}( \Xi^0_b\to\Xi^{*0} \pi^0)-\mathcal{A}(\Xi^0_b\to\Xi^{*-} \pi^+)\nonumber\\&-\sqrt{2}\,\mathcal{A}(\Xi^-_b\to\Xi^{-} \pi^0)-\mathcal{A}(\Xi^-_b\to\Xi^{*0} \pi^-)\big]=0,
\end{align}
\begin{align}\label{r3}
{ SumI_-^2}\,[\Lambda^0_b, \Delta^{++}, \overline K^0]=&2\sqrt{3}\,\big[\,\mathcal{A}( \Lambda^0_b\to\Delta^{0} \overline K^0)-\mathcal{A}(\Lambda^0_b\to\Delta^{+} K^-)\big]=0,
\end{align}
\begin{align}\label{r4}
{ SumI_-^2}\,[\Lambda^0_b, \Sigma^{*+},  \pi^+]=&2\,\big[-2\,\mathcal{A}( \Lambda^0_b\to\Sigma^{*0} \pi^0)+\mathcal{A}(\Lambda^0_b\to\Sigma^{*-} \pi^+)+\mathcal{A}(\Lambda^0_b\to\Sigma^{*+} \pi^-)\big]=0.
\end{align}

\subsection{$b\to u\overline c d/s$ modes}

\begin{align}
{ SumI_-^2}\,[\Xi^-_b, \Sigma^{+}, \overline D^0]=&2\,\big[\sqrt{2}\,\mathcal{A}( \Xi^0_b\to\Sigma^{0}\overline D^0)-\mathcal{A}( \Xi^0_b\to\Sigma^{+} D^-)\nonumber\\&-\sqrt{2}\,\mathcal{A}( \Xi^-_b\to\Sigma^{0} D^-)-\mathcal{A}( \Xi^-_b\to\Sigma^{-} \overline D^0)\big]=0,
\end{align}
\begin{align}
{ SumI_-}\,[\Xi^-_b, \Sigma^{+}, D^-_s]=-\big[\mathcal{A}( \Xi^0_b\to\Sigma^{+} D^-_s)+\sqrt{2}\mathcal{A}( \Xi^-_b\to\Sigma^{0} D^-_s)\big]=0,
\end{align}
\begin{align}
{ SumI_-}\,[\Xi^-_b, \Xi^{0}, \overline D^0]=-\mathcal{A}( \Xi^0_b\to\Xi^{0}\overline D^0)+\mathcal{A}( \Xi^-_b\to\Xi^{0} D^-)-\mathcal{A}( \Xi^-_b\to\Xi^{-} \overline D^0)=0,
\end{align}
\begin{align}
{ SumI_-}\,[\Lambda^0_b, \Sigma^{+}, \overline D^0]=-\sqrt{2}\,\mathcal{A}( \Lambda^0_b\to\Sigma^{0}\overline D^0)+\mathcal{A}( \Lambda^0_b\to\Sigma^{+} D^-)=0,
\end{align}
\begin{align}
{ SumI_-^2}\,[\Xi^-_b, \Delta^{++}, D^-_s]=2\sqrt{3}\,\big[-\mathcal{A}( \Xi^0_b\to\Delta^{+} D^-_s)+\mathcal{A}( \Xi^-_b\to\Delta^{0} D^-_s)\big]=0,
\end{align}
\begin{align}
{ SumI_-^2}\,[\Xi^-_b, \Sigma^{*+}, \overline D^0]=&2\,\big[-\sqrt{2}\,\mathcal{A}( \Xi^0_b\to\Sigma^{*0}\overline D^0)-\mathcal{A}( \Xi^0_b\to\Sigma^{*+} D^-)\nonumber\\&+\sqrt{2}\,\mathcal{A}( \Xi^-_b\to\Sigma^{*0} D^-)+\mathcal{A}( \Xi^-_b\to\Sigma^{*-} \overline D^0)\big]=0,
\end{align}
\begin{align}
{ SumI_-^2}\,[\Lambda^0_b, \Delta^{++}, \overline D^0]=2\sqrt{3}\,\big[\,\mathcal{A}( \Lambda^0_b\to\Delta^{0}\overline D^0)+\mathcal{A}( \Lambda^0_b\to\Delta^{+} D^-)\big]=0,
\end{align}
\begin{align}
{ SumI_-}\,[\Xi^-_b, \Sigma^{*+}, D^-_s]=-\mathcal{A}( \Xi^0_b\to\Sigma^{*+} D^-_s)+\sqrt{2}\,\mathcal{A}( \Xi^-_b\to\Sigma^{*0} D^-_s)=0,
\end{align}
\begin{align}
{ SumI_-}\,[\Xi^-_b, \Xi^{*0}, \overline D^0]=-\mathcal{A}( \Xi^0_b\to\Xi^{*0}\overline D^0)+\mathcal{A}( \Xi^-_b\to\Xi^{*0} D^-)+\mathcal{A}( \Xi^-_b\to\Xi^{*-} \overline D^0)=0,
\end{align}
\begin{align}
{ SumI_-}\,[\Lambda^0_b, \Sigma^{*+}, \overline D^0]=\sqrt{2}\,\mathcal{A}( \Lambda^0_b\to\Sigma^{*0}\overline D^0)+\mathcal{A}( \Lambda^0_b\to\Sigma^{*+} D^-)=0,
\end{align}
\begin{align}
{ SumI_-}\,[\Lambda^0_b, \Delta^{++}, D^-_s]=\sqrt{3}\,\mathcal{A}( \Lambda^0_b\to\Delta^{+} D^-_s)=0.
\end{align}

\section{Isospin sum rules for three-body decays}\label{I3}

\subsection{$b\to c\overline c d/s$ modes}\label{I3cc}

\begin{align}
SumI_-^3[\Xi_b^-,\Xi_c^+,\overline{D}^0,\pi^+] &=-6\big[\mathcal{A}(\Xi_b^0\to\Xi_c^0D^-\pi^+)-\sqrt{2}\mathcal{A}(\Xi_b^0\to\Xi_c^0\overline{D}^0\pi^0) \notag \\ &-\sqrt{2}\mathcal{A}(\Xi_b^0\to\Xi_c^+D^-\pi^0)-\mathcal{A}(\Xi_b^0\to\Xi_c^+\overline{D}^0\pi^-) \notag \\ &+\sqrt{2}\mathcal{A}(\Xi_b^-\to\Xi_c^0D^-\pi^0)+\mathcal{A}(\Xi_b^-\to\Xi_c^0\overline{D}^0\pi^-) \notag \\ &+\mathcal{A}(\Xi_b^-\to\Xi_c^+D^-\pi^-)\big]=0,
\end{align}
\begin{align}
SumI_-^2[\Xi_b^-,\Xi_c^+,D_s^-,K^+] &=-2\big[\mathcal{A}(\Xi_b^0\to\Xi_c^0D_s^-K^+)+\mathcal{A}(\Xi_b^0\to\Xi_c^+D_s^-K^0) \notag \\ &-\mathcal{A}(\Xi_b^-\to\Xi_c^0D_s^-K^0)\big]=0,
\end{align}
\begin{align}
SumI_-[\Lambda_b^0,\Lambda_c^+,\overline{D}^0,\overline{K}^0] &=\mathcal{A}(\Lambda_b^0\to\Lambda_c^+D^-\overline{K}^0)-\mathcal{A}(\Lambda_b^0\to\Lambda_c^+\overline{D}^0K^-)=0,
\end{align}
\begin{align}
SumI_-[\Lambda_b^0,\Lambda_c^+,D_s^-,\pi^+] &=-\sqrt{2}\mathcal{A}(\Lambda_b^0\to\Lambda_c^+D_s^-\pi^0)=0,
\end{align}
\begin{align}
SumI_-[\Lambda_b^0,\Xi_c^+,\overline{D}^0,\eta_8] &=\mathcal{A}(\Lambda_b^0\to\Xi_c^0\overline{D}^0\eta_8)+\mathcal{A}(\Lambda_b^0\to\Xi_c^+D^-\eta_8)=0,
\end{align}
\begin{align}
SumI_-[\Lambda_b^0,\Xi_c^+,D_s^-,K^+] &=\mathcal{A}(\Lambda_b^0\to\Xi_c^0D_s^-K^+)+\mathcal{A}(\Lambda_b^0\to\Xi_c^+D_s^-K^0)=0,
\end{align}
\begin{align}
SumI_-[\Xi_b^-,\Lambda_c^+,D_s^-,\overline{K}^0] &=-\mathcal{A}(\Xi_b^0\to\Lambda_c^+D_s^-\overline{K}^0)-\mathcal{A}(\Xi_b^-\to\Lambda_c^+D_s^-K^-)=0,
\end{align}
\begin{align}
SumI_-^2[\Xi_b^-,\Xi_c^+,\overline{D}^0,\overline{K}^0] &=-2\big[\mathcal{A}(\Xi_b^0\to\Xi_c^0\overline{D}^0\overline{K}^0)+\mathcal{A}(\Xi_b^0\to\Xi_c^+D^-\overline{K}^0) \notag \\ &-\mathcal{A}(\Xi_b^0\to\Xi_c^+\overline{D}^0K^-)-\mathcal{A}(\Xi_b^-\to\Xi_c^0D^-\overline{K}^0) \notag \\ &+\mathcal{A}(\Xi_b^-\to\Xi_c^0\overline{D}^0K^-)+\mathcal{A}(\Xi_b^-\to\Xi_c^+D^-K^-)\big]=0,
\end{align}
\begin{align}
SumI_-^2[\Xi_b^-,\Xi_c^+,D_s^-,\pi^+] &=-2\big[\mathcal{A}(\Xi_b^0\to\Xi_c^0D_s^-\pi^+)-\sqrt{2}\mathcal{A}(\Xi_b^0\to\Xi_c^+D_s^-\pi^0) \notag  \\&+\sqrt{2}\mathcal{A}(\Xi_b^-\to\Xi_c^0D_s^-\pi^0)+\mathcal{A}(\Xi_b^-\to\Xi_c^+D_s^-\pi^-)\big]=0,
\end{align}
\begin{align}
SumI_-[\Xi_b^-,\Xi_c^+,D_s^-,\eta_8] &=-\mathcal{A}(\Xi_b^0\to\Xi_c^+D_s^-\eta_8)+\mathcal{A}(\Xi_b^-\to\Xi_c^0D_s^-\eta_8)=0,
\end{align}
\begin{align}
SumI_-[\Xi_b^0,\Xi_c^+,\overline{D}^0,\overline{K}^0] &=\mathcal{A}(\Xi_b^0\to\Xi_c^0\overline{D}^0\overline{K}^0)+\mathcal{A}(\Xi_b^0\to\Xi_c^+D^-\overline{K}^0) \notag \\ &-\mathcal{A}(\Xi_b^0\to\Xi_c^+\overline{D}^0K^-)=0,
\end{align}
\begin{align}
SumI_-[\Xi_b^-,\Xi_c^0,\overline{D}^0,\overline{K}^0] &=-\mathcal{A}(\Xi_b^0\to\Xi_c^0\overline{D}^0\overline{K}^0)+\mathcal{A}(\Xi_b^-\to\Xi_c^0D^-\overline{K}^0) \notag \\ &-\mathcal{A}(\Xi_b^-\to\Xi_c^0\overline{D}^0K^-)=0,
\end{align}
\begin{align}
SumI_-[\Xi_b^-,\Xi_c^+,D^-,\overline{K}^0] &=-\mathcal{A}(\Xi_b^0\to\Xi_c^+D^-\overline{K}^0)+\mathcal{A}(\Xi_b^-\to\Xi_c^0D^-\overline{K}^0) \notag \\ &-\mathcal{A}(\Xi_b^-\to\Xi_c^+D^-K^-)=0,
\end{align}
\begin{align}
SumI_-[\Xi_b^-,\Xi_c^+,\overline{D}^0,K^-] &=-\mathcal{A}(\Xi_b^0\to\Xi_c^+\overline{D}^0K^-)+\mathcal{A}(\Xi_b^-\to\Xi_c^0\overline{D}^0K^-) \notag \\&+\mathcal{A}(\Xi_b^-\to\Xi_c^+D^-K^-)=0,
\end{align}
\begin{align}
SumI_-[\Xi_b^0,\Xi_c^+,D_s^-,\pi^+] &=\mathcal{A}(\Xi_b^0\to\Xi_c^0D_s^-\pi^+)-\sqrt{2}\mathcal{A}(\Xi_b^0\to\Xi_c^+D_s^-\pi^0)=0,
\end{align}
\begin{align}
SumI_-[\Xi_b^-,\Xi_c^0,D_s^-,\pi^+] &=-\mathcal{A}(\Xi_b^0\to\Xi_c^0D_s^-\pi^+)-\sqrt{2}\mathcal{A}(\Xi_b^-\to\Xi_c^0D_s^-\pi^0)=0,
\end{align}
\begin{align}
SumI_-[\Xi_b^-,\Xi_c^+,D_s^-,\pi^0] &=-\mathcal{A}(\Xi_b^0\to\Xi_c^+D_s^-\pi^0)+\mathcal{A}(\Xi_b^-\to\Xi_c^0D_s^-\pi^0) \notag \\ &+\sqrt{2}\mathcal{A}(\Xi_b^-\to\Xi_c^+D_s^-\pi^-)=0,
\end{align}
\begin{align}
SumI_-^3[\Lambda_b^0,\Sigma_c^{++},\overline{D}^0,\pi^+] &=6\mathcal{A}(\Lambda_b^0\to\Sigma_c^0D^-\pi^+)-6\big[\sqrt{2}\mathcal{A}(\Lambda_b^0\to\Sigma_c^0\overline{D}^0\pi^0) \notag \\ &+2\mathcal{A}(\Lambda_b^0\to\Sigma_c^+D^-\pi^0)+\sqrt{2}\mathcal{A}(\Lambda_b^0\to\Sigma_c^+\overline{D}^0\pi^-) \notag \\ &+\mathcal{A}(\Lambda_b^0\to\Sigma_c^{++}D^-\pi^-)\big]=0,
\end{align}
\begin{align}
SumI_-^3[\Xi_b^-,\Sigma_c^{++},\overline{D}^0,\overline{K}^0] &=-6\big[\mathcal{A}(\Lambda_b^0\to\Sigma_c^0\overline{D}^0\overline{K}^0)+\sqrt{2}\mathcal{A}(\Xi_b^0\to\Sigma_c^+D^-\overline{K}^0) \notag \\ &-\sqrt{2}\mathcal{A}(\Xi_b^0\to\Sigma_c^+\overline{D}^0K^-)-\mathcal{A}(\Xi_b^0\to\Sigma_c^{++}D^-K^-) \notag \\ &-\mathcal{A}(\Xi_b^-\to\Sigma_c^0D^-\overline{K}^0)+\mathcal{A}(\Xi_b^-\to\Sigma_c^0\overline{D}^0K^-) \notag \\ &+\sqrt{2}\mathcal{A}(\Xi_b^-\to\Sigma_c^+D^-K^-)\big]=0,
\end{align}
\begin{align}
SumI_-^3[\Xi_b^-,\Sigma_c^{++},D_s^-,\pi^+] &=-6\big[\mathcal{A}(\Xi_b^0\to\Sigma_c^0D_s^-\pi^+)-2\mathcal{A}(\Xi_b^0\to\Sigma_c^+D_s^-\pi^0) \notag \\ &-\mathcal{A}(\Xi_b^0\to\Sigma_c^{++}D_s^-\pi^-)+\sqrt{2}\big(\mathcal{A}(\Xi_b^-\to\Sigma_c^0D_s^-\pi^0) \notag \\ &+\mathcal{A}(\Xi_b^-\to\Sigma_c^+D_s^-\pi^-)\big)\big]=0,
\end{align}
\begin{align}
SumI_-^3[\Xi_b^-,\Xi_c^{*+},\overline{D}^0,\pi^+] &=-6\big[\mathcal{A}(\Xi_b^0\to\Xi_c^{*0}D^-\pi^+)-\sqrt{2}\mathcal{A}(\Xi_b^0\to\Xi_c^{*0}\overline{D}^0\pi^0) \notag \\ &-\sqrt{2}\mathcal{A}(\Xi_b^0\to\Xi_c^{*+}D^-\pi^0)-\mathcal{A}(\Xi_b^0\to\Xi_c^{*+}\overline{D}^0\pi^-) \notag \\ &+\sqrt{2}\mathcal{A}(\Xi_b^-\to\Xi_c^{*0}D^-\pi^0)+\mathcal{A}(\Xi_b^-\to\Xi_c^{*0}\overline{D}^0\pi^-) \notag \\ &+\mathcal{A}(\Xi_b^-\to\Xi_c^{*+}D^-\pi^-)\big]=0,
\end{align}
\begin{align}
SumI_-^2[\Lambda_b^0,\Sigma_c^{++},\overline{D}^0,\overline{K}^0] &=2\big[\mathcal{A}(\Lambda_b^0\to\Sigma_c^0\overline{D}^0\overline{K}^0)+\sqrt{2}\mathcal{A}(\Lambda_b^0\to\Sigma_c^+D^-\overline{K}^0) \notag \\&-\sqrt{2}\mathcal{A}(\Lambda_b^0\to\Sigma_c^+\overline{D}^0K^-)-\mathcal{A}(\Lambda_b^0\to\Sigma_c^{++}D^-K^-)\big]=0,
\end{align}
\begin{align}
SumI_-^2[\Lambda_b^0,\Sigma_c^{++},D_s^-,\pi^+] &=2\big[\mathcal{A}(\Lambda_b^0\to\Sigma_c^0D_s^-\pi^+)-2\mathcal{A}(\Lambda_b^0\to\Sigma_c^+D_s^-\pi^0) \notag \\ &-\mathcal{A}(\Lambda_b^0\to\Sigma_c^{++}D_s^-\pi^-)\big]=0,
\end{align}
\begin{align}
SumI_-[\Lambda_b^0,\Sigma_c^{++},D_s^-,\eta_8] &=\sqrt{2}\mathcal{A}(\Lambda_b^0\to\Sigma_c^+D_s^-\eta_8)=0,
\end{align}
\begin{align}
SumI_-^2[\Lambda_b^0,\Xi_c^{*+},\overline{D}^0,\pi^+] &=2\mathcal{A}(\Lambda_b^0\to\Xi_c^{*0}D^-\pi^+)-2\big[\sqrt{2}\mathcal{A}(\Lambda_b^0\to\Xi_c^{*0}\overline{D}^0\pi^0) \notag \\ &+\sqrt{2}\mathcal{A}(\Lambda_b^0\to\Xi_c^{*+}D^-\pi^0)+\mathcal{A}(\Lambda_b^0\to\Xi_c^{*+}\overline{D}^0\pi^-)\big]=0,
\end{align}
\begin{align}
SumI_-[\Lambda_b^0,\Xi_c^{*+},\overline{D}^0,\eta_8] &=\mathcal{A}(\Lambda_b^0\to\Xi_c^{*0}\overline{D}^0\eta_8)+\mathcal{A}(\Lambda_b^0\to\Xi_c^{*+}D^-\eta_8)=0,
\end{align}
\begin{align}
SumI_-[\Lambda_b^0,\Xi_c^{*+},D_s^-,K^+] &=\mathcal{A}(\Lambda_b^0\to\Xi_c^{*0}D_s^-K^+)+\mathcal{A}(\Lambda_b^0\to\Xi_c^{*+}D_s^-K^0)=0,
\end{align}
\begin{align}
SumI_-[\Lambda_b^0,\Omega_c^0,\overline{D}^0,K^+] &=\mathcal{A}(\Lambda_b^0\to\Omega_c^0D^-K^+)+\mathcal{A}(\Lambda_b^0\to\Omega_c^0\overline{D}^0K^0)=0,
\end{align}
\begin{align}
SumI_-^2[\Xi_b^-,\Sigma_c^{++},D_s^-,\overline{K}^0] &=2\big[-\sqrt{2}\mathcal{A}(\Xi_b^0\to\Sigma_c^+D_s^-\overline{K}^0)+\mathcal{A}(\Xi_b^0\to\Sigma_c^{++}D_s^-K^-) \notag \\&+\mathcal{A}(\Xi_b^-\to\Sigma_c^0D_s^-\overline{K}^0)-\sqrt{2}\mathcal{A}(\Xi_b^-\to\Sigma_c^+D_s^-K^-)\big]=0,
\end{align}
\begin{align}
SumI_-^2[\Xi_b^-,\Xi_c^{*+},\overline{D}^0,\overline{K}^0] &=-2\big[\mathcal{A}(\Xi_b^0\to\Xi_c^{*0}\overline{D}^0\overline{K}^0)+\mathcal{A}(\Xi_b^0\to\Xi_c^{*+}D^-\overline{K}^0) \notag \\ &-\mathcal{A}(\Xi_b^0\to\Xi_c^{*+}\overline{D}^0K^-)-\mathcal{A}(\Xi_b^-\to\Xi_c^{*0}D^-\overline{K}^0) \notag \\ &+\mathcal{A}(\Xi_b^-\to\Xi_c^{*0}\overline{D}^0K^-)+\mathcal{A}(\Xi_b^-\to\Xi_c^{*+}D^-K^-)\big]=0,
\end{align}
\begin{align}
SumI_-^2[\Xi_b^-,\Xi_c^{*+},D_s^-,\pi^+] &=-2\big[\mathcal{A}(\Xi_b^0\to\Xi_c^{*0}D_s^-\pi^+)-\sqrt{2}\mathcal{A}(\Xi_b^0\to\Xi_c^{*+}D_s^-\pi^0) \notag \\ &+\sqrt{2}\mathcal{A}(\Xi_b^-\to\Xi_c^{*0}D_s^-\pi^0)+\mathcal{A}(\Xi_b^-\to\Xi_c^{*+}D_s^-\pi^-)\big]=0,
\end{align}
\begin{align}
SumI_-[\Xi_b^-,\Xi_c^{*+},D_s^-,\eta_8] &=-\mathcal{A}(\Xi_b^0\to\Xi_c^{*+}D_s^-\eta_8)+\mathcal{A}(\Xi_b^-\to\Xi_c^{*0}D_s^-\eta_8)=0,
\end{align}
\begin{align}
SumI_-^2[\Xi_b^-,\Omega_c^0,\overline{D}^0,\pi^+] &=-2\big[\mathcal{A}(\Xi_b^0\to\Omega_c^0D^-\pi^+)-\sqrt{2}\mathcal{A}(\Xi_b^0\to\Omega_c^0\overline{D}^0\pi^0) \notag \\ &+\sqrt{2}\mathcal{A}(\Xi_b^-\to\Omega_c^0D^-\pi^0)+\mathcal{A}(\Xi_b^-\to\Omega_c^0\overline{D}^0\pi^-)\big]=0,
\end{align}
\begin{align}
SumI_-[\Xi_b^-,\Omega_c^0,\overline{D}^0,\eta_8] &=-\mathcal{A}(\Xi_b^0\to\Omega_c^0\overline{D}^0\eta_8)+\mathcal{A}(\Xi_b^-\to\Omega_c^0D^-\eta_8)=0,
\end{align}
\begin{align}
SumI_-[\Xi_b^-,\Omega_c^0,D_s^-,K^+] &=-\mathcal{A}(\Xi_b^0\to\Omega_c^0D_s^-K^+)+\mathcal{A}(\Xi_b^-\to\Omega_c^0D_s^-K^0)=0,
\end{align}
\begin{align}
SumI_-[\Lambda_b^0,\Sigma_c^+,\overline{D}^0,\overline{K}^0] &=\sqrt{2}\mathcal{A}(\Lambda_b^0\to\Sigma_c^0\overline{D}^0\overline{K}^0)+\mathcal{A}(\Lambda_b^0\to\Sigma_c^+D^-\overline{K}^0) \notag \\ &-\mathcal{A}(\Lambda_b^0\to\Sigma_c^+\overline{D}^0K^-)=0,
\end{align}
\begin{align}
SumI_-[\Lambda_b^0,\Sigma_c^{++},D^-,\overline{K}^0] &=\sqrt{2}\mathcal{A}(\Lambda_b^0\to\Sigma_c^+D^-\overline{K}^0)-\mathcal{A}(\Lambda_b^0\to\Sigma_c^{++}D^-K^-)=0,
\end{align}
\begin{align}
SumI_-[\Lambda_b^0,\Sigma_c^{++},\overline{D}^0,K^-] &=\sqrt{2}\mathcal{A}(\Lambda_b^0\to\Sigma_c^+\overline{D}^0K^-)+\mathcal{A}(\Lambda_b^0\to\Sigma_c^{++}D^-K^-)=0,
\end{align}
\begin{align}
SumI_-[\Lambda_b^0,\Sigma_c^+,D_s^-,\pi^+] &=\sqrt{2}\big[\mathcal{A}(\Lambda_b^0\to\Sigma_c^0D_s^-\pi^+)-\mathcal{A}(\Lambda_b^0\to\Sigma_c^+D_s^-\pi^0)\big]=0,
\end{align}
\begin{align}
SumI_-[\Lambda_b^0,\Sigma_c^{++},D_s^-,\pi^0] &=\sqrt{2}\big[\mathcal{A}(\Lambda_b^0\to\Sigma_c^+D_s^-\pi^0)+\mathcal{A}(\Lambda_b^0\to\Sigma_c^{++}D_s^-\pi^-)\big]=0,
\end{align}
\begin{align}
SumI_-[\Lambda_b^0,\Xi_c^{*0},\overline{D}^0,\pi^+] &=\mathcal{A}(\Lambda_b^0\to\Xi_c^{*0}D^-\pi^+)-\sqrt{2}\mathcal{A}(\Lambda_b^0\to\Xi_c^{*0}\overline{D}^0\pi^0)=0,
\end{align}
\begin{align}
SumI_-[\Lambda_b^0,\Xi_c^{*+},D^-,\pi^+] &=\mathcal{A}(\Lambda_b^0\to\Xi_c^{*0}D^-\pi^+)-\sqrt{2}\mathcal{A}(\Lambda_b^0\to\Xi_c^{*+}D^-\pi^0)=0,
\end{align}
\begin{align}
SumI_-[\Lambda_b^0,\Xi_c^{*+},\overline{D}^0,\pi^0] &=\mathcal{A}(\Lambda_b^0\to\Xi_c^{*0}\overline{D}^0\pi^0)+\mathcal{A}(\Lambda_b^0\to\Xi_c^{*+}D^-\pi^0) \notag \\ &+\sqrt{2}\mathcal{A}(\Lambda_b^0\to\Xi_b^{*+}\overline{D}^0\pi^-)=0,
\end{align}
\begin{align}
SumI_-[\Xi_b^0,\Sigma_c^{++},D_s^-,\overline{K}^0] &=\sqrt{2}\mathcal{A}(\Xi_b^0\to\Sigma_c^+D_s^-\overline{K}^0)-\mathcal{A}(\Xi_b^0\to\Sigma_c^{++}D_s^-K^-)=0,
\end{align}
\begin{align}
SumI_-[\Xi_b^-,\Sigma_c^+,D_s^-,\overline{K}^0] &=-\mathcal{A}(\Xi_b^0\to\Sigma_c^+D_s^-\overline{K}^0)+\sqrt{2}\mathcal{A}(\Xi_b^-\to\Sigma_c^0D_s^-\overline{K}^0) \notag \\ &-\mathcal{A}(\Xi_b^-\to\Sigma_c^+D_s^-K^-)=0,
\end{align}
\begin{align}
SumI_-[\Xi_b^-,\Sigma_c^{++},D_s^-,K^-] &=-\mathcal{A}(\Xi_b^0\to\Sigma_c^{++}D_s^-K^-)+\sqrt{2}\mathcal{A}(\Xi_b^-\to\Sigma_c^+D_s^-K^-)=0,
\end{align}
\begin{align}
SumI_-[\Xi_b^0,\Xi_c^{*+},\overline{D}^0,\overline{K}^0] &=\mathcal{A}(\Xi_b^0\to\Xi_c^{*0}\overline{D}^0\overline{K}^0)+\mathcal{A}(\Xi_b^0\to\Xi_c^{*+}D^-\overline{K}^0) \notag \\ &-\mathcal{A}(\Xi_b^0\to\Xi_c^{*+}\overline{D}^0K^-)=0,
\end{align}
\begin{align}
SumI_-[\Xi_b^-,\Xi_c^{*0},\overline{D}^0,\overline{K}^0] &=-\mathcal{A}(\Xi_b^0\to\Xi_c^{*0}\overline{D}^0\overline{K}^0)+\mathcal{A}(\Xi_b^-\to\Xi_c^{*0}D^-\overline{K}^0) \notag \\ &-\mathcal{A}(\Xi_b^-\to\Xi_c^{*0}\overline{D}^0K^-)=0,
\end{align}
\begin{align}
SumI_-[\Xi_b^-,\Xi_c^{*+},D^-,\overline{K}^0] &=-\mathcal{A}(\Xi_b^0\to\Xi_c^{*+}D^-\overline{K}^0)+\mathcal{A}(\Xi_b^-\to\Xi_c^{*0}D^-\overline{K}^0) \notag\\ &-\mathcal{A}(\Xi_b^-\to\Xi_c^{*+}D^-K^-)=0,
\end{align}
\begin{align}
SumI_-[\Xi_b^-,\Xi_c^{*+},\overline{D}^0,K^-] &=-\mathcal{A}(\Xi_b^0\to\Xi_c^{*+}\overline{D}^0K^-)+\mathcal{A}(\Xi_b^-\to\Xi_c^{*0}\overline{D}^0K^-) \notag \\ &+\mathcal{A}(\Xi_b^-\to\Xi_c^{*+}D^-K^-)=0,
\end{align}
\begin{align}
SumI_-[\Xi_b^0,\Xi_c^{*+},D_s^-,\pi^+] &=\mathcal{A}(\Xi_b^0\to\Xi_c^{*0}D_s^-\pi^+)-\sqrt{2}\mathcal{A}(\Xi_b^0\to\Xi_c^{*+}D_s^-\pi^0)=0,
\end{align}
\begin{align}
SumI_-[\Xi_b^-,\Xi_c^{*0},D_s^-,\pi^+] &=-\mathcal{A}(\Xi_b^0\to\Xi_c^{*0}D_c^-\pi^+)-\sqrt{2}\mathcal{A}(\Xi_b^-\to\Xi_c^{*0}D_s^-\pi^0)=0,
\end{align}
\begin{align}
SumI_-[\Xi_b^-,\Xi_c^{*+},D_s^-,\pi^0] &=-\mathcal{A}(\Xi_b^0\to\Xi_c^{*+}D_s^-\pi^0)+\mathcal{A}(\Xi_b^-\to\Xi_c^{*0}D_s^-\pi^0) \notag \\ &+\sqrt{2}\mathcal{A}(\Xi_b^-\to\Xi_c^{*+}D_s^-\pi^-)=0,
\end{align}
\begin{align}
SumI_-[\Xi_b^0,\Omega_c^0,\overline{D}^0,\pi^+] &=\mathcal{A}(\Xi_b^0\to\Omega_c^0D^-\pi^+)-\sqrt{2}\mathcal{A}(\Xi_b^0\to\Omega_c^0\overline{D}^0\pi^0)=0,
\end{align}
\begin{align}
SumI_-[\Xi_b^-,\Omega_c^0,D^-,\pi^+] &=-\mathcal{A}(\Xi_b^0\to\Omega_c^0D^-\pi^+)-\sqrt{2}\mathcal{A}(\Xi_b^-\to\Omega_c^0D^-\pi^0)=0,
\end{align}
\begin{align}
SumI_-[\Xi_b^-,\Omega_c^0,\overline{D}^0,\pi^0] &=-\mathcal{A}(\Xi_b^0\to\Omega_c^0\overline{D}^0\pi^0)+\mathcal{A}(\Xi_b^-\to\Omega_c^0D^-\pi^0) \notag \\ &+\sqrt{2}\mathcal{A}(\Xi_b^-\to\Omega_c^0\overline{D}^0\pi^-)=0,
\end{align}
\begin{align}
SumI_-^3[\Xi_b^-,\Sigma^+,J/\psi,\pi^+] &=6\big[-2\mathcal{A}(\Xi_b^0\to\Sigma^0J/\psi\pi^0)+\mathcal{A}(\Xi_b^0\to\Sigma^-J/\psi\pi^+) \notag \\ &+\mathcal{A}(\Xi_b^0\to\Sigma^+J/\psi\pi^-)+\sqrt{2}\big(\mathcal{A}(\Xi_b^-\to\Sigma^0J/\psi\pi^-) \notag \\ &+\mathcal{A}(\Xi_b^-\to\Sigma^-J/\psi\pi^0)\big)\big]=0,
\end{align}
\begin{align}
SumI_-[\Lambda_b^0,p,J/\psi,\overline{K}^0] &=\mathcal{A}(\Lambda_b^0\to nJ/\psi\overline{K}^0)-\mathcal{A}(\Lambda_b^0\to pJ/\psi K^-)=0,
\end{align}
\begin{align}
SumI_-[\Lambda_b^0,\Xi^0,J/\psi,K^+] &=\mathcal{A}(\Lambda_b^0\to\Xi^0J/\psi K^0)-\mathcal{A}(\Lambda_b^0\to\Xi^-J/\psi K^+)=0,
\end{align}
\begin{align}
SumI_-[\Lambda_b^0,\Lambda^0,J/\psi,\pi^+] &=-\sqrt{2}\mathcal{A}(\Lambda_b^0\to\Lambda^0J/\psi\pi^0)=0,
\end{align}
\begin{align}
SumI_-[\Lambda_b^0,\Sigma^+,J/\psi,\eta_8] &=-\sqrt{2}\mathcal{A}(\Lambda_b^0\to\Sigma^0J/\psi\eta_8)=0,
\end{align}
\begin{align}
SumI_-^2[\Xi_b^-,\Sigma^+,J/\psi,\overline{K}^0] &=2\big[\sqrt{2}\mathcal{A}(\Xi_b^0\to\Sigma^0J/\psi\overline{K}^0)+\mathcal{A}(\Xi_b^0\to\Sigma^+J/\psi K^-) \notag \\ &+\sqrt{2}\mathcal{A}(\Xi_b^-\to\Sigma^0J/\psi K^-)-\mathcal{A}(\Xi_b^-\to\Sigma^-J/\psi\overline{K}^0)\big]=0,
\end{align}
\begin{align}
SumI_-^2[\Xi_b^-,\Xi^0,J/\psi,\pi^+] &=2\big[\sqrt{2}\mathcal{A}(\Xi_b^0\to\Xi^0J/\psi\pi^0)+\mathcal{A}(\Xi_b^0\to\Xi^-J/\psi\pi^+) \notag \\ &-\mathcal{A}(\Xi_b^-\to\Xi^0J/\psi\pi^-)+\sqrt{2}\mathcal{A}(\Xi_b^-\to\Xi^-J/\psi\pi^0)\big]=0,
\end{align}
\begin{align}
SumI_-[\Xi_b^-,\Xi^0,J/\psi,\eta_8] &=-\mathcal{A}(\Xi_b^0\to\Xi^0J/\psi\eta_8)-\mathcal{A}(\Xi_b^-\to\Xi^-J/\psi\eta_8)=0,
\end{align}
\begin{align}
SumI_-[\Xi_b^-,\Lambda^0,J/\psi,\overline{K}^0] &=-\mathcal{A}(\Xi_b^0\to\Lambda^0J/\psi\overline{K}^0)-\mathcal{A}(\Xi_b^-\to\Lambda^0J/\psi K^-)=0,
\end{align}
\begin{align}
SumI_-[\Lambda_b^0,\Sigma^0,J/\psi,\pi^+] &=\sqrt{2}\big[-\mathcal{A}(\Lambda_b^0\to\Sigma^0J/\psi\pi^0)+\mathcal{A}(\Lambda_b^0\to\Sigma^-J/\psi\pi^+)\big]=0,
\end{align}
\begin{align}
SumI_-[\Lambda_b^0,\Sigma^+,J/\psi,\pi^0] &=\sqrt{2}\big[-\mathcal{A}(\Lambda_b^0\to\Sigma^0J/\psi\pi^0)+\mathcal{A}(\Lambda_b^0\to\Sigma^+J/\psi\pi^-)\big]=0,
\end{align}
\begin{align}
SumI_-[\Xi_b^0,\Sigma^+,J/\psi,\overline{K}^0] &=-\sqrt{2}\mathcal{A}(\Xi_b^0\to\Sigma^0J/\psi\overline{K}^0)-\mathcal{A}(\Xi_b^0\to\Sigma^+J/\psi K^-)\big]=0,
\end{align}
\begin{align}
SumI_-[\Xi_b^-,\Sigma^0,J/\psi,\overline{K}^0] &=-\mathcal{A}(\Xi_b^0\to\Sigma^0J/\psi\overline{K}^0)-\mathcal{A}(\Xi_b^-\to\Sigma^0J/\psi K^-) \notag \\ &+\sqrt{2}\mathcal{A}(\Xi_b^-\to\Sigma^-J/\psi\overline{K}^0)=0,
\end{align}
\begin{align}
SumI_-[\Xi_b^-,\Sigma^+,J/\psi,K^-] &=-\mathcal{A}(\Xi_b^0\to\Sigma^+J/\psi K^-)-\sqrt{2}\mathcal{A}(\Xi_b^-\to\Sigma^0J/\psi K^-)=0,
\end{align}
\begin{align}
SumI_-[\Xi_b^0,\Xi^0,J/\psi,\pi^+] &= -\sqrt{2}\mathcal{A}(\Xi_b^0\to\Xi^0J/\psi\pi^0)-\mathcal{A}(\Xi_b^0\to\Xi^-J/\psi\pi^+)=0,
\end{align}
\begin{align}
SumI_-[\Xi_b^-,\Xi^-,J/\psi,\pi^+] &=-\mathcal{A}(\Xi_b^0\to\Xi^-J/\psi\pi^+)-\sqrt{2}\mathcal{A}(\Xi_b^-\to\Xi^-J/\psi\pi^0)=0,
\end{align}
\begin{align}
SumI_-[\Xi_b^-,\Xi^0,J/\psi,\pi^0] &=-\mathcal{A}(\Xi_b^0\to\Xi^0J/\psi\pi^0)+\sqrt{2}\mathcal{A}(\Xi_b^-\to\Xi^0J/\psi\pi^-) \notag \\ &-\mathcal{A}(\Xi_b^-\to\Xi^-J/\psi\pi^0)=0,
\end{align}
\begin{align}
SumI_-^3[\Lambda_b^0,\Delta^{++},J/\psi,\pi^+] &=-6\big[\sqrt{6}\mathcal{A}(\Lambda_b^0\to\Delta^0J/\psi\pi^0)+\sqrt{3}\mathcal{A}(\Lambda_b^0\to\Delta^{+}J/\psi\pi^-) \notag \\ &-\mathcal{A}(\Lambda_b^0\to\Delta^-J/\psi\pi^+)\big]=0,
\end{align}
\begin{align}
SumI^3_-[\Xi_b^-,\Sigma^{*+},J/\psi,\pi^+] &=-6\big[-2\mathcal{A}(\Xi_b^0\to\Sigma^{*0}J/\psi\pi^0)+\mathcal{A}(\Xi_b^0\to\Sigma^{*-}J/\psi\pi^+) \notag  \\&-\mathcal{A}(\Xi_b^0\to\Sigma^{*+}J/\psi\pi^-)+\sqrt{2}\big(\mathcal{A}(\Xi_b^-\to\Sigma^{*0}J/\psi\pi^-) \notag \\ &+\mathcal{A}(\Xi_b^-\to\Sigma^{*-}J/\psi\pi^0)\big)\big]=0,
\end{align}
\begin{align}
SumI_-^3[\Xi_b^-,\Delta^{++},J/\psi,\overline{K}^0] &=6\big[-\sqrt{3}\mathcal{A}(\Xi_b^0\to\Delta^0J/\psi\overline{K}^0)+\sqrt{3}\mathcal{A}(\Xi_b^0\to\Delta^+J/\psi K^-) \notag \\ &-\sqrt{3}\mathcal{A}(\Xi_b^-\to\Delta^0J/\psi K^-)+\mathcal{A}(\Xi_b^-\to\Delta^-J/\psi\overline{K}^0)\big]=0,
\end{align}
\begin{align}
SumI_-[\Lambda_b^0,\Xi^{*0},J/\psi,K^+] &=\mathcal{A}(\Lambda_b^0\to\Xi^{*0}J/\psi K^0)+\mathcal{A}(\Lambda_b^0\to\Xi^{*-}J/\psi K^+)=0,
\end{align}
\begin{align}
SumI_-^2[\Lambda_b^0,\Sigma^{*+},J/\psi,\pi^+] &=-2\big[2\mathcal{A}(\Lambda_b^0\to\Sigma^{*0}J/\psi\pi^0)+\mathcal{A}(\Lambda_b^0\to\Sigma^{*+}J/\psi\pi^-) \notag \\ &-\mathcal{A}(\Lambda_b^0\to\Sigma^{*-}J/\psi\pi^+)\big]=0,
\end{align}
\begin{align}
SumI_-[\Lambda_b^0,\Sigma^{*+},J/\psi,\eta_8] &=\sqrt{2}\mathcal{A}(\Lambda_b^0\to\Sigma^{*0}J/\psi\eta_8)=0,
\end{align}
\begin{align}
SumI_-^2[\Lambda_b^0,\Delta^{++},J/\psi,\overline{K}^0] &=2\sqrt{3}\big[\mathcal{A}(\Lambda_b^0\to\Delta^0J/\psi\overline{K}^0)-\mathcal{A}(\Lambda_b^0\to\Delta^+J/\psi K^-)\big]=0,
\end{align}
\begin{align}
SumI_-[\Xi_b^-,\Omega^-,J/\psi,K^+] &=-\mathcal{A}(\Xi_b^0\to\Omega^-J/\psi K^+)+\mathcal{A}(\Xi_b^-\to\Omega^-J/\psi K^0)=0,
\end{align}
\begin{align}
SumI_-^2[\Xi_b^-,\Xi^{*0},J/\psi,\pi^+] &=2\sqrt{2}\mathcal{A}(\Xi_b^0\to\Xi^{*0}J/\psi\pi^0)-2\big[\mathcal{A}(\Xi_b^0\to\Xi^{*-}J/\psi\pi^+) \notag \\ &+\mathcal{A}(\Xi_b^-\to\Xi^{*0}J/\psi\pi^-)+\sqrt{2}\mathcal{A}(\Xi_b^-\to\Xi^{*-}J/\psi\pi^0)\big]=0,
\end{align}
\begin{align}
SumI_-[\Xi_b^-,\Xi^{*0},J/\psi,\eta_8] &=-\mathcal{A}(\Xi_b^0\to\Xi^{*0}J/\psi\eta_8)+\mathcal{A}(\Xi_b^-\to\Xi^{*-}J/\psi\eta_8)=0,
\end{align}
\begin{align}
SumI_-^2[\Xi_b^-,\Sigma^{*+},J/\psi,\overline{K}^0] &=2\big[-\sqrt{2}\mathcal{A}(\Xi_b^0\to\Sigma^{*0}J/\psi\overline{K}^0)+\mathcal{A}(\Xi_b^0\to\Sigma^{*+}J/\psi K^-) \notag \\ &-\sqrt{2}\mathcal{A}(\Xi_b^-\to\Sigma^{*0}J/\psi K^-)+\mathcal{A}(\Xi_b^-\to\Sigma^{*-}J/\psi\overline{K}^0)\big]=0,
\end{align}
\begin{align}
SumI_-[\Lambda_b^0,\Sigma^{*0},J/\psi,\pi^+] &=\sqrt{2}\big[-\mathcal{A}(\Lambda_b^0\to\Sigma^{*0}J/\psi\pi^0)+\mathcal{A}(\Lambda_b^0\to\Sigma^{*-}J/\psi\pi^+)\big]=0,
\end{align}
\begin{align}
SumI_-[\Lambda_b^0,\Sigma^{*+},J/\psi,\pi^0] &=\sqrt{2}\big[\mathcal{A}(\Lambda_b^0\to\Sigma^{*0}J/\psi\pi^0)+\mathcal{A}(\Lambda_b^0\to\Sigma^{*+}J/\psi\pi^-)\big]=0,
\end{align}
\begin{align}
SumI_-[\Lambda_b^0,\Delta^+,J/\psi,\overline{K}^0] &=2\mathcal{A}(\Lambda_b^0\to\Delta^0J/\psi\overline{K}^0)-\mathcal{A}(\Lambda_b^0\to\Delta^+J/\psi K^-)=0,
\end{align}
\begin{align}
SumI_-[\Lambda_b^0,\Delta^{++},J/\psi, K^-] &=\sqrt{3}\mathcal{A}(\Lambda_b^0\to\Delta^+J/\psi K^-)=0,
\end{align}
\begin{align}
SumI_-[\Xi_b^0,\Xi^{*0},J/\psi,\pi^+] &=-\sqrt{2}\mathcal{A}(\Xi_b^0\to\Xi^{*0}J/\psi\pi^0)+\mathcal{A}(\Xi_b^0\to\Xi^{*-}J/\psi\pi^+)=0,
\end{align}
\begin{align}
SumI_-[\Xi_b^-,\Xi^{*-},J/\psi,\pi^+] &=-\mathcal{A}(\Xi_b^0\to\Xi^{*-}J/\psi\pi^+)-\sqrt{2}\mathcal{A}(\Xi_b^-\to\Xi^{*-}J/\psi\pi^0)=0,
\end{align}
\begin{align}
SumI_-[\Xi_b^-,\Xi^{*0},J/\psi,\pi^0] &=-\mathcal{A}(\Xi_b^0\to\Xi^{*0}J/\psi\pi^0)+\sqrt{2}\mathcal{A}(\Xi_b^-\to\Xi^{*0}J/\psi\pi^-) \notag \\ &+\mathcal{A}(\Xi_b^-\to\Xi^{*-}J/\psi\pi^0)=0,
\end{align}
\begin{align}
SumI_-[\Xi_b^0,\Sigma^{*+},J/\psi,\overline{K}^0] &=\sqrt{2}\mathcal{A}(\Xi_b^0\to\Sigma^{*0}J/\psi\overline{K}^0)-\mathcal{A}(\Xi_b^0\to\Sigma^{*+}J/\psi K^-)=0,
\end{align}
\begin{align}
SumI_-[\Xi_b^-,\Sigma^{*0},J/\psi,\overline{K}^0] &=-\mathcal{A}(\Xi_b^0\to\Sigma^{*0}J/\psi\overline{K}^0)-\mathcal{A}(\Xi_b^-\to\Sigma^{*0}J/\psi K^-) \notag \\ &+\sqrt{2}\mathcal{A}(\Xi_b^-\to\Sigma^{*-}J/\psi\overline{K}^0)=0,
\end{align}
\begin{align}
SumI_-[\Xi_b^-,\Sigma^{*+},J/\psi,K^-] &=-\mathcal{A}(\Xi_b^0\to\Sigma^{*+}J/\psi K^-)+\sqrt{2}\mathcal{A}(\Xi_b^-\to\Sigma^{*0}J/\psi K^-)=0,
\end{align}
\begin{align}
SumI_-^2[\Lambda_b^0,\Sigma^+,D^+,\overline{D}^0] &=-2\big[\sqrt{2}\mathcal{A}(\Lambda_b^0\to\Sigma^0D^0\overline{D}^0)+\sqrt{2}\mathcal{A}(\Lambda_b^0\to\Sigma^0D^+D^-) \notag \\ &-\mathcal{A}(\Lambda_b^0\to\Sigma^+D^0D^-)+\mathcal{A}(\Lambda_b^0\to\Sigma^-D^+\overline{D}^0)\big]=0,
\end{align}
\begin{align}
SumI_-[\Lambda_b^0,p,D^+,D_s^-] &=\mathcal{A}(\Lambda_b^0\to n D^+D_s^-)+\mathcal{A}(\Lambda_b^0\to p D^0D_s^-)=0,
\end{align}
\begin{align}
SumI_-[\Lambda_b^0,\Xi^0,D_s^+,\overline{D}^0] &=\mathcal{A}(\Lambda_b^0\to\Xi^0D_s^+D^-)-\mathcal{A}(\Lambda_b^0\to\Xi^-D_s^+\overline{D}^0)=0,
\end{align}
\begin{align}
SumI_-[\Lambda_b^0,\Lambda^0,D^+,\overline{D}^0] &=\mathcal{A}(\Lambda_b^0\to\Lambda^0D^0\overline{D}^0)+\mathcal{A}(\Lambda_b^0\to\Lambda^0D^+D^-)=0,
\end{align}
\begin{align}
SumI_-[\Lambda_b^0,\Sigma^+,D_s^+,D_s^-] &=-\sqrt{2}\mathcal{A}(\Lambda_b^0\to\Sigma^0D_s^+D_s^-)=0,
\end{align}
\begin{align}
SumI_-^2[\Xi_b^-,\Sigma^+,D^+,D_s^-] &=2\sqrt{2}\mathcal{A}(\Xi_b^0\to\Sigma^0D^+D_s^-)-2\big[\mathcal{A}(\Xi_b^0\to\Sigma^+D^0D_s^-) \notag \\ &+\sqrt{2}\mathcal{A}(\Xi_b^-\to\Sigma^0D^0D_s^-)+\mathcal{A}(\Xi_b^-\to\Sigma^-D^+D_s^-)\big]=0,
\end{align}
\begin{align}
SumI_-^2[\Xi_b^-,\Xi^0,D^+,\overline{D}^0] &=-2\big[\mathcal{A}(\Xi_b^0\to\Xi^0D^0\overline{D}^0)+\mathcal{A}(\Xi_b^0\to\Xi^0D^+D^-) \notag \\ &-\mathcal{A}(\Xi_b^0\to\Xi^-D^+\overline{D}^0)-\mathcal{A}(\Xi_b^-\to\Xi^0D^0D^-) \notag \\ &+\mathcal{A}(\Xi_b^-\to\Xi^-D^0\overline{D}^0)+\mathcal{A}(\Xi_b^-\to\Xi^-D^+D^-)\big]=0,
\end{align}
\begin{align}
SumI_-[\Xi_b^-,\Xi^0,D_s^+,D_s^-] &=-\mathcal{A}(\Xi_b^0\to\Xi^0D_s^+D_s^-)-\mathcal{A}(\Xi_b^-\to\Xi^-D_s^+D_s^-)=0,
\end{align}
\begin{align}
SumI_-[\Xi_b^-,\Lambda^0,D^+,D_s^-] &=-\mathcal{A}(\Xi_b^0\to\Lambda^0D^+D_s^-)+\mathcal{A}(\Xi_b^-\to\Lambda^0D^0D_s^-)=0,
\end{align}
\begin{align}
SumI_-^3[\Xi_b^-,\Sigma^+,D^+,\overline{D}^0] &=6\big[\sqrt{2}\mathcal{A}(\Xi_b^0\to\Sigma^0D^0\overline{D}^0)+\sqrt{2}\mathcal{A}(\Xi_b^0\to\Sigma^0D^+D^-) \notag \\ &-\mathcal{A}(\Xi_b^0\to\Sigma^+D^0D^-)+\mathcal{A}(\Xi_b^0\to\Sigma^-D^+\overline{D}^0) \notag \\ &-\sqrt{2}\mathcal{A}(\Xi_b^-\to\Sigma^0D^0D^-)-\mathcal{A}(\Xi_b^-\to\Sigma^-D^0\overline{D}^0) \notag \\ &-\mathcal{A}(\Xi_b^-\to\Sigma^-D^+D^-)\big]=0,
\end{align}
\begin{align}
SumI_-[\Lambda_b^0,\Sigma^0,D^+,\overline{D}^0] &=\mathcal{A}(\Lambda_b^0\to\Sigma^0D^0\overline{D}^0)+\mathcal{A}(\Lambda_b^0\to\Sigma^0D^+D^-) \notag \\ &+\sqrt{2}\mathcal{A}(\Lambda_b^0\to\Sigma^-D^+\overline{D}^0)=0,
\end{align}
\begin{align}
SumI_-[\Lambda_b^0,\Sigma^+,D^0,\overline{D}^0] &=-\sqrt{2}\mathcal{A}(\Lambda_b^0\to\Sigma^0D^0\overline{D}^0)+\mathcal{A}(\Lambda_b^0\to\Sigma^+D^0D^-)=0,
\end{align}
\begin{align}
SumI_-[\Lambda_b^0,\Sigma^+,D^+,D^-] &=-\sqrt{2}\mathcal{A}(\Lambda_b^0\to\Sigma^0D^+D^-)+\mathcal{A}(\Lambda_b^0\to\Sigma^+D^0D^-)=0,
\end{align}
\begin{align}
SumI_-[\Xi_b^0,\Sigma^+,D^+,D_s^-] &=-\sqrt{2}\mathcal{A}(\Xi_b^0\to\Sigma^0D^+D_s^-)+\mathcal{A}(\Xi_b^0\to\Sigma^+D^0D_s^-)=0,
\end{align}
\begin{align}
SumI_-[\Xi_b^-,\Sigma^0,D^+,D_s^-] &=-\mathcal{A}(\Xi_b^0\to\Sigma^0D^+D_s^-)+\mathcal{A}(\Xi_b^-\to\Sigma^0D^0D_s^-) \notag \\ &+\sqrt{2}\mathcal{A}(\Xi_b^-\to\Sigma^-D^+D_s^-)=0,
\end{align}
\begin{align}
SumI_-[\Xi_b^-,\Sigma^+,D^0,D_s^-] &=-\mathcal{A}(\Xi_b^0\to\Sigma^+D^0D_s^-)-\sqrt{2}\mathcal{A}(\Xi_b^-\to\Sigma^0D^0D_s^-)=0,
\end{align}
\begin{align}
SumI_-[\Xi_b^0,\Xi^0,D^+,\overline{D}^0] &=\mathcal{A}(\Xi_b^0\to\Xi^0D^0\overline{D}^0)+\mathcal{A}(\Xi_b^0\to\Xi^0D^+D^-) \notag \\ &-\mathcal{A}(\Xi_b^0\to\Xi^-D^+\overline{D}^0)=0,
\end{align}
\begin{align}
SumI_-[\Xi_b^-,\Xi^-,D^+,\overline{D}^0] &=-\mathcal{A}(\Xi_b^0\to\Xi^-D^+\overline{D}^0)+\mathcal{A}(\Xi_b^-\to\Xi^-D^0\overline{D}^0) \notag \\ &+\mathcal{A}(\Xi_b^-\to\Xi^-D^+D^-)=0,
\end{align}
\begin{align}
SumI_-[\Xi_b^-,\Xi^0,D^0,\overline{D}^0] &=-\mathcal{A}(\Xi_b^0\to\Xi^0D^0\overline{D}^0)+\mathcal{A}(\Xi_b^-\to\Xi^0D^0D^-) \notag \\ &-\mathcal{A}(\Xi_b^-\to\Xi^-D^0\overline{D}^0)=0,
\end{align}
\begin{align}
SumI_-[\Xi_b^-,\Xi^0,D^+,D^-] &=-\mathcal{A}(\Xi_b^0\to\Xi^0D^+D^-)+\mathcal{A}(\Xi_b^-\to\Xi^0D^0D^-) \notag \\ &-\mathcal{A}(\Xi_b^-\to\Xi^-D^+D^-)=0,
\end{align}
\begin{align}
SumI_-^3[\Lambda_b^0,\Delta^{++},D^+,\overline{D}^0] &=6\big[\sqrt{3}\mathcal{A}(\Lambda_b^0\to\Delta^0D^0\overline{D}^0)+\sqrt{3}\mathcal{A}(\Lambda_b^0\to\Delta^0D^+D^-) \notag \\ &+\sqrt{3}\mathcal{A}(\Lambda_b^0\to\Delta^+D^0D^-)+\mathcal{A}(\Lambda_b^0\to\Delta^-D^+\overline{D}^0)\big]=0,
\end{align}
\begin{align}
SumI_-^3[\Xi_b^-,\Sigma^{*+},D^+,\overline{D}^0] &=-6\big[\sqrt{2}\mathcal{A}(\Xi_b^0\to\Sigma^{*0}D^0\overline{D}^0)+\sqrt{2}\mathcal{A}(\Xi_b^0\to\Sigma^{*0}D^+D^-) \notag \\ &+\mathcal{A}(\Xi_b^0\to\Sigma^{*+}D^0D^-)+\mathcal{A}(\Xi_b^0\to\Sigma^{*-}D^+\overline{D}^0) \notag \\ &-\sqrt{2}\mathcal{A}(\Xi_b^-\to\Sigma^{*0}D^0D^-)-\mathcal{A}(\Xi_b^-\to\Sigma^{*-}D^0\overline{D}^0) \notag \\ &-\mathcal{A}(\Xi_b^-\to\Sigma^{*-}D^+D^-)\big]=0,
\end{align}
\begin{align}
SumI_-^3[\Xi_b^-,\Delta^{++},D^+,D_s^-] &=6\big[-\sqrt{3}\mathcal{A}(\Xi_b^0\to\Delta^0D^+D_s^-)-\sqrt{3}\mathcal{A}(\Xi_b^0\to\Delta^+D^0D_s^-) \notag \\ &+\sqrt{3}\mathcal{A}(\Xi_b^-\to\Delta^0D^0D_s^-))+\mathcal{A}(\Xi_b^-\to\Delta^-D^+D_s^-)\big]=0,
\end{align}
\begin{align}
SumI_-[\Lambda_b^0,\Xi^{*0},D_s^+,\overline{D}^0] &=\mathcal{A}(\Lambda_b^0\to\Xi^{*0}D_s^+D^-)+\mathcal{A}(\Lambda_b^0\to\Xi^{*-}D_s^+\overline{D}^0)=0,
\end{align}
\begin{align}
SumI_-^2[\Lambda_b^0,\Sigma^{*+},D^+,\overline{D}^0] &=2\big[\sqrt{2}\mathcal{A}(\Lambda_b^0\to\Sigma^{*0}D^0\overline{D}^0)+\sqrt{2}\mathcal{A}(\Lambda_b^0\to\Sigma^{*0}D^+D^-) \notag \\ &+\mathcal{A}(\Lambda_b^0\to\Sigma^{*+}D^0D^-)+\mathcal{A}(\Lambda_b^0\to\Sigma^{*-}D^+\overline{D}^0)\big]=0,
\end{align}
\begin{align}
SumI_-[\Lambda_b^0,\Sigma^{*+},D_s^+,D_s^-] &=\sqrt{2}\mathcal{A}(\Lambda_b^0\to\Sigma^{*0}D_s^+D_s^-)=0,
\end{align}
\begin{align}
SumI_-^2[\Lambda_b^0,\Delta^{++},D^+,D_s^-] &=2\sqrt{3}\big[\mathcal{A}(\Lambda_b^0\to\Delta^0D^+D_s^-)+\mathcal{A}(\Lambda_b^0\to\Delta^+D^0D_s^-)\big]=0,
\end{align}
\begin{align}
SumI_-[\Xi_b^-,\Omega^-,D_s^+,\overline{D}^0] &=-\mathcal{A}(\Xi_b^0\to\Omega^-D_s^+\overline{D}^0)+\mathcal{A}(\Xi_b^-\to\Omega^-D_s^+D^-)=0,
\end{align}
\begin{align}
SumI_-^2[\Xi_b^-,\Xi^{*0},D^+,\overline{D}^0] &=-2\big[\mathcal{A}(\Xi_b^0\to\Xi^{*0}D^0\overline{D}^0)+\mathcal{A}(\Xi_b^0\to\Xi^{*0}D^+D^-) \notag \\ &+\mathcal{A}(\Xi_b^0\to\Xi^{*-}D^+\overline{D}^0)-\mathcal{A}(\Xi_b^-\to\Xi^{*0}D^0D^-) \notag \\ &-\mathcal{A}(\Xi_b^-\to\Xi^{*-}D^0\overline{D}^0)-\mathcal{A}(\Xi_b^-\to\Xi^{*-}D^+D^-)\big]=0,
\end{align}
\begin{align}
SumI_-[\Xi_b^-,\Xi^{*0},D_s^+,D_s^-] &=-\mathcal{A}(\Xi_b^0\to\Xi^{*0}D_s^+D_s^-)+\mathcal{A}(\Xi_b^-\to\Xi^{*-}D_s^+D_s^-)=0,
\end{align}
\begin{align}
SumI_-^2[\Xi_b^-,\Sigma^{*+},D^+,D_s^-] &=2\big[-\sqrt{2}\mathcal{A}(\Xi_b^0\to\Sigma^{*0}D^+D_s^-)-\mathcal{A}(\Xi_b^0\to\Sigma^{*+}D^0D_s^-) \notag \\ &+\sqrt{2}\mathcal{A}(\Xi_b^-\to\Sigma^{*0}D^0D_s^-)+\mathcal{A}(\Xi_b^-\to\Sigma^{*-}D^+D_s^-)\big]=0,
\end{align}
\begin{align}
SumI_-[\Lambda_b^0,\Sigma^{*0},D^+,\overline{D}^0] &=\mathcal{A}(\Lambda_b^0\to\Sigma^{*0}D^0\overline{D}^0)+\mathcal{A}(\Lambda_b^0\to\Sigma^{*0}D^+D^-) \notag \\ &+\sqrt{2}\mathcal{A}(\Lambda_b^0\to\Sigma^{*-}D^+\overline{D}^0)=0,
\end{align}
\begin{align}
SumI_-[\Lambda_b^0,\Sigma^{*+},D^0,\overline{D}^0] &=\sqrt{2}\mathcal{A}(\Lambda_b^0\to\Sigma^{*0}D^0\overline{D}^0)+\mathcal{A}(\Lambda_b^0\to\Sigma^{*+}D^0D^-)=0,
\end{align}
\begin{align}
SumI_-[\Lambda_b^0,\Sigma^{*+},D^+,D^-] &=\sqrt{2}\mathcal{A}(\Lambda_b^0\to\Sigma^{*0}D^+D^-)+\mathcal{A}(\Lambda_b^0\to\Sigma^{*+}D^0D^-)=0,
\end{align}
\begin{align}
SumI_-[\Lambda_b^0,\Delta^+,D^+,D_s^-] &=2\mathcal{A}(\Lambda_b^0\to\Delta^0D^+D_s^-)+\mathcal{A}(\Lambda_b^0\to\Delta^+D^0D_s^-)=0,
\end{align}
\begin{align}
SumI_-[\Lambda_b^0,\Delta^{++},D^0,D_s^-] &=\sqrt{3}\mathcal{A}(\Lambda_b^0\to\Delta^+D^0D_s^-)=0,
\end{align}
\begin{align}
SumI_-[\Xi_b^0,\Xi^{*0},D^+,\overline{D}^0] &=\mathcal{A}(\Xi_b^0\to\Xi^{*0}D^0\overline{D}^0)+\mathcal{A}(\Xi_b^0\to\Xi^{*0}D^+D^-) \notag \\ &+\mathcal{A}(\Xi_b^0\to\Xi^{*-}D^+\overline{D}^0)=0,
\end{align}
\begin{align}
SumI_-[\Xi_b^-,\Xi^{*-},D^+,\overline{D}^0] &=-\mathcal{A}(\Xi_b^0\to\Xi^{*-}D^+\overline{D}^0)+\mathcal{A}(\Xi_b^-\to\Xi^{*-}D^0\overline{D}^0) \notag \\ &+\mathcal{A}(\Xi_b^-\to\Xi^{*-}D^+D^-)=0,
\end{align}
\begin{align}
SumI_-[\Xi_b^-,\Xi^{*0},D^0,\overline{D}^0] &=-\mathcal{A}(\Xi_b^0\to\Xi^{*0}D^0\overline{D}^0)+\mathcal{A}(\Xi_b^-\to\Xi^{*0}D^0D^-) \notag \\ &+\mathcal{A}(\Xi_b^-\to\Xi^{*-}D^0\overline{D}^0)=0,
\end{align}
\begin{align}
SumI_-[\Xi_b^-,\Xi^{*0},D^+,D^-] &=-\mathcal{A}(\Xi_b^0\to\Xi^{*0}D^+D^-)+\mathcal{A}(\Xi_b^-\to\Xi^{*0}D^0D^-) \notag \\ &+\mathcal{A}(\Xi_b^-\to\Xi^{*-}D^+D^-)=0,
\end{align}
\begin{align}
SumI_-[\Xi_b^0,\Sigma^{*+},D^+,D_s^-] &=\sqrt{2}\mathcal{A}(\Xi_b^0\to\Sigma^{*0}D^+D_s^-)+\mathcal{A}(\Xi_b^0\to\Sigma^{*+}D^0D_s^-)=0,
\end{align}
\begin{align}
SumI_-[\Xi_b^-,\Sigma^{*0},D^+,D_s^-] &=-\mathcal{A}(\Xi_b^0\to\Sigma^{*0}D^+D_s^-)+\mathcal{A}(\Xi_b^-\to\Sigma^{*0}D^0D_s^-) \notag \\ &+\sqrt{2}\mathcal{A}(\Xi_b^-\to\Sigma^{*-}D^+D_s^-)=0,
\end{align}
\begin{align}
SumI_-[\Xi_b^-,\Sigma^{*+},D^0,D_s^-] &=-\mathcal{A}(\Xi_b^0\to\Sigma^{*+}D^0D_s^-)+\sqrt{2}\mathcal{A}(\Xi_b^-\to\Sigma^{*0}D^0D_s^-)=0.
\end{align}

\subsection{$b\to c\overline u d/s$ modes}
\begin{align}
SumI_-^3[\Lambda_b^0,\Lambda_c^+,\pi^+,\pi^+] &= 6\sqrt{2}\big[\mathcal{A}(\Lambda_b^0\to\Lambda_c^+\pi^0\pi^-)+\mathcal{A}(\Lambda_b^0\to\Lambda_c^+\pi^-\pi^0)\big]=0,
\end{align}
\begin{align}
SumI_-^3[\Lambda_b^0,\Xi_c^+,\pi^+,K^+] &=-6\big[\sqrt{2}\mathcal{A}(\Lambda_b^0\to\Xi_c^0\pi^0K^0)+\mathcal{A}(\Lambda_b^0\to\Xi_c^0\pi^-K^+) \notag \\ &+\mathcal{A}(\Lambda_b^0\to\Xi_c^+\pi^-K^0)\big]=0,
\end{align}
\begin{align}
SumI_-^3[\Xi_b^-,\Lambda_c^+,\pi^+,\overline{K}^0] &=6\big[-\sqrt{2}\mathcal{A}(\Xi_b^0\to\Lambda_c^+\pi^0K^-)+\mathcal{A}(\Xi_b^0\to\Lambda_c^+\pi^-\overline{K}^0) \notag \\ &+\mathcal{A}(\Xi_b^-\to\Lambda_c^+\pi^-K^-)\big]=0,
\end{align}
\begin{align}
SumI_-^4[\Xi_b^-,\Xi_c^+,\pi^+,\pi^+] &=24[-2\mathcal{A}(\Xi_b^0\to\Xi_c^0\pi^0\pi^0)+\mathcal{A}(\Xi_b^0\to\Xi_c^0\pi^+\pi^-) \notag \\ &+\mathcal{A}(\Xi_b^0\to\Xi_c^0\pi^-\pi^+)-\sqrt{2}\mathcal{A}(\Xi_b^0\to\Xi_c^+\pi^0\pi^-) \notag \\ &-\sqrt{2}\mathcal{A}(\Xi_b^0\to\Xi_c^+\pi^-\pi^0)+\sqrt{2}\mathcal{A}(\Xi_b^-\to\Xi_c^0\pi^0\pi^-) \notag \\ &+\sqrt{2}\mathcal{A}(\Xi_b^-\to\Xi_c^0\pi^-\pi^0)+\mathcal{A}(\Xi_b^-\to\Xi_c^+\pi^-\pi^-)\big]=0,
\end{align}
\begin{align}
SumI_-^3[\Xi_b^-,\Xi_c^+,\pi^+,\eta_8] &=6\big[\sqrt{2}\mathcal{A}(\Xi_b^0\to\Xi_c^0\pi^0\eta_8)+\mathcal{A}(\Xi_b^0\to\Xi_c^+\pi^-\eta_8) \notag \\ &-\mathcal{A}(\Xi_b^-\to\Xi_c^0\pi^-\eta_8)\big]=0,
\end{align}
\begin{align}
SumI_-^3[\Xi_b^-,\Xi_c^+,K^+,\overline{K}^0] &=-6\big[\mathcal{A}(\Xi_b^0\to\Xi_c^0K^0\overline{K}^0)-\mathcal{A}(\Xi_b^0\to\Xi_c^0K^+K^-) \notag \\ &-\mathcal{A}(\Xi_b^0\to\Xi_c^+K^0K^-)+\mathcal{A}(\Xi_b^-\to\Xi_c^0K^0K^-)\big]=0,
\end{align}
\begin{align}
SumI_-^2[\Lambda_b^0,\Lambda_c^+,\pi^+,\overline{K}^0] &= 2\sqrt{2}\mathcal{A}(\Lambda_b^0\to\Lambda_c^+\pi^0K^-)-2\mathcal{A}(\Lambda_b^0\to\Lambda_c^+\pi^-\overline{K}^0)=0,
\end{align}
\begin{align}
SumI_-^3[\Lambda_b^0,\Xi_c^+,\pi^+,\pi^+] &=6\big[2\mathcal{A}(\Lambda_b^0\to\Xi_c^0\pi^0\pi^0)-\mathcal{A}(\Lambda_b^0\to\Xi_c^0\pi^+\pi^-) \notag \\ &-\mathcal{A}(\Lambda_b^0\to\Xi_c^0\pi^-\pi^+)+\sqrt{2}\big(\mathcal{A}(\Lambda_b^0\to\Xi_c^+\pi^0\pi^-) \notag \\ &+\mathcal{A}(\Lambda_b^0\to\Xi_c^+\pi^-\pi^0)\big)\big]=0,
\end{align}
\begin{align}
SumI_-^2[\Lambda_b^0,\Xi_c^+,\pi^+,\eta_8] &=-2\big[\sqrt{2}\mathcal{A}(\Lambda_b^0\to\Xi_c^0\pi^0\eta_8)+\mathcal{A}(\Lambda_b^0\to\Xi_c^+\pi^-\eta_8)\big]=0,
\end{align}
\begin{align}
SumI_-^2[\Lambda_b^0,\Xi_c^+,K^+,\overline{K}^0] &=2\big[\mathcal{A}(\Lambda_b^0\to\Xi_c^0K^0\overline{K}^0)-\mathcal{A}(\Lambda_b^0\to\Xi_c^0K^+K^-) \notag \\ &-\mathcal{A}(\Lambda_b^0\to\Xi_c^+K^0K^-)\big]=0,
\end{align}
\begin{align}
SumI_-^2[\Xi_b^-,\Lambda_c^+,\overline{K}^0,\overline{K}^0] &=2\big[\mathcal{A}(\Xi_b^0\to\Lambda_c^+K^-\overline{K}^0)+\mathcal{A}(\Xi_b^0\to\Lambda_c^+\overline{K}^0K^-) \notag \\ &+\mathcal{A}(\Xi_b^-\to\Lambda_c^+K^-K^-)\big]=0,
\end{align}
\begin{align}
SumI_-^3[\Xi_b^-,\Xi_c^+,\pi^+,\overline{K}^0] &=6\big[\sqrt{2}\mathcal{A}(\Xi_b^0\to\Xi_c^0\pi^0\overline{K}^0)+\mathcal{A}(\Xi_b^0\to\Xi_c^0\pi^+K^-) \notag \\ &-\sqrt{2}\mathcal{A}(\Xi_b^0\to\Xi_c^+\pi^0K^-)+\mathcal{A}(\Xi_b^0\to\Xi_c^+\pi^-\overline{K}^0) \notag \\ &+\sqrt{2}\mathcal{A}(\Xi_b^-\to\Xi_c^0\pi^0K^-)-\mathcal{A}(\Xi_b^-\to\Xi_c^0\pi^-\overline{K}^0) \notag \\ &+\mathcal{A}(\Xi_b^-\to\Xi_c^+\pi^-K^-)\big]=0,
\end{align}
\begin{align}
SumI_-^2[\Xi_b^-,\Xi_c^+,\overline{K}^0,\eta_8] &=-2\big[\mathcal{A}(\Xi_b^0\to\Xi_c^0\overline{K}^0\eta_8)-\mathcal{A}(\Xi_b^0\to\Xi_c^+K^-\eta_8) \notag \\ &+\mathcal{A}(\Xi_b^-\to\Xi_c^0K^-\eta_8)\big]=0,
\end{align}
\begin{align}
SumI_-^3[\Xi_b^0,\Xi_c^+,\pi^+,\pi^+] &=6\big[2\mathcal{A}(\Xi_b^0\to\Xi_c^0\pi^0\pi^0)-\mathcal{A}(\Xi_b^0\to\Xi_c^0\pi^+\pi^-) \notag \\ &-\mathcal{A}(\Xi_b^0\to\Xi_c^0\pi^-\pi^+)+\sqrt{2}\big(\mathcal{A}(\Xi_b^0\to\Xi_c^+\pi^0\pi^-) \notag \\&+\mathcal{A}(\Xi_b^0\to\Xi_c^+\pi^-\pi^0)\big)\big]=0,
\end{align}
\begin{align}
SumI_-^3[\Xi_b^-,\Xi_c^0,\pi^+,\pi^+] &=6\big[-2\mathcal{A}(\Xi_b^0\to\Xi_c^0\pi^0\pi^0)+\mathcal{A}(\Xi_b^0\to\Xi_c^0\pi^+\pi^-) \notag \\ &+\mathcal{A}(\Xi_b^0\to\Xi_c^0\pi^-\pi^+)+\sqrt{2}\big(\mathcal{A}(\Xi_b^-\to\Xi_c^0\pi^0\pi^-) \notag \\ &+\mathcal{A}(\Xi_b^-\to\Xi_c^0\pi^-\pi^0)\big)\big]=0,
\end{align}
\begin{align}
SumI_-^3[\Xi_b^-,\Xi_c^+,\pi^0,\pi^+] &=6\sqrt{2}\mathcal{A}(\Xi_b^0\to\Xi_c^0\pi^0\pi^0)-6\big[\sqrt{2}\mathcal{A}(\Xi_b^0\to\Xi_c^0\pi^-\pi^+) \notag \\ &-\mathcal{A}(\Xi_b^0\to\Xi_c^+\pi^0\pi^-)-2\mathcal{A}(\Xi_b^0\to\Xi_c^+\pi^-\pi^0) \notag \\ &+\mathcal{A}(\Xi_b^-\to\Xi_c^0\pi^0\pi^-)+2\mathcal{A}(\Xi_b^-\to\Xi_c^0\pi^-\pi^0) \notag \\ &+\sqrt{2}\mathcal{A}(\Xi_b^-\to\Xi_c^+\pi^-\pi^-)\big]=0,
\end{align}
\begin{align}
SumI_-^2[\Xi_b^0,\Xi_c^+,\pi^+,\overline{K}^0] &=-2\big[\sqrt{2}\mathcal{A}(\Xi_b^0\to\Xi_c^0\pi^0\overline{K}^0)+\mathcal{A}(\Xi_b^0\to\Xi_c^0\pi^+K^-) \notag \\ &-\sqrt{2}\mathcal{A}(\Xi_b^0\to\Xi_c^+\pi^0K^-)+\mathcal{A}(\Xi_b^0\to\Xi_c^+\pi^-\overline{K}^0)\big]=0,
\end{align}
\begin{align}
SumI_-^2[\Xi_b^-,\Xi_c^0,\pi^+,\overline{K}^0] &=2\big[\sqrt{2}\mathcal{A}(\Xi_b^0\to\Xi_c^0\pi^0\overline{K}^0)+\mathcal{A}(\Xi_b^0\to\Xi_c^0\pi^+K^-) \notag \\ &+\sqrt{2}\mathcal{A}(\Xi_b^-\to\Xi_c^0\pi^0K^-)-\mathcal{A}(\Xi_b^-\to\Xi_c^0\pi^-\overline{K}^0)\big]=0,
\end{align}
\begin{align}
SumI_-^2[\Xi_b^-,\Xi_c^+,\pi^0,\overline{K}^0] &=-2\big[\mathcal{A}(\Xi_b^0\to\Xi_c^0\pi^0\overline{K}^0)-\mathcal{A}(\Xi_b^0\to\Xi_c^+\pi^0K^-) \notag \\ &+\sqrt{2}\mathcal{A}(\Xi_b^0\to\Xi_c^+\pi^-\overline{K}^0)+\mathcal{A}(\Xi_b^-\to\Xi_c^0\pi^0K^-) \notag \\ &+\sqrt{2}\big(-\mathcal{A}(\Xi_b^-\to\Xi_c^0\pi^-\overline{K}^0)+\mathcal{A}(\Xi_b^-\to\Xi_c^+\pi^-K^-)\big)\big]=0,
\end{align}
\begin{align}
SumI_-^2[\Xi_b^-,\Xi_c^+,\pi^+,K^-] &=-2\big[\mathcal{A}(\Xi_b^0\to\Xi_c^0\pi^+K^-)-\sqrt{2}\mathcal{A}(\Xi_b^0\to\Xi_c^+\pi^0K^-) \notag \\ &+\sqrt{2}\mathcal{A}(\Xi_b^-\to\Xi_c^0\pi^0K^-)+\mathcal{A}(\Xi_b^-\to\Xi_c^+\pi^-K^-)\big]=0,
\end{align}
\begin{align}
SumI_-^4[\Lambda_b^0,\Sigma^{++},\pi^+,\pi^+] &=24\big[2\mathcal{A}(\Lambda_b^0\to\Sigma_c^0\pi^0\pi^0)-\mathcal{A}(\Lambda_b^0\to\Sigma_c^0\pi^+\pi^-) \notag \\ &-\mathcal{A}(\Lambda_b^0\to\Sigma_c^0\pi^-\pi^+)+2\big(\mathcal{A}(\Lambda_b^0\to\Sigma_c^+\pi^0\pi^-) \notag \\ &+\mathcal{A}(\Lambda_b^0\to\Sigma_c^+\pi^-\pi^0)\big)+\mathcal{A}(\Lambda_b^0\to\Sigma_c^{++}\pi^-\pi^-)\big]=0,
\end{align}
\begin{align}
SumI_-^3[\Lambda_b^0,\Sigma_c^{++},\pi^+,\eta_8] &=-6\sqrt{2}\big[\mathcal{A}(\Lambda_b^0\to\Sigma_c^0\pi^0\eta_8)+\mathcal{A}(\Lambda_b^0\to\Sigma_c^+\pi^-\eta_8)\big]=0,
\end{align}
\begin{align}
SumI_-^3[\Lambda_b^0,\Sigma_c^{++},K^+,\overline{K}^0] &=-6\big[-\mathcal{A}(\Lambda_b^0\to\Sigma_c^0K^0\overline{K}^0)+\mathcal{A}(\Lambda_b^0\to\Sigma_c^0K^+K^-) \notag \\ &+\sqrt{2}\mathcal{A}(\Lambda_b^0\to\Sigma_c^+K^0K^-)\big]=0,
\end{align}
\begin{align}
SumI_-^3[\Lambda_b^0,\Xi_c^{*+},\pi^+,K^+] &=-6\big[\sqrt{2}\mathcal{A}(\Lambda_b^0\to\Xi_c^{*0}\pi^0K^0)+\mathcal{A}(\Lambda_b^0\to\Xi_c^{*0}\pi^-K^+) \notag \\ &+\mathcal{A}(\Lambda_b^0\to\Xi_c^+\pi^-K^0)\big]=0,
\end{align}
\begin{align}
SumI_-^4[\Xi_b^-,\Sigma_c^{++},\pi^+,\overline{K}^0] &=24\big[\sqrt{2}\mathcal{A}(\Xi_b^0\to\Sigma_c^0\pi^0\overline{K}^0)+\mathcal{A}(\Xi_b^0\to\Sigma_c^0\pi^+K^-) \notag \\ &-2\mathcal{A}(\Xi_b^0\to\Sigma_c^+\pi^0K^-)+\sqrt{2}\mathcal{A}(\Xi_b^0\to\Sigma_c^+\pi^-\overline{K}^0) \notag \\ &-\mathcal{A}(\Xi_b^0\to\Sigma_c^{++}\pi^-K^-)+\sqrt{2}\mathcal{A}(\Xi_b^-\to\Sigma_c^0\pi^0K^-) \notag \\ &-\mathcal{A}(\Xi_b^-\to\Sigma_c^0\pi^-\overline{K}^0)+\sqrt{2}\mathcal{A}(\Xi_b^-\to\Sigma_c^+\pi^-K^-)\big]=0,
\end{align}
\begin{align}
SumI_-^3[\Xi_b^-,\Sigma_c^{++},\overline{K}^0,\eta_8] &=-6\big[\mathcal{A}(\Xi_b^0\to\Sigma_c^0\overline{K}^0\eta_8)-\sqrt{2}\mathcal{A}(\Xi_b^0\to\Sigma_c^+K^-\eta_8) \notag \\ &+\mathcal{A}(\Xi_b^-\to\Sigma_c^0K^-\eta_8)\big]=0,
\end{align}
\begin{align}
SumI_-^4[\Xi_b^-,\Xi_c^{*+},\pi^+,\pi^+] &=24\big[-2\mathcal{A}(\Xi_b^0\to\Xi_c^{*0}\pi^0\pi^0)+\mathcal{A}(\Xi_b^0\to\Xi_c^{*0}\pi^-\pi^+) \notag \\ &+\mathcal{A}(\Xi_b^0\to\Xi_c^{*0}\pi^+\pi^-)-\sqrt{2}\mathcal{A}(\Xi_b^0\to\Xi_c^{*+}\pi^0\pi^-) \notag \\ &-\sqrt{2}\mathcal{A}(\Xi_b^0\to\Xi_c^{*+}\pi^-\pi^0)+\sqrt{2}\mathcal{A}(\Xi_b^-\to\Xi_c^{*0}\pi^0\pi^-) \notag \\ &+\sqrt{2}\mathcal{A}(\Xi_b^-\to\Xi_c^{*0}\pi^-\pi^0)+\mathcal{A}(\Xi_b^-\to\Xi_c^{*+}\pi^-\pi^-)\big]=0,
\end{align}
\begin{align}
SumI_-^3[\Xi_b^-,\Xi_c^{*+},\pi^+,\eta_8] &=6\big[\sqrt{2}\mathcal{A}(\Xi_b^0\to\Xi_c^{*0}\pi^0\eta_8)+\mathcal{A}(\Xi_b^0\to\Xi_c^{*+}\pi^-\eta_8) \notag \\ &-\mathcal{A}(\Xi_b^-\to\Xi_c^{*0}\pi^-\eta_8)\big]=0,
\end{align}
\begin{align}
SumI_-^3[\Xi_b^-,\Xi_c^{*+},K^+,\overline{K}^0] &=-6\big[\mathcal{A}(\Xi_b^0\to\Xi_c^{*0}K^0\overline{K}^0)-\mathcal{A}(\Xi_b^0\to\Xi_c^{*0}K^+K^-) \notag \\ &-\mathcal{A}(\Xi_b^0\to\Xi_c^{*+}K^0K^-)+\mathcal{A}(\Xi_b^-\to\Xi_c^{*0}K^0K^-)\big]=0,
\end{align}
\begin{align}
SumI^3_-[\Xi_b^-,\Sigma_c^{++},\overline{K}^0,\eta_8] &=-6\big[\mathcal{A}(\Xi_b^0\to\Sigma_c^0\overline{K}^0\eta_8)-\sqrt{2}\mathcal{A}(\Xi_b^0\to\Sigma_c^+K^-\eta_8) \notag \\ &+\mathcal{A}(\Xi_b^-\to\Sigma_c^0K^-\eta_8)\big]=0,
\end{align}
\begin{align}
SumI_-^3[\Xi_b^-,\Omega_c^0,\pi^+,K^+] &=6\big[\sqrt{2}\mathcal{A}(\Xi_b^0\to\Omega_c^0\pi^0K^0)+\mathcal{A}(\Xi_b^0\to\Omega
_c^0\pi^-K^+) \notag \\ &-\mathcal{A}(\Xi_b^-\to\Omega_c^0\pi^-K^0)\big]=0,
\end{align}
\begin{align}
SumI_-^3[\Lambda_b^0,\Sigma_c^{++},\pi^+,\overline{K}^0] &=-6\big[\sqrt{2}\mathcal{A}(\Lambda_b^0\to\Sigma_c^0\pi^0\overline{K}^0)+\mathcal{A}(\Lambda_b^0\to\Sigma_c^0\pi^+K^-) \notag \\ &-2\mathcal{A}(\Lambda_b^0\to\Sigma_c^+\pi^0K^-)+\sqrt{2}\mathcal{A}(\Lambda_b^0\to\Sigma_c^+\pi^-\overline{K}^0) \notag \\ &-\mathcal{A}(\Lambda_b^0\to\Sigma_c^{++}\pi^-K^-)\big]=0,
\end{align}
\begin{align}
SumI_-^2[\Lambda_b^0,\Sigma_c^{++},\overline{K}^0,\eta_8] &=2\big[\mathcal{A}(\Lambda_b^0\to\Sigma_c^0\overline{K}^0\eta_8)-\sqrt{2}\mathcal{A}(\Lambda_b^0\to\Sigma_c^+K^-\eta_8)\big]=0,
\end{align}
\begin{align}
SumI_-^3[\Lambda_b^0,\Xi_c^{*+},\pi^+,\pi^+] &=6\big[2\mathcal{A}(\Lambda_b^0\to\Xi_c^{*0}\pi^0\pi^0)-\mathcal{A}(\Lambda_b^0\to\Xi_c^{*0}\pi^+\pi^-) \notag \\ &-\mathcal{A}(\Lambda_b^0\to\Xi_c^{*0}\pi^-\pi^+)+\sqrt{2}\big(\mathcal{A}(\Lambda_b^0\to\Xi_c^{*+}\pi^0\pi^-) \notag \\ &+\mathcal{A}(\Lambda_b^0\to\Xi_c^{*+}\pi^-\pi^0)\big)\big]=0,
\end{align}
\begin{align}
SumI_-^2[\Lambda_b^0,\Xi_c^{*+},\pi^+,\eta_8] &=-2\big[\sqrt{2}\mathcal{A}(\Lambda_b^0\to\Xi_c^{*0}\pi^0\eta_8)+\mathcal{A}(\Lambda_b^0\to\Xi_c^{*+}\pi^-\eta_8)\big]=0,
\end{align}
\begin{align}
SumI_-^2[\Lambda_b^0,\Xi_c^{*+},K^+,\overline{K}^0] &=2\big[\mathcal{A}(\Lambda_b^0\to\Xi_c^{*0}K^0\overline{K}^0)-\mathcal{A}(\Lambda_b^0\to\Xi_c^{*0}K^+K^-) \notag \\ &-\mathcal{A}(\Lambda_b^0\to\Xi_c^{*+}K^0K^-)\big]=0,
\end{align}
\begin{align}
SumI_-^2[\Lambda_b^0,\Omega_c^0,\pi^+,K^+] &=-2\big[\sqrt{2}\mathcal{A}(\Lambda_b^0\to\Omega_c^0\pi^0K^0)+\mathcal{A}(\Lambda_b^0\to\Omega_c^0\pi^-K^+)\big]=0,
\end{align}
\begin{align}
SumI_-^3[\Xi_b^-,\Sigma_c^{++},\overline{K}^0,\overline{K}^0] &=-6\big[\mathcal{A}(\Xi_b^0\to\Xi_c^0\overline{K}^0
\overline{K}^0)-\sqrt{2}\mathcal{A}(\Xi_b^0\to\Sigma_c^+K^-\overline{K}^0) \notag \\ &-\sqrt{2}\mathcal{A}(\Xi_b^0\to\Sigma_c^+\overline{K}^0K^-)+\mathcal{A}(\Xi_b^0\to\Sigma_c^{++}K^-K^-) \notag \\ &+\mathcal{A}(\Xi_b^-\to\Sigma_c^0K^-\overline{K}^0)+\mathcal{A}(\Xi_b^-\to\Sigma_c^0\overline{K}^0K^-) \notag \\ &-\sqrt{2}\mathcal{A}(\Xi_b^-\to\Sigma_c^+K^-K^-)\big]=0,
\end{align}
\begin{align}
SumI_-^3[\Xi_b^-,\Xi_c^{*+},\pi^+,\overline{K}^0] &=6\big[\sqrt{2}\mathcal{A}(\Xi_b^0\to\Xi_c^{*0}\pi^0\overline{K}^0)+\mathcal{A}(\Xi_b^0\to\Xi_c^{*0}\pi^+K^-) \notag \\ &-\sqrt{2}\mathcal{A}(\Xi_b^0\to\Xi_c^{*+}\pi^0K^-)+\mathcal{A}(\Xi_b^0\to\Xi_c^{*+}\pi^-\overline{K}^0) \notag \\ &+\sqrt{2}\mathcal{A}(\Xi_b^-\to\Xi_c^{*}\pi^0K^-)-\mathcal{A}(\Xi_b^-\to\Xi_c^{*0}\pi^-\overline{K}^0) \notag \\ &+\mathcal{A}(\Xi_b^-\to\Xi_c^{*+}\pi^-K^-)\big]=0,
\end{align}
\begin{align}
SumI_-^2[\Xi_b^-,\Xi_c^{*+},\overline{K}^0,\eta_8] &=-2\big[\mathcal{A}(\Xi_b^0\to\Xi_c^{*0}\overline{K}^0\eta_8)-\mathcal{A}(\Xi_b^0\to\Xi_c^{*+}K^-\eta_8) \notag \\ &+\mathcal{A}(\Xi_b^-\to\Xi_c^{*0}K^-\eta_8)\big]=0,
\end{align}
\begin{align}
SumI_-^3[\Xi_b^-,\Omega_c^0,\pi^+,\pi^+] &=6\big[-2\mathcal{A}(\Xi_b^0\to\Omega_c^0\pi^0\pi^0)+\mathcal{A}(\Xi_b^0\to\Omega_c^0\pi^+\pi^-) \notag \\ &+\mathcal{A}(\Xi_b^0\to\Omega_c^0\pi^-\pi^+)+\sqrt{2}\big(\mathcal{A}(\Xi_b^-\to\Omega_c^0\pi^0\pi^-) \notag \\ &+\mathcal{A}(\Xi_b^-\to\Omega_c^0\pi^-\pi^0)\big)\big]=0,
\end{align}
\begin{align}
SumI_-^2[\Xi_b^-,\Omega_c^0,\pi^+,\eta_8] &=2\sqrt{2}\mathcal{A}(\Xi_b^0\to\Omega_c^0\pi^0\eta_8)-2\mathcal{A}(\Xi_b^-\to\Omega_c^0\pi^-\eta_8)=0,
\end{align}
\begin{align}
SumI_-^2[\Xi_b^-,\Omega_c^0,K^+,\overline{K}^0] &=-2\big[\mathcal{A}(\Xi_b^0\to\Omega_c^0K^0\overline{K}^0)-\mathcal{A}(\Xi_b^0\to\Omega_c^0K^+K^-) \notag \\ &+\mathcal{A}(\Xi_b^-\to\Omega_c^0K^0K^-)\big]=0,
\end{align}
\begin{align}
SumI_-^3[\Lambda_b^0,\Sigma_c^+,\pi^+,\pi^+] &=6\sqrt{2}\big[2\mathcal{A}(\Lambda_b^0\to\Sigma_c^0\pi^0\pi^0)-\mathcal{A}(\Lambda_b^0\to\Sigma_c^0\pi^+\pi^-) \notag \\ &-\mathcal{A}(\Lambda_b^0\to\Sigma_c^0\pi^-\pi^+)+\mathcal{A}(\Lambda_b^0\to\Sigma_c^+\pi^0\pi^-) \notag \\ &+\mathcal{A}(\Lambda_b^0\to\Sigma_c^+\pi^-\pi^0)\big]=0,
\end{align}
\begin{align}
SumI_-^3[\Lambda_b^0,\Sigma_c^{++},\pi^0,\pi^+] &=-6\sqrt{2}\big[\mathcal{A}(\Lambda_b^0\to\Sigma_c^0\pi^0\pi^0)-\mathcal{A}(\Lambda_b^0\to\Sigma_c^0\pi^-\pi^+) \notag \\ &+\mathcal{A}(\Lambda_b^0\to\Sigma_c^+\pi^0\pi^-)+2\mathcal{A}(\Lambda_b^0\to\Sigma_c^+\pi^-\pi^0) \notag \\ &+\mathcal{A}(\Lambda_b^0\to\Sigma_c^{++}\pi^-\pi^-)\big]=0,
\end{align}
\begin{align}
SumI_-^3[\Xi_b^0,\Sigma_c^{++},\pi^+,\overline{K}^0] &=-6\big[\sqrt{2}\mathcal{A}(\Xi_b^0\to\Sigma_c^0\pi^0\overline{K}^0)+\mathcal{A}(\Xi_b^0\to\Sigma_c^0\pi^+K^-) \notag \\ &-2\mathcal{A}(\Xi_b^0\to\Sigma_c^+\pi^0K^-)+\sqrt{2}\mathcal{A}(\Xi_b^0\to\Sigma_c^+\pi^-\overline{K}^0) \notag \\ &-\mathcal{A}(\Xi_b^0\to\Sigma_c^{++}\pi^-K^-)\big]=0,
\end{align}
\begin{align}
SumI_-^3[\Xi_b^-,\Sigma_c^+,\pi^+,\overline{K}^0] &=6\big[2\mathcal{A}(\Xi_b^0\to\Sigma_c^0\pi^0\overline{K}^0)+\sqrt{2}\mathcal{A}(\Xi_b^0\to\Sigma_c^0\pi^+K^-) \notag \\ &-\sqrt{2}\mathcal{A}(\Xi_b^0\to\Sigma_c^+\pi^0K^-)+\mathcal{A}(\Xi_b^0\to\Sigma_c^+\pi^-\overline{K}^0) \notag \\ &+2\mathcal{A}(\Xi_b^-\to\Sigma_c^0\pi^0K^-)-\sqrt{2}\mathcal{A}(\Xi_b^-\to\Sigma_c^0\pi^-\overline{K}^0) \notag \\ &+\mathcal{A}(\Xi_b^-\to\Sigma_c^+\pi^-K^-)\big]=0,
\end{align}
\begin{align}
SumI_-^3[\Xi_b^-,\Sigma_c^{++},\pi^0,\overline{K}^0] &=-6\big[\mathcal{A}(\Xi_b^0\to\Sigma_c^0\pi^0\overline{K}^0)-\sqrt{2}\mathcal{A}(\Xi_b^0\to\Sigma_c^+\pi^0K^-) \notag \\ &+2\mathcal{A}(\Xi_b^0\to\Sigma_c^+\pi^-\overline{K}^0)-\sqrt{2}\mathcal{A}(\Xi_b^0\to\Sigma_c^{++}\pi^-K^-) \notag \\ &+\mathcal{A}(\Xi_b^-\to\Sigma_c^0\pi^0K^-)-\sqrt{2}\mathcal{A}(\Xi_b^-\to\Sigma_c^0\pi^-\overline{K}^0) \notag \\ &+2\mathcal{A}(\Xi_b^-\to\Sigma_c^+\pi^-K^-)\big]=0,
\end{align}
\begin{align}
SumI_-^3[\Xi_b^-,\Sigma_c^{++},\pi^+,K^-] &=-6\big[\mathcal{A}(\Xi_b^0\to\Sigma_c^0\pi^+K^-)-2\mathcal{A}(\Xi_b^0\to\Sigma_c^+\pi^0K^-) \notag \\ &-\mathcal{A}(\Xi_b^0\to\Sigma_c^{++}\pi^-K^-)+\sqrt{2}\big(\mathcal{A}(\Xi_b^-\to\Sigma_c^0\pi^0K^-) \notag \\ &+\mathcal{A}(\Xi_b^-\to\Sigma_c^+\pi^-K^-)\big)\big]=0,
\end{align}
\begin{align}
SumI_-^3[\Xi_b^0,\Xi_c^{*+},\pi^+,\pi^+] &=6\big[2\mathcal{A}(\Xi_b^0\to\Xi_c^{*0}\pi^0\pi^0)-\mathcal{A}(\Xi_b^0\to\Xi_c^{*0}\pi^-\pi^+) \notag \\ &-\mathcal{A}(\Xi_b^0\to\Xi_c^{*0}\pi^+\pi^-)+\sqrt{2}\big(\mathcal{A}(\Xi_b^0\to\Xi_c^{*+}\pi^0\pi^-) \notag \\ &+\mathcal{A}(\Xi_b^0\to\Xi_c^{*+}\pi^-\pi^0)\big)\big]=0,
\end{align}
\begin{align}
SumI_-^3[\Xi_b^-,\Xi_c^{*0},\pi^+,\pi^+] &=6\big[-2\mathcal{A}(\Xi_b^0\to\Xi_c^{*0}\pi^0\pi^0)+\mathcal{A}(\Xi_b^0\to\Xi_c^{*0}\pi^-\pi^+) \notag \\ &+\mathcal{A}(\Xi_b^0\to\Xi_c^{*0}\pi^+\pi^-)+\sqrt{2}\big(\mathcal{A}(\Xi_b^-\to\Xi_c^{*0}\pi^0\pi^-) \notag \\ &+\mathcal{A}(\Xi_b^-\to\Xi_c^{*0}\pi^-\pi^0)\big)\big]=0,
\end{align}
\begin{align}
SumI_-^3[\Xi_b^-,\Xi_c^{*+},\pi^0,\pi^+] &=6\sqrt{2}\mathcal{A}(\Xi_b^0\to\Xi_c^{*0}\pi^0\pi^0)-6\big[\sqrt{2}\mathcal{A}(\Xi_b^0\to\Xi_c^{*0}\pi^-\pi^+) \notag \\ &-\mathcal{A}(\Xi_b^0\to\Xi_c^{*+}\pi^0\pi^-)-2\mathcal{A}(\Xi_b^0\to\Xi_c^{*+}\pi^-\pi^0) \notag \\ &+\mathcal{A}(\Xi_b^-\to\Xi_c^{*0}\pi^0\pi^-)+2\mathcal{A}(\Xi_b^-\to\Xi_c^{*0}\pi^-\pi^0) \notag \\ &+\sqrt{2}\mathcal{A}(\Xi_b^-\to\Xi_c^{*+}\pi^-\pi^-)\big]=0,
\end{align}
\begin{align}
SumI_-^2[\Lambda_b^0,\Sigma_c^+,\pi^+,\overline{K}^0] &=-2\big[2\mathcal{A}(\Lambda_b^0\to\Sigma_c^0\pi^0\overline{K}^0)+\sqrt{2}\mathcal{A}(\Lambda_b^0\to\Sigma_c^0\pi^+K^-) \notag \\ &-\sqrt{2}\mathcal{A}(\Lambda_b^0\to\Sigma_c^+\pi^0K^-)+\mathcal{A}(\Lambda_b^0\to\Sigma_c^+\pi^-\overline{K}^0)\big]=0,
\end{align}
\begin{align}
SumI_-^2[\Lambda_b^0,\Sigma_c^{++},\pi^0,\overline{K}^0] &=2\big[\mathcal{A}(\Lambda_b^0\to\Sigma_c^0\pi^0\overline{K}^0)-\sqrt{2}\mathcal{A}(\Lambda_b^0\to\Sigma_c^+\pi^0K^-) \notag \\ &+2\mathcal{A}(\Lambda_b^0\to\Sigma_c^+\pi^-\overline{K}^0)-\sqrt{2}\mathcal{A}(\Lambda_b^0\to\Sigma_c^{++}\pi^-K^-)\big]=0,
\end{align}
\begin{align}
SumI_-^2[\Lambda_b^0,\Sigma_c^{++},\pi^+,K^-] &=2\big[\mathcal{A}(\Lambda_b^0\to\Sigma_c^0\pi^+K^-)-2\mathcal{A}(\Lambda_b^0\to\Sigma_c^+\pi^0K^-) \notag \\ &-\mathcal{A}(\Lambda_b^0\to\Sigma_c^{++}\pi^-K^-)\big]=0,
\end{align}
\begin{align}
SumI_-^2[\Lambda_b^0,\Xi_c^{*0},\pi^+,\pi^+] &=4\mathcal{A}(\Lambda_b^0\to\Xi_c^{*0}\pi^0\pi^0)-2\big[\mathcal{A}(\Lambda_b^0\to\Xi_c^{*0}\pi^+\pi^-) \notag \\ &+\mathcal{A}(\Lambda_b^0\to\Xi_c^{*0}\pi^-\pi^+)\big]=0,
\end{align}
\begin{align}
SumI_-^2[\Lambda_b^0,\Xi_c^{*+},\pi^0,\pi^+] &=-2\big[\sqrt{2}\mathcal{A}(\Lambda_b^0\to\Xi_c^{*0}\pi^0\pi^0)-\sqrt{2}\mathcal{A}(\Lambda_b^0\to\Xi_c^{*0}\pi^-\pi^+) \notag \\ &+\mathcal{A}(\Lambda_b^0\to\Xi_c^{*+}\pi^0\pi^-)+2\mathcal{A}(\Lambda_b^0\to\Xi_c^{*+}\pi^-\pi^0)\big]=0,
\end{align}
\begin{align}
SumI_-^2[\Xi_b^0,\Sigma_c^{++},\overline{K}^0,\overline{K}^0] &=2\big[\mathcal{A}(\Xi_b^0\to\Sigma_c^0\overline{K}^0\overline{K}^0)-\sqrt{2}\mathcal{A}(\Xi_b^0\to\Sigma_c^+K^-\overline{K}^0) \notag \\ &-\sqrt{2}\mathcal{A}(\Xi_b^0\to\Sigma_c^+\overline{K}^0K^-)+\mathcal{A}(\Xi_b^0\to\Sigma_c^{++}K^-K^-)\big]=0,
\end{align}
\begin{align}
SumI_-^2[\Xi_b^-,\Sigma_c^{++},K^-,\overline{K}^0] &=2\big[-\sqrt{2}\mathcal{A}(\Xi_b^0\to\Sigma_c^+K^-\overline{K}^0)+\mathcal{A}(\Xi_b^0\to\Sigma_c^{++}K^-K^-) \notag \\ &+\mathcal{A}(\Xi_b^-\to\Sigma_c^0K^-\overline{K}^0)-\sqrt{2}\mathcal{A}(\Xi_b^-\to\Sigma_c^+K^-K^-)\big]=0,
\end{align}
\begin{align}
SumI_-^2[\Xi_b^0,\Xi_c^{*+},\pi^+,\overline{K}^0] &=-2\big[\sqrt{2}\mathcal{A}(\Xi_b^0\to\Xi_c^{*0}\pi^0\overline{K}^0)+\mathcal{A}(\Xi_b^0\to\Xi_c^{*0}\pi^+K^-) \notag \\ &-\sqrt{2}\mathcal{A}(\Xi_b^0\to\Xi_c^{*+}\pi^0K^-)+\mathcal{A}(\Xi_b^0\to\Xi_c^{*+}\pi^-\overline{K}^0)\big]=0,
\end{align}
\begin{align}
SumI_-^2[\Xi_b^-,\Xi_c^{*0},\pi^+,\overline{K}^0] &=2\big[\sqrt{2}\mathcal{A}(\Xi_b^0\to\Xi_c^{*0}\pi^0\overline{K}^0)+\mathcal{A}(\Xi_b^0\to\Xi_c^{*0}\pi^+K^-) \notag \\ &+\sqrt{2}\mathcal{A}(\Xi_b^-\to\Xi_c^{*0}\pi^0K^-)-\mathcal{A}(\Xi_b^-\to\Xi_c^{*0}\pi^-\overline{K}^0)\big]=0,
\end{align}
\begin{align}
SumI_-^2[\Xi_b^-,\Xi_c^{*+},\pi^0,\overline{K}^0] &=-2\big[\mathcal{A}(\Xi_b^0\to\Xi_c^{*0}\pi^0\overline{K}^0)-\mathcal{A}(\Xi_b^0\to\Xi_c^{*+}\pi^0K^-) \notag \\ &+\sqrt{2}\mathcal{A}(\Xi_b^0\to\Xi_c^{*+}\pi^-\overline{K}^0)+\mathcal{A}(\Xi_b^-\to\Xi_c^{*0}\pi^0K^-) \notag \\ &+\sqrt{2}\big(-\mathcal{A}(\Xi_b^-\to\Xi_c^{*0}\pi^-\overline{K}^0)+\mathcal{A}(\Xi_b^-\to\Xi_c^{*+}\pi^-K^-)\big)\big]=0,
\end{align}
\begin{align}
SumI_-^2[\Xi_b^-,\Xi_c^{*+},\pi^+,K^-] &=-2\big[\mathcal{A}(\Xi_b^0\to\Xi_c^{*0}\pi^+K^-)-\sqrt{2}\mathcal{A}(\Xi_b^0\to\Xi_c^{*+}\pi^0K^-) \notag \\ &+\sqrt{2}\mathcal{A}(\Xi_b^-\to\Xi_c^{*0}\pi^0K^-)+\mathcal{A}(\Xi_b^-\to\Xi_c^{*+}\pi^-K^-)\big]=0,
\end{align}
\begin{align}
SumI_-^2[\Xi_b^0,\Omega_c^0,\pi^+,\pi^+] &=4\mathcal{A}(\Xi_b^0\to\Omega_c^0\pi^0\pi^0)-2\big[\mathcal{A}(\Xi_b^0\to\Omega_c^0\pi^+\pi^-) \notag \\ &+\mathcal{A}(\Xi_b^0\to\Omega_c^0\pi^-\pi^+)\big]=0,
\end{align}
\begin{align}
SumI_-^2[\Xi_b^-,\Omega_c^0,\pi^0,\pi^+] &=2\sqrt{2}\mathcal{A}(\Xi_b^0\to\Omega_c^0\pi^0\pi^0)-2\big[\sqrt{2}\mathcal{A}(\Xi_b^0\to\Omega_c^0\pi^-\pi^+) \notag \\ &+\mathcal{A}(\Xi_b^-\to\Omega_c^0\pi^0\pi^-)+2\mathcal{A}(\Xi_b^-\to\Omega_c^0\pi^-\pi^0)\big]=0,
\end{align}
\begin{align}
SumI_-^3[\Lambda_b^0,p,D^+,\pi^+] &=-6\big[\sqrt{2}\mathcal{A}(\Lambda_b^0\to nD^0\pi^0)+\mathcal{A}(\Lambda_b^0\to nD^+\pi^-) \notag \\ &+\mathcal{A}(\Lambda_b^0\to pD^0\pi^-)\big]=0,
\end{align}
\begin{align}
SumI_-^3[\Lambda_b^0,\Sigma^+,D^+,K^+] &=-6\big[\sqrt{2}\mathcal{A}(\Lambda_b^0\to\Sigma^0D^0K^0)+\mathcal{A}(\Lambda_b^0\to\Sigma^-D^0K^+) \notag \\ &+\mathcal{A}(\Lambda_b^0\to\Sigma^-D^+K^0)\big]=0,
\end{align}
\begin{align}
SumI_-^3[\Lambda_b^0,\Sigma^+,D_s^+,\pi^+] &=6\sqrt{2}\big[\mathcal{A}(\Lambda_b^0\to\Sigma^0D_s^+\pi^-)+\mathcal{A}(\Lambda_b^0\to\Sigma^-D_s^+\pi^0)\big]=0
\end{align}
\begin{align}
SumI_-^4[\Xi_b^-,\Sigma^+,D^+,\pi^+] &=-24\big[2\mathcal{A}(\Xi_b^0\to\Sigma^0D^0\pi^0)+\sqrt{2}\mathcal{A}(\Xi_b^0\to\Sigma^0D^+\pi^-) \notag \\ &-\mathcal{A}(\Xi_b^0\to\Sigma^-D^0\pi^+)-\mathcal{A}(\Xi_b^0\to\Sigma^+D^0\pi^-) \notag \\ &+\sqrt{2}\mathcal{A}(\Xi_b^0\to\Sigma^-D^+\pi^0)-\sqrt{2}\mathcal{A}(\Xi_b^-\to\Sigma^0D^0\pi^-) \notag \\ &-\sqrt{2}\mathcal{A}(\Xi_b^-\to\Sigma^-D^0\pi^0)-\mathcal{A}(\Xi_b^-\to\Sigma^-D^+\pi^-)\big]=0,
\end{align}
\begin{align}
SumI_-^3[\Xi_b^-,p,D^+,\overline{K}^0] &=-6\big[\mathcal{A}(\Xi_b^0\to nD^0\overline{K}^0)-\mathcal{A}(\Xi_b^0\to nD^+K^-) \notag \\ &-\mathcal{A}(\Xi_b^0\to pD^0K^-)+\mathcal{A}(\Xi_b^-\to nD^0K^-)\big]=0,
\end{align}
\begin{align}
SumI_-^3[\Xi_b^-,\Sigma^+,D^+,\eta_8] &=6\big[\sqrt{2}\mathcal{A}(\Xi_b^0\to\Sigma^0D^0\eta_8)+\mathcal{A}(\Xi_b^0\to\Sigma^-D^+\eta_8) \notag \\ &-\mathcal{A}(\Xi_b^-\to\Sigma^-D^0\eta_8)\big]=0,
\end{align}
\begin{align}
SumI_-^3[\Xi_b^-,\Sigma^+,D_s^+,\overline{K}^0] &=6\big[-\sqrt{2}\mathcal{A}(\Xi_b^0\to\Sigma^0D_s^+K^-)+\mathcal{A}(\Xi_b^0\to\Sigma^-D_s^+\overline{K}^0) \notag \\ &+\mathcal{A}(\Xi_b^-\to\Sigma^-D_s^+K^-)\big]=0,
\end{align}
\begin{align}
SumI_-^3[\Xi_b^-,\Xi^0,D^+,K^+] &=-6\big[\mathcal{A}(\Xi_b^0\to\Xi^0D^0K^0)-\mathcal{A}(\Xi_b^0\to\Xi^-D^0K^+) \notag \\ &-\mathcal{A}(\Xi_b^0\to\Xi^-D^+K^0)+\mathcal{A}(\Xi_b^-\to\Xi^-D^0K^0)\big]=0,
\end{align}
\begin{align}
SumI_-^3[\Xi_b^-,\Xi^0,D_s^+,\pi^+] &=6\big[\mathcal{A}(\Xi_b^0\to\Xi^0D_s^+\pi^-)-\sqrt{2}\mathcal{A}(\Xi_b^0\to\Xi^-D_s^+\pi^0) \notag \\ &+\mathcal{A}(\Xi_b^-\to\Xi^-D_s^+\pi^-)\big]=0,
\end{align}
\begin{align}
SumI_-^3[\Xi_b^-,\Lambda^0,D^+,\pi^+] &=6\big[\sqrt{2}\mathcal{A}(\Xi_b^0\to\Lambda^0D^0\pi^0)+\mathcal{A}(\Xi_b^0\to\Lambda^0D^+\pi^-) \notag \\ &-\mathcal{A}(\Xi_b^-\to\Lambda^0D^0\pi^-)\big]=0,
\end{align}
\begin{align}
SumI_-^3[\Lambda_b^0,\Sigma^+,D^+,\pi^+] &=6\big[2\mathcal{A}(\Lambda_b^0\to\Sigma^0D^0\pi^0)+\sqrt{2}\mathcal{A}(\Lambda_b^0\to\Sigma^0D^+\pi^-) \notag \\ &-\mathcal{A}(\Lambda_b^0\to\Sigma^-D^0\pi^+)-\mathcal{A}(\Lambda_b^0\to\Sigma^+D^0\pi^-) \notag \\ &+\sqrt{2}\mathcal{A}(\Lambda_b^0\to\Sigma^-D^+\pi^0)\big]=0,
\end{align}
\begin{align}
SumI_-^2[\Lambda_b^0,p,D^+,\overline{K}^0] &=2\big[\mathcal{A}(\Lambda_b^0\to nD^0\overline{K}^0)-\mathcal{A}(\Lambda_b^0\to nD^+K^-) \notag \\ &-\mathcal{A}(\Lambda_b^0\to pD^0K^-)\big]=0,
\end{align}
\begin{align}
SumI_-^2[\Lambda_b^0,\Sigma^+,D^+,\eta_8] &=-2\big[\sqrt{2}\mathcal{A}(\Lambda_b^0\to\Sigma^0D^0\eta_8)+\mathcal{A}(\Lambda_b^0\to\Sigma^-D^+\eta_8)\big]=0,
\end{align}
\begin{align}
SumI_-^2[\Lambda_b^0,\Sigma^+,D_s^+,\overline{K}^0] &=2\sqrt{2}\mathcal{A}(\Lambda_b^0\to\Sigma^0D_s^+K^-)-2\mathcal{A}(\Lambda_b^0\to\Sigma^-D_s^+\overline{K}^0)=0,
\end{align}
\begin{align}
SumI_-^2[\Lambda_b^0,\Xi^0,D^+,K^+] &=2\big[\mathcal{A}(\Lambda_b^0\to\Xi^0D^0K^0)-\mathcal{A}(\Lambda_b^0\to\Xi^-D^0K^+) \notag \\ &-\mathcal{A}(\Lambda_b^0\to\Xi^-D^+K^0)\big]=0,
\end{align}
\begin{align}
SumI_-^2[\Lambda_b^0,\Xi^0,D_s^+,\pi^+] &=-2\mathcal{A}(\Lambda_b^0\to\Xi^0D_s^+\pi^-)+2\sqrt{2}\mathcal{A}(\Lambda_b^0\to\Xi^-D_s^+\pi^0)=0,
\end{align}
\begin{align}
SumI_-^2[\Lambda_b^0,\Lambda^0,D^+,\pi^+] &=-2\big[\sqrt{2}\mathcal{A}(\Lambda_b^0\to\Lambda^0D^0\pi^0)+\mathcal{A}(\Lambda_b^0\to\Lambda^0D^+\pi^-)\big]=0,
\end{align}
\begin{align}
SumI_-^3[\Xi_b^-,\Sigma^+,D^+,\overline{K}^0] &=6\big[\sqrt{2}\mathcal{A}(\Xi_b^0\to\Sigma^0D^0\overline{K}^0)-\sqrt{2}\mathcal{A}(\Xi_b^0\to\Sigma^0D^+K^-) \notag \\ &+\mathcal{A}(\Xi_b^0\to\Sigma^+D^0K^-)+\mathcal{A}(\Xi_b^0\to\Sigma^-D^+\overline{K}^0) \notag \\ &+\sqrt{2}\mathcal{A}(\Xi_b^-\to\Sigma^0D^0K^-) -\mathcal{A}(\Xi_b^-\to\Sigma^-D^0\overline{K}^0) \notag \\ &+\mathcal{A}(\Xi_b^-\to\Sigma^-D^+K^-)\big]=0,
\end{align}
\begin{align}
SumI_-^3[\Xi_b^-,\Xi^0,D^+,\pi^+] &=6\big[\sqrt{2}\mathcal{A}(\Xi_b^0\to\Xi^0D^0\pi^0)+\mathcal{A}(\Xi_b^0\to\Xi^0D^+\pi^-) \notag \\ &+\mathcal{A}(\Xi_b^0\to\Xi^-D^0\pi^+)-\sqrt{2}\mathcal{A}(\Xi_b^0\to\Xi^-D^+\pi^0)-\mathcal{A}(\Xi_b^-\to\Xi^0D^0\pi^-) \notag \\ &+\sqrt{2}\mathcal{A}(\Xi_b^-\to\Xi^-D^0\pi^0)+\mathcal{A}(\Xi_b^-\to\Xi^-D^+\pi^-)\big]=0,
\end{align}
\begin{align}
SumI_-^2[\Xi_b^-,\Xi^0,D^+,\eta_8] &=-2\big[\mathcal{A}(\Xi_b^0\to\Xi^0D^0\eta_8)-\mathcal{A}(\Xi_b^0\to\Xi^-D^+\eta_8) \notag \\ &+\mathcal{A}(\Xi_b^-\to\Xi^-D^0\eta_8)\big]=0,
\end{align}
\begin{align}
SumI_-^2[\Xi_b^-,\Xi^0,D_s^+,\overline{K}^0] &=2\big[\mathcal{A}(\Xi_b^0\to\Xi^0D_s^+K^-)+\mathcal{A}(\Xi_b^0\to\Xi^-D_s^+\overline{K}^0) \notag \\ &+\mathcal{A}(\Xi_b^-\to\Xi^-D_s^+K^-)\big]=0,
\end{align}
\begin{align}
SumI_-^2[\Xi_b^-,\Lambda^0,D^+,\overline{K}^0] &=-2\big[\mathcal{A}(\Xi_b^0\to\Lambda^0D^0\overline{K}^0)-\mathcal{A}(\Xi_b^0\to\Lambda^0D^+K^-) \notag \\ &+\mathcal{A}(\Xi_b^-\to\Lambda^0D^0K^-)\big]=0,
\end{align}
\begin{align}
SumI_-^3[\Xi_b^0,\Sigma^+,D^+,\pi^+] &=6\big[2\mathcal{A}(\Xi_b^0\to\Sigma^0D^0\pi^0)+\sqrt{2}\mathcal{A}(\Xi_b^0\to\Sigma^0D^+\pi^-) \notag \\ &-\mathcal{A}(\Xi_b^0\to\Sigma^-D^0\pi^+)-\mathcal{A}(\Xi_b^0\to\Sigma^+D^0\pi^-) \notag \\ &+\sqrt{2}\mathcal{A}(\Xi_b^0\to\Sigma^-D^+\pi^0)\big]=0,
\end{align}
\begin{align}
SumI_-^3[\Xi_b^-,\Sigma^0,D^+,\pi^+] &=6\big[\sqrt{2}\mathcal{A}(\Xi_b^0\to\Sigma^0D^0\pi^0)+\mathcal{A}(\Xi_b^0\to\Sigma^0D^+\pi^-) \notag \\ &-\sqrt{2}\mathcal{A}(\Xi_b^0\to\Sigma^-D^0\pi^+)+2\mathcal{A}(\Xi_b^0\to\Sigma^-D^+\pi^0) \notag \\ &-\mathcal{A}(\Xi_b^-\to\Sigma^0D^0\pi^-)-2\mathcal{A}(\Xi_b^-\to\Sigma^-D^0\pi^0) \notag \\ &-\sqrt{2}\mathcal{A}(\Xi_b^-\to\Sigma^-D^+\pi^-)\big]=0,
\end{align}
\begin{align}
SumI_-^3[\Xi_b^-,\Sigma^+,D^0,\pi^+] &=6\big[-2\mathcal{A}(\Xi_b^0\to\Sigma^0D^0\pi^0)+\mathcal{A}(\Xi_b^0\to\Sigma^-D^0\pi^+) \notag \\ &+\mathcal{A}(\Xi_b^0\to\Sigma^+D^0\pi^-)+\sqrt{2}\big(\mathcal{A}(\Xi_b^-\to\Sigma^0D^0\pi^-) \notag \\ &+\mathcal{A}(\Xi_b^-\to\Sigma^-D^0\pi^0)\big)\big]=0,
\end{align}
\begin{align}
SumI_-^3[\Xi_b^-,\Sigma^+,D^+,\pi^0] &=6\big[\sqrt{2}\mathcal{A}(\Xi_b^0\to\Sigma^0D^0\pi^0)+2\mathcal{A}(\Xi_b^0\to\Sigma^0D^+\pi^-) \notag \\ &-\sqrt{2}\mathcal{A}(\Xi_b^0\to\Sigma^+D^0\pi^-)+\mathcal{A}(\Xi_b^0\to\Sigma^-D^+\pi^0) \notag \\ &-2\mathcal{A}(\Xi_b^-\to\Sigma^0D^0\pi^-)-\mathcal{A}(\Xi_b^-\to\Sigma^-D^0\pi^0) \notag \\ &-\sqrt{2}\mathcal{A}(\Xi_b^-\to\Sigma^-D^+\pi^-)\big]=0,
\end{align}
\begin{align}
SumI_-^2[\Lambda_b^0,\Sigma^0,D^+,\pi^+] &=-2\big[\sqrt{2}\mathcal{A}(\Lambda_b^0\to\Sigma^0D^0\pi^0)+\mathcal{A}(\Lambda_b^0\to\Sigma^0D^+\pi^-) \notag \\ &-\sqrt{2}\mathcal{A}(\Lambda_b^0\to\Sigma^-D^0\pi^+)+2\mathcal{A}(\Lambda_b^0\to\Sigma^-D^+\pi^0)\big]=0,
\end{align}
\begin{align}
SumI_-^2[\Lambda_b^0,\Sigma^+,D^0,\pi^+] &=4\mathcal{A}(\Lambda_b^0\to\Sigma^0D^0\pi^0)-2\big[\mathcal{A}(\Lambda_b^0\to\Sigma^-D^0\pi^+) \notag \\ &+\mathcal{A}(\Lambda_b^0\to\Sigma^+D^0\pi^-)\big]=0,
\end{align}
\begin{align}
SumI_-^2[\Lambda_b^0,\Sigma^+,D^+,\pi^0] &=-2\big[\sqrt{2}\mathcal{A}(\Lambda_b^0\to\Sigma^0D^0\pi^0)+2\mathcal{A}(\Lambda_b^0\to\Sigma^0D^+\pi^-) \notag \\ &-\sqrt{2}\mathcal{A}(\Lambda_b^0\to\Sigma^+D^0\pi^-)+\mathcal{A}(\Lambda_b^0\to\Sigma^-D^+\pi^0)\big]=0,
\end{align}
\begin{align}
SumI_-^2[\Xi_b^0,\Sigma^+,D^+,\overline{K}^0] &=-2\big[\sqrt{2}\mathcal{A}(\Xi_b^0\to\Sigma^0D^0\overline{K}^0)-\sqrt{2}\mathcal{A}(\Xi_b^0\to\Sigma^0D^+K^-) \notag \\ &+\mathcal{A}(\Xi_b^0\to\Sigma^+D^0K^-)+\mathcal{A}(\Xi_b^0\to\Sigma^-D^+\overline{K}^0)\big]=0,
\end{align}
\begin{align}
SumI_-^2[\Xi_b^-,\Sigma^0,D^+,\overline{K}^0] &=-2\big[\mathcal{A}(\Xi_b^0\to\Sigma^0D^0\overline{K}^0)-\mathcal{A}(\Xi_b^0\to\Sigma^0D^+K^-) \notag \\ &+\sqrt{2}\mathcal{A}(\Xi_b^0\to\Sigma^-D^+\overline{K}^0)+\mathcal{A}(\Xi_b^-\to\Sigma^0D^0K^-) \notag \\ &+\sqrt{2}\big(-\mathcal{A}(\Xi_b^-\to\Sigma^-D^0\overline{K}^0)+\mathcal{A}(\Xi_b^-\to\Sigma^-D^+K^-)\big)\big]=0,
\end{align}
\begin{align}
SumI_-^2[\Xi_b^-,\Sigma^+,D^0,\overline{K}^0] &=2\big[\sqrt{2}\mathcal{A}(\Xi_b^0\to\Sigma^0D^0\overline{K}^0)+\mathcal{A}(\Xi_b^0\to\Sigma^+D^0K^-) \notag \\ &+\sqrt{2}\mathcal{A}(\Xi_b^-\to\Sigma^0D^0K^-)-\mathcal{A}(\Xi_b^-\to\Sigma^-D^0\overline{K}^0)\big]=0,
\end{align}
\begin{align}
SumI_-^2[\Xi_b^-,\Sigma^+,D^+,K^-] &=2\sqrt{2}\mathcal{A}(\Xi_b^0\to\Sigma^0D^+K^-)-2\big[\mathcal{A}(\Xi_b^0\to\Sigma^+D^0K^-) \notag \\ &+\sqrt{2}\mathcal{A}(\Xi_b^-\to\Sigma^0D^0K^-)+\mathcal{A}(\Xi_b^-\to\Sigma^-D^+K^-)\big]=0,
\end{align}
\begin{align}
SumI_-^2[\Xi_b^0,\Xi^0,D^+,\pi^+] &=-2\big[\sqrt{2}\mathcal{A}(\Xi_b^0\to\Xi^0D^0\pi^0)+\mathcal{A}(\Xi_b^0\to\Xi^0D^+\pi^-) \notag \\ &+\mathcal{A}(\Xi_b^0\to\Xi^-D^0\pi^+)-\sqrt{2}\mathcal{A}(\Xi_b^0\to\Xi^-D^+\pi^0)\big]=0,
\end{align}
\begin{align}
SumI_-^2[\Xi_b^-,\Xi^-,D^+,\pi^+] &=-2\big[\mathcal{A}(\Xi_b^0\to\Xi^-D^0\pi^+)-\sqrt{2}\mathcal{A}(\Xi_b^0\to\Xi^-D^+\pi^0) \notag \\ &+\sqrt{2}\mathcal{A}(\Xi_b^-\to\Xi^-D^0\pi^0)+\mathcal{A}(\Xi_b^-\to\Xi^-D^+\pi^-)\big]=0,
\end{align}
\begin{align}
SumI_-^2[\Xi_b^-,\Xi^0,D^0,\pi^+] &=2\big[\sqrt{2}\mathcal{A}(\Xi_b^0\to\Xi^0D^0\pi^0)+\mathcal{A}(\Xi_b^0\to\Xi^-D^0\pi^+) \notag \\ &-\mathcal{A}(\Xi_b^-\to\Xi^0D^0\pi^-)+\sqrt{2}\mathcal{A}(\Xi_b^-\to\Xi^-D^0\pi^0)\big]=0,
\end{align}
\begin{align}
SumI_-^2[\Xi_b^-,\Xi^0,D^+,\pi^0] &=-2\big[\mathcal{A}(\Xi_b^0\to\Xi^0D^0\pi^0)+\sqrt{2}\mathcal{A}(\Xi_b^0\to\Xi^0D^+\pi^-) \notag \\ &-\mathcal{A}(\Xi_b^0\to\Xi^-D^+\pi^0)-\sqrt{2}\mathcal{A}(\Xi_b^-\to\Xi^0D^0\pi^-) \notag \\ &+\mathcal{A}(\Xi_b^-\to\Xi^-D^0\pi^0)+\sqrt{2}\mathcal{A}(\Xi_b^-\to\Xi^-D^+\pi^-)\big]=0,
\end{align}
\begin{align}
SumI_-^3[\Lambda_b^0,\Sigma^{*+},D^+,K^+] &=6\big[\sqrt{2}\mathcal{A}(\Lambda_b^0\to\Sigma^{*0}D^0K^0)+\mathcal{A}(\Lambda_b^0\to\Sigma^{*-}D^0K^+) \notag \\ &+\mathcal{A}(\Lambda_b^0\to\Sigma^{*-}D^+K^0)\big]=0,
\end{align}
\begin{align}
SumI_-^3[\Lambda_b^0,\Sigma^{*+},D_s^+,\pi^+] &=-6\sqrt{2}\big[\mathcal{A}(\Lambda_b^0\to\Sigma^{*0}D_s^+\pi^-)+\mathcal{A}(\Lambda_b^0\to\Sigma^{*-}D_s^+\pi^0)\big]=0,
\end{align}
\begin{align}
SumI_-^3[\Lambda_b^0,\Delta^{++},D^+,\pi^+] &=-24\big[\sqrt{6}\mathcal{A}(\Lambda_b^0\to\Delta^0D^0\pi^0)+\sqrt{3}\mathcal{A}(\Lambda_b^0\to\Delta^0D^+\pi^-) \notag \\ &-\mathcal{A}(\Lambda_b^0\to\Delta^-D^0\pi^+)+\sqrt{3}\mathcal{A}(\Lambda_b^0\to\Delta^+D^0\pi^-) \notag \\ &+\sqrt{2}\mathcal{A}(\Lambda_b^0\to\Delta^-D^+\pi^0)\big]=0,
\end{align}
\begin{align}
SumI_-^3[\Lambda_b^0,\Delta^{++},D^+,\eta_8] &=6\big[\sqrt{3}\mathcal{A}(\Lambda_b^0\to\Delta^0D^0\eta_8)+\mathcal{A}(\Lambda_b^0\to\Delta^-D^+\eta_8)\big]=0,
\end{align}
\begin{align}
SumI_-^3[\Lambda_b^0,\Delta^{++},D_s^+,\overline{K}^0] &=-6\sqrt{3}\mathcal{A}(\Lambda_b^0\to\Delta^0D_s^+K^-)+6\mathcal{A}(\Lambda_b^0\to\Delta^-D_s^+\overline{K}^0)=0,
\end{align}
\begin{align}
SumI_-^3[\Xi_b^-,\Xi^{*0},D^+,K^+] &=-6\big[\mathcal{A}(\Xi_b^0\to\Xi^{*0}D^0K^0)+\mathcal{A}(\Xi_b^0\to\Xi^{*-}D^0K^+) \notag \\ &+\mathcal{A}(\Xi_b^0\to\Xi^{*-}D^+K^0)-\mathcal{A}(\Xi_b^-\to\Xi^{*-}D^0K^0)\big]=0,
\end{align}
\begin{align}
SumI_-^3[\Xi_b^-,\Xi^{*0},D_s^+,\pi^+] &=6\big[\mathcal{A}(\Xi_b^0\to\Xi^{*0}D_s^+\pi^-)+\sqrt{2}\mathcal{A}(\Xi_b^0\to\Xi^{*-}D_s^+\pi^0) \notag \\ &-\mathcal{A}(\Xi_b^-\to\Xi^{*-}D_s^+\pi^-)\big]=0,
\end{align}
\begin{align}
SumI_-^4[\Xi_b^-,\Sigma^{*+},D^+,\pi^+] &=24\big[2\mathcal{A}(\Xi_b^0\to\Sigma^{*0}D^0\pi^0)+\sqrt{2}\mathcal{A}(\Xi_b^0\to\Sigma^{*0}D^+\pi^-) \notag \\ &-\mathcal{A}(\Xi_b^0\to\Sigma^{*-}D^0\pi^+)+\mathcal{A}(\Xi_b^0\to\Sigma^{*+}D^0\pi^-) \notag \\ &+\sqrt{2}\mathcal{A}(\Xi_b^0\to\Sigma^{*-}D^+\pi^0)-\sqrt{2}\mathcal{A}(\Xi_b^-\to\Sigma^{*0}D^0\pi^-) \notag \\ &-\sqrt{2}\mathcal{A}(\Xi_b^-\to\Sigma^{*-}D^0\pi^0)-\mathcal{A}(\Xi_b^-\to\Sigma^{*-}D^+\pi^-)\big]=0,
\end{align}
\begin{align}
SumI_-^3[\Xi_b^-,\Sigma^{*+},D^+,\eta_8] &=-6\big[\sqrt{2}\mathcal{A}(\Xi_b^0\to\Sigma^{*0}D^0\eta_8)+\mathcal{A}(\Xi_b^0\to\Sigma^{*-}D^+\eta_8) \notag \\ &-\mathcal{A}(\Xi_b^-\to\Sigma^{*-}D^0\eta_8)\big]=0,
\end{align}
\begin{align}
SumI_-^3[\Xi_b^-,\Sigma^{*+},D_s^+,\overline{K}^0] &=6\sqrt{2}\mathcal{A}(\Xi_b^0\to\Sigma^{*0}D_s^+K^-)-6\big[\mathcal{A}(\Xi_b^0\to\Sigma^{*-}D_s^+\overline{K}^0) \notag \\ &+\mathcal{A}(\Xi_b^-\to\Sigma^{*-}D_s^+K^-)\big]=0,
\end{align}
\begin{align}
SumI_-^4[\Xi_b^-,\Delta^{++},D^+,\overline{K}^0] &=-24\big[\sqrt{3}\mathcal{A}(\Xi_b^0\to\Delta^0D^0\overline{K}^0)-\sqrt{3}\mathcal{A}(\Xi_b^0\to\Delta^0D^+K^-) \notag \\ &-\sqrt{3}\mathcal{A}(\Xi_b^0\to\Delta^+D^0K^-)+\mathcal{A}(\Xi_b^0\to\Delta^-D^+\overline{K}^0) \notag \\ &+\sqrt{3}\mathcal{A}(\Xi_b^-\to\Delta^0D^0K^-)-\mathcal{A}(\Xi_b^-\to\Delta^-D^0\overline{K}^0) \notag \\ &+\mathcal{A}(\Xi_b^-\to\Delta^-D^+K^-)\big]=0,
\end{align}
\begin{align}
SumI_-^2[\Lambda_b^0,\Xi^{*0},D^+,K^+] &=2\big[\mathcal{A}(\Lambda_b^0\to\Xi^{*0}D^0K^0)+\mathcal{A}(\Lambda_b^0\to\Xi^{*-}D^0K^+) \notag \\ &+\mathcal{A}(\Lambda_b^0\to\Xi^{*-}D^+K^0)\big]=0,
\end{align}
\begin{align}
SumI_-^2[\Lambda_b^0,\Xi^{*0},D_s^+,\pi^+] &=-2\big[\mathcal{A}(\Lambda_b^0\to\Xi^{*0}D_s^+\pi^-)+\sqrt{2}\mathcal{A}(\Lambda_b^0\to\Xi^{*-}D_s^+\pi^0)\big]=0,
\end{align}
\begin{align}
SumI_-^3[\Lambda_b^0,\Sigma^{*+},D^+,\pi^+] &=-6\big[2\mathcal{A}(\Lambda_b^0\to\Sigma^{*0}D^0\pi^0)+\sqrt{2}\mathcal{A}(\Lambda_b^0\to\Sigma^{*0}D^+\pi^-) \notag \\ &-\mathcal{A}(\Lambda_b^0\to\Sigma^{*-}D^0\pi^+)+\mathcal{A}(\Lambda_b^0\to\Sigma^{*+}D^0\pi^-) \notag \\ &+\sqrt{2}\mathcal{A}(\Lambda_b^0\to\Sigma^{*-}D^+\pi^0)\big]=0,
\end{align}
\begin{align}
SumI_-^2[\Lambda_b^0,\Sigma^{*+},D^+,\eta_8] &=2\big[\sqrt{2}\mathcal{A}(\Lambda_b^0\to\Sigma^{*0}D^0\eta_8)+\mathcal{A}(\Lambda_b^0\to\Sigma^{*-}D^+\eta_8)\big]=0,
\end{align}
\begin{align}
SumI_-^2[\Lambda_b^0,\Sigma^{*+},D_s^+,\overline{K}^0] &=-2\sqrt{2}\mathcal{A}(\Lambda_b^0\to\Sigma^{*0}D_s^+K^-)+2\mathcal{A}(\Lambda_b^0\to\Sigma^{*-}D_s^+\overline{K}^0)=0,
\end{align}
\begin{align}
SumI_-^3[\Lambda_b^0,\Delta^{++},D^+,\overline{K}^0] &=6\big[\sqrt{3}\mathcal{A}(\Lambda_b^0\to\Delta^0D^0\overline{K}^0)-\sqrt{3}\mathcal{A}(\Lambda_b^0\to\Delta^0D^+K^-) \notag \\ &-\sqrt{3}\mathcal{A}(\Lambda_b^0\to\Delta^+D^0K^-)+\mathcal{A}(\Lambda_b^0\to\Delta^-D^+\overline{K}^0)\big]=0,
\end{align}
\begin{align}
SumI_-^2[\Xi_b^-,\Omega^-,D^+,K^+] &=-2\big[\mathcal{A}(\Xi_b^0\to\Omega^-D^0K^+)+\mathcal{A}(\Xi_b^0\to\Omega^-D^+K^0) \notag \\ &-\mathcal{A}(\Xi_b^-\to\Omega^-D^0K^0)\big]=0,
\end{align}
\begin{align}
SumI_-^2[\Xi_b^-,\Omega^-,D_s^+,\pi^+] &=2\sqrt{2}\mathcal{A}(\Xi_b^0\to\Omega^-D_s^+\pi^0)-2\mathcal{A}(\Xi_b^-\to\Omega^-D_s^+\pi^-)=0,
\end{align}
\begin{align}
SumI_-^2[\Xi_b^-,\Xi^{*0},D^+,\eta_8] &=-2\big[\mathcal{A}(\Xi_b^0\to\Xi^{*0}D^0\eta_8)+\mathcal{A}(\Xi_b^0\to\Xi^{*-}D^+\eta_8) \notag \\ &-\mathcal{A}(\Xi_b^-\to\Xi^{*-}D^0\eta_8)\big]=0,
\end{align}
\begin{align}
SumI_-^2[\Xi_b^-,\Xi^{*0},D_s^+,\overline{K}^0] &=2\big[\mathcal{A}(\Xi_b^0\to\Xi^{*0}D_s^+K^-)-\mathcal{A}(\Xi_b^0\to\Xi^{*-}D_s^+\overline{K}^0) \notag \\ &-\mathcal{A}(\Xi_b^-\to\Xi^{*-}D_s^+K^-)\big]=0,
\end{align}
\begin{align}
SumI_-^3[\Xi_b^-,\Sigma^{*+},D^+,\overline{K}^0] &=-6\big[\sqrt{2}\mathcal{A}(\Xi_b^0\to\Sigma^{*0}D^0\overline{K}^0)-\sqrt{2}\mathcal{A}(\Xi_b^0\to\Sigma^{*0}D^+K^-) \notag \\ &-\mathcal{A}(\Xi_b^0\to\Sigma^{*+}D^0K^-)+\mathcal{A}(\Xi_b^0\to\Sigma^{*-}D^+\overline{K}^0) \notag \\ &+\sqrt{2}\mathcal{A}(\Xi_b^-\to\Sigma^{*0}D^0K^-)-\mathcal{A}(\Xi_b^-\to\Sigma^{*-}D^0\overline{K}^0) \notag \\ &+\mathcal{A}(\Xi_b^-\to\Sigma^{*-}D^+K^-)\big]=0,
\end{align}
\begin{align}
SumI_-^3[\Lambda_b^0,\Delta^+,D^+,\pi^+] &=-6\big[2\sqrt{2}\mathcal{A}(\Lambda_b^0\to\Delta^0D^0\pi^0)+2\mathcal{A}(\Lambda_b^0\to\Delta^0D^+\pi^-) \notag \\ &-\sqrt{3}\mathcal{A}(\Lambda_b^0\to\Delta^-D^0\pi^+)+\mathcal{A}(\Lambda_b^0\to\Delta^+D^0\pi^-) \notag \\ &+\sqrt{6}\mathcal{A}(\Lambda_b^0\to\Delta^-D^+\pi^0)\big]=0,
\end{align}
\begin{align}
SumI_-^3[\Lambda_b^0,\Delta^{++},D^0,\pi^+] &=-6\big[\sqrt{6}\mathcal{A}(\Lambda_b^0\to\Delta^0D^0\pi^0)-\mathcal{A}(\Lambda_b^0\to\Delta^-D^0\pi^+) \notag \\ &+\sqrt{3}\mathcal{A}(\Lambda_b^0\to\Delta^+D^0\pi^-)\big]=0,
\end{align}
\begin{align}
SumI_-^3[\Lambda_b^0,\Delta^{++},D^+,\pi^0] &=6\big[\sqrt{3}\mathcal{A}(\Lambda_b^0\to\Delta^0D^0\pi^0)+\sqrt{6}\mathcal{A}(\Lambda_b^0\to\Delta^0D^+\pi^-) \notag \\ &+\sqrt{6}\mathcal{A}(\Lambda_b^0\to\Delta^+D^0\pi^-)+\mathcal{A}(\Lambda_b^0\to\Delta^-D^+\pi^0)\big]=0,
\end{align}
\begin{align}
SumI_-^3[\Xi_b^0,\Sigma^{*+},D^+,\pi^+] &=-6\big[2\mathcal{A}(\Xi_b^0\to\Sigma^{*0}D^0\pi^0)+\sqrt{2}\mathcal{A}(\Xi_b^0\to\Sigma^{*0}D^+\pi^-) \notag \\ &-\mathcal{A}(\Xi_b^0\to\Sigma^{*-}D^0\pi^+)+\mathcal{A}(\Xi_b^0\to\Sigma^{*+}D^0\pi^-) \notag \\ &+\sqrt{2}\mathcal{A}(\Xi_b^0\to\Sigma^{*-}D^+\pi^0)\big]=0,
\end{align}
\begin{align}
SumI_-^3[\Xi_b^-,\Sigma^{*0},D^+,\pi^+] &=6\big[\sqrt{2}\mathcal{A}(\Xi_b^0\to\Sigma^{*0}D^0\pi^0)+\mathcal{A}(\Xi_b^0\to\Sigma^{*0}D^+\pi^-) \notag \\ &-\sqrt{2}\mathcal{A}(\Xi_b^0\to\Sigma^{*-}D^0\pi^+)+2\mathcal{A}(\Xi_b^0\to\Sigma^{*-}D^+\pi^0) \notag \\ &-\mathcal{A}(\Xi_b^-\to\Sigma^{*0}D^0\pi^-)-2\mathcal{A}(\Xi_b^-\to\Sigma^{*-}D^0\pi^0) \notag \\ &-\sqrt{2}\mathcal{A}(\Xi_b^-\to\Sigma^{*-}D^+\pi^-)\big]=0,
\end{align}
\begin{align}
SumI_-^3[\Xi_b^-,\Sigma^{*+},D^0,\pi^+] &=-6\big[-2\mathcal{A}(\Xi_b^0\to\Sigma^{*0}D^0\pi^0)+\mathcal{A}(\Xi_b^0\to\Sigma^{*-}D^0\pi^+) \notag \\ &-\mathcal{A}(\Xi_b^0\to\Sigma^{*+}D^0\pi^-)+\sqrt{2}\big(\mathcal{A}(\Xi_b^-\to\Sigma^{*0}D^0\pi^-) \notag \\ &+\mathcal{A}(\Xi_b^-\to\Sigma^{*-}D^0\pi^0)\big)\big]=0,
\end{align}
\begin{align}
SumI_-^3[\Xi_b^-,\Sigma^{*+},D^+,\pi^0] &=-6\big[\sqrt{2}\mathcal{A}(\Xi_b^0\to\Sigma^{*0}D^0\pi^0)+2\mathcal{A}(\Xi_b^0\to\Sigma^{*0}D^+\pi^-) \notag \\ &+\sqrt{2}\mathcal{A}(\Xi_b^0\to\Sigma^{*+}D^0\pi^-)+\mathcal{A}(\Xi_b^0\to\Sigma^{*-}D^+\pi^0) \notag \\ &-2\mathcal{A}(\Xi_b^-\to\Sigma^{*0}D^0\pi^-)-\mathcal{A}(\Xi_b^-\to\Sigma^{*-}D^0\pi^0) \notag \\ &-\sqrt{2}\mathcal{A}(\Xi_b^-\to\Sigma^{*-}D^+\pi^-)\big]=0,
\end{align}
\begin{align}
SumI_-^3[\Xi_b^0,\Delta^{++},D^+,\overline{K}^0] &=6\big[\sqrt{3}\mathcal{A}(\Xi_b^0\to\Delta^0D^0\overline{K}^0)-\sqrt{3}\mathcal{A}(\Xi_b^0\to\Delta^0D^+K^-) \notag \\ &-\sqrt{3}\mathcal{A}(\Xi_b^0\to\Delta^+D^0K^-)+\mathcal{A}(\Xi_b^0\to\Delta^-D^+\overline{K}^0)\big]=0,
\end{align}
\begin{align}
SumI_-^3[\Xi_b^-,\Delta^+,D^+,\overline{K}^0] &=6\big[-2\mathcal{A}(\Xi_b^0\to\Delta^0D^0\overline{K}^0)+2\mathcal{A}(\Xi_b^0\to\Delta^0D^+K^-) \notag \\ &+\mathcal{A}(\Xi_b^0\to\Delta^+D^0K^-)-\sqrt{3}\mathcal{A}(\Xi_b^0\to\Delta^-D^+\overline{K}^0) \notag \\ &-2\mathcal{A}(\Xi_b^-\to\Delta^0D^0K^-)+\sqrt{3}\big(\mathcal{A}(\Xi_b^-\to\Delta^-D^0\overline{K}^0) \notag \\ &-\mathcal{A}(\Xi_b^-\to\Delta^-D^+K^-)\big)\big]=0,
\end{align}
\begin{align}
SumI_-^3[\Xi_b^-,\Delta^{++},D^0,\overline{K}^0] &=6\big[-\sqrt{3}\mathcal{A}(\Xi_b^0\to\Delta^0D^0\overline{K}^0)+\sqrt{3}\mathcal{A}(\Xi_b^0\to\Delta^+D^0K^-) \notag \\ &-\sqrt{3}\mathcal{A}(\Xi_b^-\to\Delta^0D^0K^-)+\mathcal{A}(\Xi_b^-\to\Delta^-D^0\overline{K}^0)\big]=0,
\end{align}
\begin{align}
SumI_-^3[\Xi_b^-,\Delta^{++},D^+,K^-] &=6\big[-\sqrt{3}\mathcal{A}(\Xi_b^0\to\Delta^0D^+K^-)-\sqrt{3}\mathcal{A}(\Xi_b^0\to\Delta^+D^0K^-) \notag \\ &+\sqrt{3}\mathcal{A}(\Xi_b^-\to\Delta^0D^0K^-)+\mathcal{A}(\Xi_b^-\to\Delta^-D^+K^-)\big]=0,
\end{align}
\begin{align}
SumI_-^2[\Lambda_b^0,\Sigma^{*0},D^+,\pi^+] &=-2\big[\sqrt{2}\mathcal{A}(\Lambda_b^0\to\Sigma^{*0}D^0\pi^0)+\mathcal{A}(\Lambda_b^0\to\Sigma^{*0}D^+\pi^-) \notag \\ &-\sqrt{2}\mathcal{A}(\Lambda_b^0\to\Sigma^{*-}D^0\pi^+)+2\mathcal{A}(\Lambda_b^0\to\Sigma^{*-}D^+\pi^0)\big]=0,
\end{align}
\begin{align}
SumI_-^2[\Lambda_b^0,\Sigma^{*+},D^0,\pi^+] &=-2\big[2\mathcal{A}(\Lambda_b^0\to\Sigma^{*0}D^0\pi^0)-\mathcal{A}(\Lambda_b^0\to\Sigma^{*-}D^0\pi^+) \notag \\ &+\mathcal{A}(\Lambda_b^0\to\Sigma^{*+}D^0\pi^-)\big]=0,
\end{align}
\begin{align}
SumI_-^2[\Lambda_b^0,\Sigma^{*+},D^+,\pi^0] &=2\big[\sqrt{2}\mathcal{A}(\Lambda_b^0\to\Sigma^{*0}D^0\pi^0)+2\mathcal{A}(\Lambda_b^0\to\Sigma^{*0}D^+\pi^-) \notag \\ &+\sqrt{2}\mathcal{A}(\Lambda_b^0\to\Sigma^{*+}D^0\pi^-)+\mathcal{A}(\Lambda_b^0\to\Sigma^{*-}D^+\pi^0)\big]=0,
\end{align}
\begin{align}
SumI_-^2[\Lambda_b^0,\Delta^+,D^+,\overline{K}^0] &=4\mathcal{A}(\Lambda_b^0\to\Delta^0D^0\overline{K}^0)-4\mathcal{A}(\Lambda_b^0\to\Delta^0D^+K^-) \notag \\ &-2\mathcal{A}(\Lambda_b^0\to\Delta^+D^0K^-)+2\sqrt{3}\mathcal{A}(\Lambda_b^0\to\Delta^-D^+\overline{K}^0)=0,
\end{align}
\begin{align}
SumI_-^2[\Lambda_b^0,\Delta^{++},D^0,\overline{K}^0] &=2\sqrt{3}\big[\mathcal{A}(\Lambda_b^0\to\Delta^0D^0\overline{K}^0)-\mathcal{A}(\Lambda_b^0\to\Delta^+D^0K^-)\big]=0,
\end{align}
\begin{align}
SumI_-^2[\Lambda_b^0,\Delta^{++},D^+,K^-] &=2\sqrt{3}\big[\mathcal{A}(\Lambda_b^0\to\Delta^0D^+K^-)+\mathcal{A}(\Lambda_b^0\to\Delta^+D^0K^-)\big]=0,
\end{align}
\begin{align}
SumI_-^2[\Xi_b^0,\Sigma^{*+},D^+,\overline{K}^0] &=2\big[\sqrt{2}\mathcal{A}(\Xi_b^0\to\Sigma^{*0}D^0\overline{K}^0)-\sqrt{2}\mathcal{A}(\Xi_b^0\to\Sigma^{*0}D^+K^-) \notag \\ &-\mathcal{A}(\Xi_b^0\to\Sigma^{*+}D^0K^-)+\mathcal{A}(\Xi_b^0\to\Sigma^{*-}D^+\overline{K}^0)\big]=0,
\end{align}
\begin{align}
SumI_-^2[\Xi_b^-,\Sigma^{*0},D^+,\overline{K}^0] &=-2\big[\mathcal{A}(\Xi_b^0\to\Sigma^{*0}D^0\overline{K}^0)-\mathcal{A}(\Xi_b^0\to\Sigma^{*0}D^+K^-) \notag \\ &+\sqrt{2}\mathcal{A}(\Xi_b^0\to\Sigma^{*-}D^+\overline{K}^0)+\mathcal{A}(\Xi_b^-\to\Sigma^{*0}D^0K^-) \notag \\ &+\sqrt{2}\big(-\mathcal{A}(\Xi_b^-\to\Sigma^{*-}D^0\overline{K}^0)+\mathcal{A}(\Xi_b^-\to\Sigma^{*-}D^+K^-)\big)\big]=0,
\end{align}
\begin{align}
SumI_-^2[\Xi_b^-,\Sigma^{*+},D^0,\overline{K}^0] &=2\big[-\sqrt{2}\mathcal{A}(\Xi_b^0\to\Sigma^{*0}D^0\overline{K}^0)+\mathcal{A}(\Xi_b^0\to\Sigma^{*+}D^0K^-) \notag \\ &-\sqrt{2}\mathcal{A}(\Xi_b^-\to\Sigma^{*0}D^0K^-)+\mathcal{A}(\Xi_b^-\to\Sigma^{*-}D^0\overline{K}^0)\big]=0,
\end{align}
\begin{align}
SumI_-^2[\Xi_b^-,\Sigma^{*+},D^+,K^-] &=2\big[-\sqrt{2}\mathcal{A}(\Xi_b^0\to\Sigma^{*0}D^+K^-)-\mathcal{A}(\Xi_b^0\to\Sigma^{*+}D^0K^-) \notag \\ &+\sqrt{2}\mathcal{A}(\Xi_b^-\to\Sigma^{*0}D^0K^-)
+\mathcal{A}(\Xi_b^-\to\Sigma^{*-}D^+K^-)\big]=0.
\end{align}

\subsection{$b\to u\overline u d/s$ modes}

\begin{align}
SumI_-^3[\Lambda_b^0,\Sigma^{*+},\pi^+,K^+] &=-6\big[2\mathcal{A}(\Lambda_b^0\to\Sigma^{*0}\pi^0K^0) +\sqrt{2}\mathcal{A}(\Lambda_b^0\to\Sigma^{*0}\pi^-K^+) \notag\\ &+\sqrt{2}\mathcal{A}(\Lambda_b^0\to\Sigma^{*-}\pi^0K^+) -\mathcal{A}(\Lambda_b^0\to\Sigma^{*-}\pi^+K^0) \notag\\ &-\mathcal{A}(\Lambda_b^0\to\Sigma^{*+}\pi^-K^0)\big]=0,
\end{align}
\begin{align}
SumI_-^4[\Lambda_b^0,\Delta^{++},\pi^+,\pi^+] &=24\big[2\sqrt{3}\mathcal{A}(\Lambda_b^0\to\Delta^0\pi^0\pi^0)-\sqrt{3}\mathcal{A}(\Lambda_b^0\to\Delta^0\pi^-\pi^+)\notag \\ &-\sqrt{3}\mathcal{A}(\Lambda_b^0\to\Delta^0\pi^+\pi^-)-\sqrt{2}\mathcal{A}(\Lambda_b^0\to\Delta^-\pi^0\pi^+)\notag \\ &+\sqrt{6}\mathcal{A}(\Lambda_b^0\to\Delta^+\pi^0\pi^-)+\mathcal{A}(\Lambda_b^0\to\Delta^{++}\pi^-\pi^-)\notag \\ &-\sqrt{2}\mathcal{A}(\Lambda_b^0\to\Delta^-\pi^+\pi^0)+\sqrt{6}\mathcal{A}(\Lambda_b^0\to\Delta^+\pi^-\pi^0)\big]=0,
\end{align}
\begin{align}
SumI_-^3[\Lambda_b^0,\Delta^+,\pi^+,\pi^+] &=6\big[4\mathcal{A}(\Lambda_b^0\to\Delta^0\pi^0\pi^0)-2\mathcal{A}(\Lambda_b^0\to\Delta^0\pi^-\pi^+) \notag \\ &-2\mathcal{A}(\Lambda_b^0\to\Delta^0\pi^+\pi^-)+\sqrt{2}\big(-\sqrt{3}\mathcal{A}(\Lambda_b^0\to\Delta^-\pi^0\pi^+)+\mathcal{A}(\Lambda_b^0\to\Delta^+\pi^0\pi^-) \notag \\ &-\sqrt{3}\mathcal{A}(\Lambda_b^0\to\Delta^-\pi^+\pi^0)+\mathcal{A}(\Lambda_b^0\to\Delta^+\pi^-\pi^0)\big)\big]=0,
\end{align}
\begin{align}
SumI_-^3[\Lambda_b^0,\Delta^{++},\pi^0,\pi^+] &=-6\big[\sqrt{6}\mathcal{A}(\Lambda_b^0\to\Delta^0\pi^0\pi^0)-\sqrt{6}\mathcal{A}(\Lambda_b^0\to\Delta^0\pi^-\pi^+) \notag \\ &-\mathcal{A}(\Lambda_b^0\to\Delta^-\pi^0\pi^+)+\sqrt{3}\mathcal{A}(\Lambda_b^0\to\Delta^+\pi^0\pi^-)\notag \\ &+\sqrt{2}\mathcal{A}(\Lambda_b^0\to\Delta^{++}\pi^-\pi^-)+2\sqrt{3}\mathcal{A}(\Lambda_b^0\to\Delta^+\pi^-\pi^0)\big]=0,
\end{align}
\begin{align}
SumI_-^3[\Lambda_b^0,\Delta^{++},\pi^+,\eta_8] &=-6\big[\sqrt{6}\mathcal{A}(\Lambda_b^0\to\Delta^0\pi^0\eta_8)-\mathcal{A}(\Lambda_b^0\to\Delta^-\pi^+\eta_8) \notag \\ &+\sqrt{3}\mathcal{A}(\Lambda_b^0\to\Delta^+\pi^-\eta_8)\big]=0,
\end{align}
\begin{align}
SumI_-^3[\Lambda_b^0,\Delta^{++},K^+,\overline{K}^0] &=6\big[\sqrt{3}\mathcal{A}(\Lambda_b^0\to\Delta^0K^0\overline{K}^0)-\sqrt{3}\mathcal{A}(\Lambda_b^0\to\Delta^0K^+K^-) \notag \\ &-\sqrt{3}\mathcal{A}(\Lambda_b^0\to\Delta^+K^0K^-)+\mathcal{A}(\Lambda_b^0\to\Delta^-K^+\overline{K}^0)\big]=0,
\end{align}
\begin{align}
SumI_-^3[\Xi_b^-,\Xi^{*0},\pi^+,K^+] &=6\big[\sqrt{2}\mathcal{A}(\Xi_b^0\to\Xi^{*0}\pi^0K^0)+\mathcal{A}(\Xi_b^0\to\Xi^{*0}\pi^-K^+) \notag \\ &+\sqrt{2}\mathcal{A}(\Xi_b^0\to\Xi^{*-}\pi^0K^+)-\mathcal{A}(\Xi_b^0\to\Xi^{*-}\pi^+K^0)\notag \\ &-\mathcal{A}(\Xi_b^-\to\Xi^{*0}\pi^-K^0)-\sqrt{2}\mathcal{A}(\Xi_b^-\to\Xi^{*-}\pi^0K^0) \notag \\ &-\mathcal{A}(\Xi_b^-\to\Xi^{*-}\pi^-K^+)\big]=0,
\end{align}
\begin{align}
SumI_-^4[\Xi_b^-,\Sigma^{*+},\pi^+,\pi^+]
&=24\big[-2\sqrt{2}\mathcal{A}(\Xi_b^0\to\Sigma^{*0}\pi^0\pi^0)+\sqrt{2}\mathcal{A}(\Xi_b^0\to\Sigma^{*0}\pi^+\pi^-) \notag \\
&+\sqrt{2}\mathcal{A}(\Xi_b^0\to\Sigma^{*0}\pi^-\pi^+)+\sqrt{2}\mathcal{A}(\Xi_b^0\to\Sigma^{*-}\pi^0\pi^+) \notag \\
&-\sqrt{2}\mathcal{A}(\Xi_b^0\to\Sigma^{*+}\pi^0\pi^-)+\sqrt{2}\mathcal{A}(\Xi_b^0\to\Sigma^{*-}\pi^+\pi^0) \notag \\
&-\sqrt{2}\mathcal{A}(\Xi_b^0\to\Sigma^{*+}\pi^-\pi^0)+2\mathcal{A}(\Xi_b^-\to\Sigma^{*0}\pi^0\pi^-) \notag \\
&+2\mathcal{A}(\Xi_b^-\to\Sigma^{*0}\pi^-\pi^0)+2\mathcal{A}(\Xi_b^-\to\Sigma^{*-}\pi^0\pi^0) \notag \\
&-\mathcal{A}(\Xi_b^-\to\Sigma^{*-}\pi^-\pi^+)-\mathcal{A}(\Xi_b^-\to\Sigma^{*-}\pi^+\pi^-) \notag \\
&+\mathcal{A}(\Xi_b^-\to\Sigma^{*+}\pi^-\pi^-)\big]=0,
\end{align}
\begin{align}
SumI_-^3[\Xi_b^0,\Sigma^{*+},\pi^+,\pi^+] &=-6\sqrt{2}\big[-2\mathcal{A}(\Xi_b^0\to\Sigma^{*0}\pi^0\pi^0)+\mathcal{A}(\Xi_b^0\to\Sigma^{*0}\pi^+\pi^-) \notag \\ &+\mathcal{A}(\Xi_b^0\to\Sigma^{*0}\pi^-\pi^+)+\mathcal{A}(\Xi_b^0\to\Sigma^{*-}\pi^0\pi^+) \notag \\ &-\mathcal{A}(\Xi_b^0\to\Sigma^{*+}\pi^0\pi^-)+\mathcal{A}(\Xi_b^0\to\Sigma^{*-}\pi^+\pi^0) \notag \\ &-\mathcal{A}(\Xi_b^0\to\Sigma^{*+}\pi^-\pi^0)\big]=0,
\end{align}
\begin{align}
SumI_-^3[\Xi_b^-,\Sigma^{*0},\pi^+,\pi^+]
&=6\big[-2\mathcal{A}(\Xi_b^0\to\Sigma^{*0}\pi^0\pi^0)+\mathcal{A}(\Xi_b^0\to\Sigma^{*0}\pi^+\pi^-) \notag \\
&+\mathcal{A}(\Xi_b^0\to\Sigma^{*0}\pi^-\pi^+)+2\mathcal{A}(\Xi_b^0\to\Sigma^{*-}\pi^0\pi^+) \notag \\
&+2\mathcal{A}(\Xi_b^0\to\Sigma^{*-}\pi^+\pi^0)+\sqrt{2}\big(\mathcal{A}(\Xi_b^-\to\Sigma^{*0}\pi^0\pi^-) \notag \\
&+\mathcal{A}(\Xi_b^-\to\Sigma^{*0}\pi^-\pi^0)+2\mathcal{A}(\Xi_b^-\to\Sigma^{*-}\pi^0\pi^0) \notag \\
&-\mathcal{A}(\Xi_b^-\to\Sigma^{*-}\pi^-\pi^+)-\mathcal{A}(\Xi_b^-\to\Sigma^{*-}\pi^+\pi^-)\big)\big]=0,
\end{align}
\begin{align}
SumI_-^3[\Xi_b^-,\Sigma^{*+},\pi^+,\pi^0]
&=-6\big[-2\mathcal{A}(\Xi_b^0\to\Sigma^{*0}\pi^0\pi^0)+2\mathcal{A}(\Xi_b^0\to\Sigma^{*0}\pi^+\pi^-) \notag \\
&-2\mathcal{A}(\Xi_b^0\to\Sigma^{*+}\pi^0\pi^-)+\mathcal{A}(\Xi_b^0\to\Sigma^{*-}\pi^+\pi^0) \notag \\
&-\mathcal{A}(\Xi_b^0\to\Sigma^{*+}\pi^-\pi^0)+\sqrt{2}\big(2\mathcal{A}(\Xi_b^-\to\Sigma^{*0}\pi^0\pi^-) \notag \\
&+\mathcal{A}(\Xi_b^-\to\Sigma^{*0}\pi^-\pi^0)+\mathcal{A}(\Xi_b^-\to\Sigma^{*-}\pi^0\pi^0) \notag \\
&-\mathcal{A}(\Xi_b^-\to\Sigma^{*-}\pi^+\pi^-)+\mathcal{A}(\Xi_b^-\to\Sigma^{*+}\pi^-\pi^-)\big)\big]=0,
\end{align}
\begin{align}
SumI_-^3[\Xi_b^-,\Sigma^{*+},\pi^+,\eta_8] &=-6\big[-2\mathcal{A}(\Xi_b^0\to\Sigma^{*0}\pi^0\eta_8)+\mathcal{A}(\Xi_b^0\to\Sigma^{*-}\pi^+\eta_8) \notag \\ &-\mathcal{A}(\Xi_b^0\to\Sigma^{*+}\pi^-\eta_8)+\sqrt{2}\big(\mathcal{A}(\Xi_b^-\to\Sigma^{*0}\pi^-\eta_8) \notag \\ &+\mathcal{A}(\Xi_b^-\to\Sigma^{*-}\pi^0\eta_8)\big)\big]=0,
\end{align}
\begin{align}
SumI_-^3[\Xi_b^-,\Sigma^{*+},K^+,\overline{K}^0] &=-6\big[\sqrt2\mathcal{A}(\Xi_b^0\to\Sigma^{*0}K^0\overline{K}^0)-\sqrt{2}\mathcal{A}(\Xi_b^0\to\Sigma^{*0}K^+K^-) \notag \\ &-\mathcal{A}(\Xi_b^0\to\Sigma^{*+}K^0K^-)+\mathcal{A}(\Xi_b^0\to\Sigma^{*-}K^+\overline{K}^0) \notag \\ &+\sqrt{2}\mathcal{A}(\Xi_b^-\to\Sigma^{*0}K^0K^-)-\mathcal{A}(\Xi_b^-\to\Sigma^{*-}K^0\overline{K}^0) \notag \\ &+\mathcal{A}(\Xi_b^-\to\Sigma^{*-}K^+K^-)\big]=0,
\end{align}
\begin{align}
SumI_-^4[\Xi_b^-,\Delta^{++},\pi^+,\overline{K}^0]
&=24\big[\sqrt{6}\mathcal{A}(\Xi_b^0\to\Delta^0\pi^0\overline{K}^0)+\sqrt{3}\mathcal{A}(\Xi_b^0\to\Delta^0\pi^+K^-) \notag \\
&-\sqrt{6}\mathcal{A}(\Xi_b^0\to\Delta^+\pi^0K^-)-\mathcal{A}(\Xi_b^0\to\Delta^{++}\pi^-K^-) \notag \\
&-\mathcal{A}(\Xi_b^0\to\Delta^-\pi^+\overline{K}^0)+\sqrt{3}\mathcal{A}(\Xi_b^0\to\Delta^+\pi^-\overline{K}^0) \notag \\
&+\sqrt{6}\mathcal{A}(\Xi_b^-\to\Delta^0\pi^0K^-)-\sqrt{3}\mathcal{A}(\Xi_b^-\to\Delta^0\pi^-\overline{K}^0) \notag \\
&-\sqrt{2}\mathcal{A}(\Xi_b^-\to\Delta^-\pi^0\overline{K}^0)-\mathcal{A}(\Xi_b^-\to\Delta^-\pi^+K^-) \notag \\
&+\sqrt{3}\mathcal{A}(\Xi_b^-\to\Delta^+\pi^-K^-)\big]=0,
\end{align}
\begin{align}
SumI_-^3[\Xi_b^0,\Delta^{++},\pi^+,\overline{K}^0] &=6\big[-\sqrt{6}\mathcal{A}(\Xi_b^0\to\Delta^0\pi^0\overline{K}^0)-\sqrt{3}\mathcal{A}(\Xi_b^0\to\Delta^0\pi^+K^-) \notag \\ &+\sqrt{6}\mathcal{A}(\Xi_b^0\to\Delta^+\pi^0K^-)+\mathcal{A}(\Xi_b^0\to\Delta^{++}\pi^-K^-) \notag \\ &+\mathcal{A}(\Xi_b^0\to\Delta^-\pi^+\overline{K}^0)-\sqrt{3}\mathcal{A}(\Xi_b^0\to\Delta^+\pi^-\overline{K}^0)\big]=0,
\end{align}
\begin{align}
SumI_-^3[\Xi_b^-,\Delta^+,\pi^+,\overline{K}^0]
&=6\big[2\sqrt{2}\mathcal{A}(\Xi_b^0\to\Delta^0\pi^0\overline{K}^0)+2\mathcal{A}(\Xi_b^0\to\Delta^0\pi^+K^-) \notag \\
&-\sqrt{2}\mathcal{A}(\Xi_b^0\to\Delta^+\pi^0K^-)-\sqrt{3}\mathcal{A}(\Xi_b^0\to\Delta^-\pi^+\overline{K}^0) \notag \\
&+\mathcal{A}(\Xi_b^0\to\Delta^+\pi^-\overline{K}^0)+2\sqrt{2}\mathcal{A}(\Xi_b^-\to\Delta^0\pi^0K^-) \notag \\
&-2\mathcal{A}(\Xi_b^-\to\Delta^0\pi^-\overline{K}^0)-\sqrt{6}\mathcal{A}(\Xi_b^-\to\Delta^-\pi^0\overline{K}^0) \notag \\
&-\sqrt{3}\mathcal{A}(\Xi_b^-\to\Delta^-\pi^+K^-)+\mathcal{A}(\Xi_b^-\to\Delta^+\pi^-K^-)\big]=0,
\end{align}
\begin{align}
SumI_-^3[\Xi_b^-,\Delta^{++},\pi^0,\overline{K}^0] &=6\big[-\sqrt{3}\mathcal{A}(\Xi_b^0\to\Delta^0\pi^0\overline{K}^0)+\sqrt{3}\mathcal{A}(\Xi_b^0\to\Delta^+\pi^0K^-) \notag \\ &+\sqrt{2}\mathcal{A}(\Xi_b^0\to\Delta^{++}\pi^-K^-)-\sqrt{6}\mathcal{A}(\Xi_b^0\to\Delta^+\pi^-\overline{K}^0) \notag \\ &-\sqrt{3}\mathcal{A}(\Xi_b^-\to\Delta^0\pi^0K^-)+\sqrt{6}\mathcal{A}(\Xi_b^-\to\Delta^0\pi^-\overline{K}^0) \notag \\ &+\mathcal{A}(\Xi_b^-\to\Delta^-\pi^0\overline{K}^0)-\sqrt{6}\mathcal{A}(\Xi_b^-\to\Delta^+\pi^-K^-)\big]=0,
\end{align}
\begin{align}
SumI_-^3[\Xi_b^-,\Delta^{++},\pi^+,K^-] &=6\big[-\sqrt{3}\mathcal{A}(\Xi_b^0\to\Delta^0\pi^+K^-)+\sqrt{6}\mathcal{A}(\Xi_b^0\to\Delta^+\pi^0K^-) \notag \\ &+\mathcal{A}(\Xi_b^0\to\Delta^{++}\pi^-K^-)-\sqrt{6}\mathcal{A}(\Xi_b^-\to\Delta^0\pi^0K^-) \notag \\ &+\mathcal{A}(\Xi_b^-\to\Delta^-\pi^+K^-)-\sqrt{3}\mathcal{A}(\Xi_b^-\to\Delta^+\pi^-K^-)\big]=0,
\end{align}
\begin{align}
SumI_-^3[\Xi_b^-,\Delta^{++},\overline{K}^0,\eta_8] &=6\big[-\sqrt{3}\mathcal{A}(\Xi_b^0\to\Delta^0\overline{K}^0\eta_8)+\sqrt{3}\mathcal{A}(\Xi_b^0\to\Delta^+K^-\eta_8) \notag \\ &-\sqrt{3}\mathcal{A}(\Xi_b^-\to\Delta^0K^-\eta_8)+\mathcal{A}(\Xi_b^-\to\Delta^-\overline{K}^0\eta_8)\big]=0,
\end{align}
\begin{align}
SumI_-^3[\Lambda_b^0,\Sigma^+,\pi^+,K^+] &=6\big[2\mathcal{A}(\Lambda_b^0\to\Sigma^0\pi^0K^0)+\sqrt{2}\mathcal{A}(\Lambda_b^0\to\Sigma^0\pi^-K^+) \notag \\ &+\sqrt{2}\mathcal{A}(\Lambda_b^0\to\Sigma^-\pi^0K^+)-\mathcal{A}(\Lambda_b^0\to\Sigma^-\pi^+K^0) \notag \\ &-\mathcal{A}(\Lambda_b^0\to\Sigma^+\pi^-K^0)\big]=0,
\end{align}
\begin{align}
SumI_-^3[\Lambda_b^0,p,\pi^+,\pi^+] &=6\big[2\mathcal{A}(\Lambda_b^0\to\ n\pi^0\pi^0)-\mathcal{A}(\Lambda_b^0\to\ n\pi^+\pi^-) \notag \\ &-\mathcal{A}(\Lambda_b^0\to\ n\pi^-\pi^+)+\sqrt{2}\big(\mathcal{A}(\Lambda_b^0\to\ p\pi^0\pi^-) \notag \\ &+\mathcal{A}(\Lambda_b^0\to\ p\pi^-\pi^0)\big)\big]=0,
\end{align}
\begin{align}
SumI_-^4[\Xi_b^-,\Sigma^+,\pi^+,\pi^+] &=-24\big[-2\sqrt{2}\mathcal{A}(\Xi_b^0\to\Sigma^0\pi^0\pi^0)+\sqrt{2}\mathcal{A}(\Xi_b^0\to\Sigma^0\pi^+\pi^-) \notag \\ &+\sqrt{2}\mathcal{A}(\Xi_b^0\to\Sigma^0\pi^-\pi^+)+\sqrt{2}\mathcal{A}(\Xi_b^0\to\Sigma^-\pi^0\pi^+) \notag \\ &+\sqrt{2}\mathcal{A}(\Xi_b^0\to\Sigma^+\pi^0\pi^-)+\sqrt{2}\mathcal{A}(\Xi_b^0\to\Sigma^-\pi^+\pi^0) \notag \\ &+\sqrt{2}\mathcal{A}(\Xi_b^0\to\Sigma^+\pi^-\pi^0)+2\mathcal{A}(\Xi_b^-\to\Sigma^0\pi^0\pi^-) \notag \\ &+2\mathcal{A}(\Xi_b^-\to\Sigma^0\pi^-\pi^0)+2\mathcal{A}(\Xi_b^-\to\Sigma^-\pi^0\pi^0) \notag \\
&-\mathcal{A}(\Xi_b^-\to\Sigma^-\pi^-\pi^+)-\mathcal{A}(\Xi_b^-\to\Sigma^-\pi^+\pi^-) \notag \\ &-\mathcal{A}(\Xi_b^-\to\Sigma^+\pi^-\pi^-)\big]=0,
\end{align}
\begin{align}
SumI_-^3[\Xi_b^0,\Sigma^+,\pi^+,\pi^+] &=6\sqrt{2}\big[-2\mathcal{A}(\Xi_b^0\to\Sigma^0\pi^0\pi^0)+\mathcal{A}(\Xi_b^0\to\Sigma^0\pi^+\pi^-) \notag \\ &+\mathcal{A}(\Xi_b^0\to\Sigma^0\pi^-\pi^+)+\mathcal{A}(\Xi_b^0\to\Sigma^-\pi^0\pi^+) \notag \\ &+\mathcal{A}(\Xi_b^0\to\Sigma^+\pi^0\pi^-)+\mathcal{A}(\Xi_b^0\to\Sigma^-\pi^+\pi^0) \notag \\ &+\mathcal{A}(\Xi_b^0\to\Sigma^+\pi^-\pi^0)\big]=0,
\end{align}
\begin{align}
SumI_-^3[\Xi_b^-,\Sigma^+,\pi^+,\eta_8] &=6\big[-2\mathcal{A}(\Xi_b^0\to\Sigma^0\pi^0\eta_8)+\mathcal{A}(\Xi_b^0\to\Sigma^-\pi^+\eta_8) \notag \\ &+\mathcal{A}(\Xi_b^0\to\Sigma^+\pi^-\eta_8)+\sqrt{2}\big(\mathcal{A}(\Xi_b^-\to\Sigma^0\pi^-\eta_8) \notag \\ &+\mathcal{A}(\Xi_b^-\to\Sigma^-\pi^0\eta_8)\big)\big]=0,
\end{align}
\begin{align}
SumI_-^3[\Xi_b^-,\Sigma^+ ,K^+,\overline{K}^0] &=6\big[\sqrt{2}\mathcal{A}(\Xi_b^0\to\Sigma^0K^0\overline{K}^0)-\sqrt{2}\mathcal{A}(\Xi_b^0\to\Sigma^0K^+K^-) \notag \\ &+\mathcal{A}(\Xi_b^0\to\Sigma^+K^0K^-)+\mathcal{A}(\Xi_b^0\to\Sigma^-K^+\overline{K}^0) \notag \\ &+\sqrt{2}\mathcal{A}(\Xi_b^-\to\Sigma^0K^0K^-)-\mathcal{A}(\Xi_b^-\to\Sigma^-K^0\overline{K}^0) \notag \\ &+\mathcal{A}(\Xi_b^-\to\Sigma^-K^+K^-)\big]=0,
\end{align}
\begin{align}
SumI_-^3[\Xi_b^-,p,\pi^+,\overline{K}^0] &=6\big[\sqrt{2}\mathcal{A}(\Xi_b^0\to\ n\pi^0\overline{K}^0)+\mathcal{A}(\Xi_b^0\to\ n\pi^+K^-) \notag \\ &-\sqrt{2}\mathcal{A}(\Xi_b^0\to\ p\pi^0K^-)+\mathcal{A}(\Xi_b^0\to\ p\pi^-\overline{K}^0) \notag \\ &+\sqrt{2}\mathcal{A}(\Xi_b^-\to\ n\pi^0K^-)-\mathcal{A}(\Xi_b^-\to\ n\pi^-\overline{K}^0) \notag \\ &+\mathcal{A}(\Xi_b^-\to\ p\pi^-K^-)\big]=0,
\end{align}
\begin{align}
SumI_-^3[\Xi_b^-,\Xi^0,\pi^+,K^+] &=6\big[\sqrt{2}\mathcal{A}(\Xi_b^0\to\Xi^0\pi^0K^0)+\mathcal{A}(\Xi_b^0\to\Xi^0\pi^-K^+) \notag \\ &-\sqrt{2}\mathcal{A}(\Xi_b^0\to\Xi^-\pi^0K^+)+\mathcal{A}(\Xi_b^0\to\Xi^-\pi^+K^0) \notag \\ &-\mathcal{A}(\Xi_b^-\to\Xi^0\pi^-K^0)+\sqrt{2}\mathcal{A}(\Xi_b^-\to\Xi^-\pi^0K^0) \notag \\ &+\mathcal{A}(\Xi_b^-\to\Xi^-\pi^-K^+)\big]=0,
\end{align}
\begin{align}
SumI_-^3[\Xi_b^-,\Lambda^0,\pi^+,\pi^+] &=6\big[-2\mathcal{A}(\Xi_b^0\to\Lambda^0\pi^0\pi^0)+\mathcal{A}(\Xi_b^0\to\Lambda^0\pi^+\pi^-) \notag \\ &+\mathcal{A}(\Xi_b^0\to\Lambda^0\pi^-\pi^+)+\sqrt{2}\big(\mathcal{A}(\Xi_b^-\to\Lambda^0\pi^0\pi^-) \notag \\ &+\mathcal{A}(\Xi_b^-\to\Lambda^0\pi^-\pi^0)\big)\big]=0,
\end{align}
\begin{align}
SumI_-^3[\Xi_b^-,\Xi^0,\pi^+,\pi^+] &=6\big[-2\mathcal{A}(\Xi_b^0\to\Xi^0\pi^0\pi^0)+\mathcal{A}(\Xi_b^0\to\Xi^0\pi^-\pi^+) \notag \\ &+\mathcal{A}(\Xi_b^0\to\Xi^0\pi^+\pi^-)-\sqrt{2}\mathcal{A}(\Xi_b^0\to\Xi^-\pi^0\pi^+) \notag \\ &-\sqrt{2}\mathcal{A}(\Xi_b^0\to\Xi^-\pi^+\pi^0)+\sqrt{2}\mathcal{A}(\Xi_b^-\to\Xi^0\pi^0\pi^-) \notag \\ &+\sqrt{2}\mathcal{A}(\Xi_b^-\to\Xi^0\pi^-\pi^0)-2\mathcal{A}(\Xi_b^-\to\Xi^-\pi^0\pi^0) \notag \\ &+\mathcal{A}(\Xi_b^-\to\Xi^-\pi^+\pi^-)+\mathcal{A}(\Xi_b^-\to\Xi^-\pi^-\pi^+)\big]=0,
\end{align}
\begin{align}
SumI_-^3[\Xi_b^-,\Sigma^0,\pi^+,\pi^+] &=6\big[-2\mathcal{A}(\Xi_b^0\to\Sigma^0\pi^0\pi^0)+\mathcal{A}(\Xi_b^0\to\Sigma^0\pi^+\pi^-) \notag \\ &+\mathcal{A}(\Xi_b^0\to\Sigma^0\pi^-\pi^+)+2\mathcal{A}(\Xi_b^0\to\Sigma^-\pi^0\pi^+) \notag \\ &+2\mathcal{A}(\Xi_b^0\to\Sigma^-\pi^+\pi^0)+\sqrt{2}\big(\mathcal{A}(\Xi_b^-\to\Sigma^0\pi^0\pi^-) \notag \\ &+\mathcal{A}(\Xi_b^-\to\Sigma^0\pi^-\pi^0)+2\mathcal{A}(\Xi_b^-\to\Sigma^-\pi^0\pi^0) \notag \\ &-\mathcal{A}(\Xi_b^-\to\Sigma^-\pi^-\pi^+)-\mathcal{A}(\Xi_b^-\to\Sigma^-\pi^+\pi^-)\big)\big]=0,
\end{align}
\begin{align}
SumI_-^3[\Xi_b^-,\Sigma^+,\pi^0,\pi^+] &=6\big[-2\mathcal{A}(\Xi_b^0\to\Sigma^0\pi^0\pi^0)+2\mathcal{A}(\Xi_b^0\to\Sigma^0\pi^-\pi^+) \notag \\ &+\mathcal{A}(\Xi_b^0\to\Sigma^-\pi^0\pi^+)+\mathcal{A}(\Xi_b^0\to\Sigma^+\pi^0\pi^-) \notag \\ &+2\mathcal{A}(\Xi_b^0\to\Sigma^+\pi^-\pi^0)+\sqrt{2}\big(\mathcal{A}(\Xi_b^-\to\Sigma^0\pi^0\pi^-) \notag \\ &+2\mathcal{A}(\Xi_b^-\to\Sigma^0\pi^-\pi^0)+\mathcal{A}(\Xi_b^-\to\Sigma^-\pi^0\pi^0) \notag \\ &-\mathcal{A}(\Xi_b^-\to\Sigma^-\pi^-\pi^+)-\mathcal{A}(\Xi_b^-\to\Sigma^+\pi^-\pi^-)\big)\big]=0,
\end{align}
\begin{align}
SumI_-^2[\Lambda_b^0,\Xi^{*0},\pi^+,K^+] &=-2\big[\sqrt{2}\mathcal{A}(\Lambda_b^0\to\Xi^{*0}\pi^0K^0)+\mathcal{A}(\Lambda_b^0\to\Xi^{*0}\pi^-K^+) \notag \\ &+\sqrt{2}\mathcal{A}(\Lambda_b^0\to\Xi^{*-}\pi^0K^+)-\mathcal{A}(\Lambda_b^0\to\Xi^{*-}\pi^+K^0)\big]=0,
\end{align}
\begin{align}
SumI_-^3[\Lambda_b^0,\Sigma^{*+},\pi^+,\pi^+] &=-6\sqrt{2}\big[-2\mathcal{A}(\Lambda_b^0\to\Sigma^{*0}\pi^0\pi^0)+\mathcal{A}(\Lambda_b^0\to\Sigma^{*0}\pi^-\pi^+) \notag \\ &+\mathcal{A}(\Lambda_b^0\to\Sigma^{*0}\pi^+\pi^-)-\mathcal{A}(\Lambda_b^0\to\Sigma^{*+}\pi^0\pi^-) \notag \\ &+\mathcal{A}(\Lambda_b^0\to\Sigma^{*-}\pi^0\pi^+)-\mathcal{A}(\Lambda_b^0\to\Sigma^{*+}\pi^-\pi^0) \notag \\ &+\mathcal{A}(\Lambda_b^0\to\Sigma^{*-}\pi^+\pi^0)\big]=0,
\end{align}
\begin{align}
SumI_-^2[\Lambda_b^0,\Sigma^{*0},\pi^+,\pi^+] &=4\mathcal{A}(\Lambda_b^0\to\Sigma^{*0}\pi^0\pi^0)-2\big[\mathcal{A}(\Lambda_b^0\to\Sigma^{*0}\pi^-\pi^+) \notag \\ &+\mathcal{A}(\Lambda_b^0\to\Sigma^{*0}\pi^+\pi^-)+2\big(\mathcal{A}(\Lambda_b^0\to\Sigma^{*-}\pi^0\pi^+) \notag \\ &+\mathcal{A}(\Lambda_b^0\to\Sigma^{*-}\pi^+\pi^0)\big)\big]=0,
\end{align}
\begin{align}
SumI_-^2[\Lambda_b^0,\Sigma^{*+},\pi^0,\pi^+] &=-2\big[2\mathcal{A}(\Lambda_b^0\to\Sigma^{*0}\pi^0\pi^0)-2\mathcal{A}(\Lambda_b^0\to\Sigma^{*0}\pi^-\pi^+)\notag \\ &+\mathcal{A}(\Lambda_b^0\to\Sigma^{*+}\pi^0\pi^-)-\mathcal{A}(\Lambda_b^0\to\Sigma^{*-}\pi^0\pi^+) \notag \\ &+2\mathcal{A}(\Lambda_b^0\to\Sigma^{*+}\pi^-\pi^0)\big]=0,
\end{align}
\begin{align}
SumI_-^2[\Lambda_b^0,\Sigma^{*+},\pi^+,\eta_8] &=-2\big[2\mathcal{A}(\Lambda_b^0\to\Sigma^{*0}\pi^0\eta_8)+\mathcal{A}(\Lambda_b^0\to\Sigma^{*+}\pi^-\eta_8) \notag \\ &-\mathcal{A}(\Lambda_b^0\to\Sigma^{*-}\pi^+\eta_8)\big]=0,
\end{align}
\begin{align}
SumI_-^2[\Lambda_b^0,\Sigma^{*+},K^+,\overline{K}^0] &=2\big[\sqrt{2}\mathcal{A}(\Lambda_b^0\to\Sigma^{*0}K^0\overline{K}^0)-\sqrt{2}\mathcal{A}(\Lambda_b^0\to\Sigma^{*0}K^+K^-) \notag \\ &-\mathcal{A}(\Lambda_b^0\to\Sigma^{*+}K^0K^-)+\mathcal{A}(\Lambda_b^0\to\Sigma^{*-}K^+\overline{K}^0)\big]=0,
\end{align}
\begin{align}
SumI_-^3[\Lambda_b^0,\Delta^{++},\pi^+,\overline{K}^0] &=6\big[-\sqrt{6}\mathcal{A}(\Lambda_b^0\to\Delta^0\pi^0\overline{K}^0)-\sqrt{3}\mathcal{A}(\Lambda_b^0\to\Delta^0\pi^+K^-) \notag \\ &+\sqrt{6}\mathcal{A}(\Lambda_b^0\to\Delta^+\pi^0K^-)+\mathcal{A}(\Lambda_b^0\to\Delta^{++}\pi^-K^-) \notag \\ &-\sqrt{3}\mathcal{A}(\Lambda_b^0\to\Delta^+\pi^-\overline{K}^0)+\mathcal{A}(\Lambda_b^0\to\Delta^-\pi^+\overline{K}^0)\big]=0,
\end{align}
\begin{align}
SumI_-^2[\Lambda_b^0,\Delta^+,\pi^+,\overline{K}^0] &=-2\big[2\sqrt{2}\mathcal{A}(\Lambda_b^0\to\Delta^0\pi^0\overline{K}^0)+2\mathcal{A}(\Lambda_b^0\to\Delta^0\pi^+K^-) \notag \\ &-\sqrt{2}\mathcal{A}(\Lambda_b^0\to\Delta^+\pi^0K^-)+\mathcal{A}(\Lambda_b^0\to\Delta^+\pi^-\overline{K}^0) \notag \\ &-\sqrt{3}\mathcal{A}(\Lambda_b^0\to\Delta^-\pi^+\overline{K}^0)\big]=0,
\end{align}
\begin{align}
SumI_-^2[\Lambda_b^0,\Delta^{++},\pi^0,\overline{K}^0] &=2\big[\sqrt{3}\mathcal{A}(\Lambda_b^0\to\Delta^0\pi^0\overline{K}^0)-\sqrt{3}\mathcal{A}(\Lambda_b^0\to\Delta^+\pi^0K^-) \notag \\ &-\sqrt{2}\mathcal{A}(\Lambda_b^0\to\Delta^{++}\pi^-K^-)+\sqrt{6}\mathcal{A}(\Lambda_b^0\to\Delta^+\pi^-\overline{K}^0)\big]=0,
\end{align}
\begin{align}
SumI_-^2[\Lambda_b^0,\Delta^{++},\pi^+,K^-] &=2\sqrt{3}\mathcal{A}(\Lambda_b^0\to\Delta^0\pi^+K^-)-2\big[\sqrt{6}\mathcal{A}(\Lambda_b^0\to\Delta^+\pi^0K^-) \notag \\ &+\mathcal{A}(\Lambda_b^0\to\Delta^{++}\pi^-K^-)\big]=0,
\end{align}
\begin{align}
SumI_-^2[\Lambda_b^0,\Delta^{++},\overline{K}^0,\eta_8] &=2\sqrt{3}\big[\mathcal{A}(\Lambda_b^0\to\Delta^0\overline{K}^0\eta_8)-\mathcal{A}(\Lambda_b^0\to\Delta^+K^-\eta_8)\big]=0,
\end{align}
\begin{align}
SumI_-^2[\Xi_b^-,\Omega^-,\pi^+,K^+] &=2\sqrt{2}\mathcal{A}(\Xi_b^0\to\Omega^-\pi^0K^+)-2\big[\mathcal{A}(\Xi_b^0\to\Omega^-\pi^+K^0) \notag \\ &+\sqrt{2}\mathcal{A}(\Xi_b^-\to\Omega^-\pi^0K^0)+\mathcal{A}(\Xi_b^-\to\Omega^-\pi^-K^+)\big]=0,
\end{align}
\begin{align}
SumI_-^3[\Xi_b^-,\Xi^{*0},\pi^+,\pi^+] &=6\big[-2\mathcal{A}(\Xi_b^0\to\Xi^{*0}\pi^0\pi^0)+\mathcal{A}(\Xi_b^0\to\Xi^{*0}\pi^+\pi^-) \notag \\ &+\mathcal{A}(\Xi_b^0\to\Xi^{*0}\pi^-\pi^+)+\sqrt{2}\mathcal{A}(\Xi_b^0\to\Xi^{*-}\pi^0\pi^+) \notag \\ &+\sqrt{2}\mathcal{A}(\Xi_b^0\to\Xi^{*-}\pi^+\pi^0)+\sqrt{2}\mathcal{A}(\Xi_b^-\to\Xi^{*0}\pi^0\pi^-) \notag \\ &+\sqrt{2}\mathcal{A}(\Xi_b^-\to\Xi^{*0}\pi^-\pi^0)+2\mathcal{A}(\Xi_b^-\to\Xi^{*-}\pi^0\pi^0) \notag \\ &-\mathcal{A}(\Xi_b^-\to\Xi^{*-}\pi^+\pi^-)-\mathcal{A}(\Xi_b^-\to\Xi^{*-}\pi^-\pi^+)\big]=0,
\end{align}
\begin{align}
SumI_-^2[\Xi_b^0,\Xi^{*0},\pi^+,\pi^+] &=-2\big[-2\mathcal{A}(\Xi_b^0\to\Xi^{*0}\pi^0\pi^0)+\mathcal{A}(\Xi_b^0\to\Xi^{*0}\pi^+\pi^-) \notag \\ &+\mathcal{A}(\Xi_b^0\to\Xi^{*0}\pi^-\pi^+)+\sqrt{2}\big(\mathcal{A}(\Xi_b^0\to\Xi^{*-}\pi^0\pi^+) \notag \\ &+\mathcal{A}(\Xi_b^0\to\Xi^{*-}\pi^+\pi^0)\big)\big]=0,
\end{align}
\begin{align}
SumI_-^2[\Xi_b^-,\Xi^{*-},\pi^+,\pi^+] &=2\big[\sqrt{2}\mathcal{A}(\Xi_b^0\to\Xi^{*-}\pi^0\pi^+)+\sqrt{2}\mathcal{A}(\Xi_b^0\to\Xi^{*-}\pi^+\pi^0) \notag \\ &+2\mathcal{A}(\Xi_b^-\to\Xi^{*-}\pi^0\pi^0)-\mathcal{A}(\Xi_b^-\to\Xi^{*-}\pi^+\pi^-) \notag \\ &-\mathcal{A}(\Xi_b^-\to\Xi^{*-}\pi^-\pi^+)\big]=0,
\end{align}
\begin{align}
SumI_-^2[\Xi_b^-,\Xi^{*0},\pi^0,\pi^+] &=2\sqrt{2}\mathcal{A}(\Xi_b^0\to\Xi^{*0}\pi^0\pi^0)-2\big[\sqrt{2}\mathcal{A}(\Xi_b^0\to\Xi^{*0}\pi^-\pi^+) \notag \\ &+\mathcal{A}(\Xi_b^0\to\Xi^{*-}\pi^0\pi^+)+\mathcal{A}(\Xi_b^-\to\Xi^{*0}\pi^0\pi^-) \notag \\ &+2\mathcal{A}(\Xi_b^-\to\Xi^{*0}\pi^-\pi^0)+\sqrt{2}\big(\mathcal{A}(\Xi_b^-\to\Xi^{*-}\pi^0\pi^0) \notag \\ &-\mathcal{A}(\Xi_b^-\to\Xi^{*-}\pi^-\pi^+)\big)\big]=0,
\end{align}
\begin{align}
SumI_-^2[\Xi_b^-,\Xi^{*0},\pi^+,\eta_8] &=2\sqrt{2}\mathcal{A}(\Xi_b^0\to\Xi^{*0}\pi^0\eta_8)-2\big[\mathcal{A}(\Xi_b^0\to\Xi^{*-}\pi^+\eta_8) \notag \\ &+\mathcal{A}(\Xi_b^-\to\Xi^{*0}\pi^-\eta_8)+\sqrt{2}\mathcal{A}(\Xi_b^-\to\Xi^{*-}\pi^0\eta_8)\big]=0,
\end{align}
\begin{align}
SumI_-^2[\Xi_b^-,\Xi^{*0},K^+,\overline{K}^0] &=-2\big[\mathcal{A}(\Xi_b^0\to\Xi^{*0}K^0\overline{K}^0)-\mathcal{A}(\Xi_b^0\to\Xi^{*0}K^+K^-) \notag \\ &+\mathcal{A}(\Xi_b^0\to\Xi^{*-}K^+\overline{K}^0)+\mathcal{A}(\Xi_b^-\to\Xi^{*0}K^0K^-) \notag \\ &-\mathcal{A}(\Xi_b^-\to\Xi^{*-}K^0\overline{K}^0)+\mathcal{A}(\Xi_b^-\to\Xi^{*-}K^+K^-)\big]=0,
\end{align}
\begin{align}
SumI_-^3[\Xi_b^-,\Sigma^{*+},\pi^+,\overline{K}^0] &=6\big[2\mathcal{A}(\Xi_b^0\to\Sigma^{*0}\pi^0\overline{K}^0)+\sqrt{2}\mathcal{A}(\Xi_b^0\to\Sigma^{*0}\pi^+K^-) \notag \\ &-\sqrt{2}\mathcal{A}(\Xi_b^0\to\Sigma^{*+}\pi^0K^-)-\mathcal{A}(\Xi_b^0\to\Sigma^{*-}\pi^+\overline{K}^0) \notag \\ &+\mathcal{A}(\Xi_b^0\to\Sigma^{*+}\pi^-\overline{K}^0)+2\mathcal{A}(\Xi_b^-\to\Sigma^{*0}\pi^0K^-) \notag \\ &-\sqrt{2}\mathcal{A}(\Xi_b^-\to\Sigma^{*0}\pi^-\overline{K}^0)-\sqrt{2}\mathcal{A}(\Xi_b^-\to\Sigma^{*-}\pi^0\overline{K}^0) \notag \\ &-\mathcal{A}(\Xi_b^-\to\Sigma^{*-}\pi^+K^-)+\mathcal{A}(\Xi_b^-\to\Sigma^{*+}\pi^-K^-)\big]=0,
\end{align}
\begin{align}
SumI_-^2[\Xi_b^0,\Sigma^{*+},\pi^+,\overline{K}^0] &=-2\big[2\mathcal{A}(\Xi_b^0\to\Sigma^{*0}\pi^0\overline{K}^0)+\sqrt{2}\mathcal{A}(\Xi_b^0\to\Sigma^{*0}\pi^+K^-) \notag \\ &-\sqrt{2}\mathcal{A}(\Xi_b^0\to\Sigma^{*+}\pi^0K^-)-\mathcal{A}(\Xi_b^0\to\Sigma^{*-}\pi^+\overline{K}^0) \notag \\&+\mathcal{A}(\Xi_b^0\to\Sigma^{*+}\pi^-\overline{K}^0\big]=0,
\end{align}
\begin{align}
SumI_-^2[\Xi_b^-,\Sigma^{*0},\pi^+,\overline{K}^0] &=2\big[\sqrt{2}\mathcal{A}(\Xi_b^0\to\Sigma^{*0}\pi^0\overline{K}^0)+\mathcal{A}(\Xi_b^0\to\Sigma^{*0}\pi^+K^-) \notag \\ &-\sqrt{2}\mathcal{A}(\Xi_b^0\to\Sigma^{*-}\pi^+\overline{K}^0)+\sqrt{2}\mathcal{A}(\Xi_b^-\to\Sigma^{*0}\pi^0K^-) \notag \\ &-\mathcal{A}(\Xi_b^-\to\Sigma^{*0}\pi^-\overline{K}^0)-2\mathcal{A}(\Xi_b^-\to\Sigma^{*-}\pi^0\overline{K}^0) \notag \\ &-\sqrt{2}\mathcal{A}(\Xi_b^-\to\Sigma^{*-}\pi^+K^-)\big]=0,
\end{align}
\begin{align}
SumI_-^2[\Xi_b^-,\Sigma^{*+},\pi^0,\overline{K}^0] &=-2\big[\sqrt{2}\mathcal{A}(\Xi_b^0\to\Sigma^{*0}\pi^0\overline{K}^0)-\mathcal{A}(\Xi_b^0\to\Sigma^{*+}\pi^0K^-) \notag \\ &+\sqrt{2}\mathcal{A}(\Xi_b^0\to\Sigma^{*+}\pi^-\overline{K}^0)+\sqrt{2}\mathcal{A}(\Xi_b^-\to\Sigma^{*0}\pi^0K^-) \notag \\ &-2\mathcal{A}(\Xi_b^-\to\Sigma^{*0}\pi^-\overline{K}^0)-\mathcal{A}(\Xi_b^-\to\Sigma^{*-}\pi^0\overline{K}^0) \notag \\ &+\sqrt{2}\mathcal{A}(\Xi_b^-\to\Sigma^{*+}\pi^-K^-)\big]=0,
\end{align}
\begin{align}
SumI_-^2[\Xi_b^-,\Sigma^{*+},\pi^+,K^-] &=-2\big[\sqrt{2}\mathcal{A}(\Xi_b^0\to\Sigma^{*0}\pi^+K^-)-\sqrt{2}\mathcal{A}(\Xi_b^0\to\Sigma^{*+}\pi^0K^-) \notag \\ &+2\mathcal{A}(\Xi_b^-\to\Sigma^{*0}\pi^0K^-)-\mathcal{A}(\Xi_b^-\to\Sigma^{*-}\pi^+K^-) \notag \\ &+\mathcal{A}(\Xi_b^-\to\Sigma^{*+}\pi^-K^-)\big]=0,
\end{align}
\begin{align}
SumI_-^2[\Xi_b^-,\Sigma^{*+},\overline{K}^0,\eta_8] &=2\big[-\sqrt{2}\mathcal{A}(\Xi_b^0\to\Sigma^{*0}\overline{K}^0\eta_8)+\mathcal{A}(\Xi_b^0\to\Sigma^{*+}K^-\eta_8) \notag \\ &-\sqrt{2}\mathcal{A}(\Xi_b^-\to\Sigma^{*0}K^-\eta_8)+\mathcal{A}(\Xi_b^-\to\Sigma^{*-}\overline{K}^0\eta_8)\big]=0,
\end{align}
\begin{align}
SumI_-^3[\Xi_b^-,\Delta^{++},\overline{K}^0,\overline{K}^{0}] &=-6\big[\sqrt{3}\mathcal{A}(\Xi_b^0\to\Delta^0\overline{K}^0\overline{K}^0)+\mathcal{A}(\Xi_b^0\to\Delta^{++}K^-K^-) \notag \\ &-\sqrt{3}\mathcal{A}(\Xi_b^0\to\Delta^+K^-\overline{K}^0)-\sqrt{3}\mathcal{A}(\Xi_b^0\to\Delta^+\overline{K}^0K^-) \notag \\ &+\sqrt{3}\mathcal{A}(\Xi_b^-\to\Delta^0K^-\overline{K}^0)+\sqrt{3}\mathcal{A}(\Xi_b^-\to\Delta^0\overline{K}^0K^-) \notag \\ &-\sqrt{3}\mathcal{A}(\Xi_b^-\to\Delta^+K^-K^-)-\mathcal{A}(\Xi_b^-\to\Delta^-\overline{K}^0\overline{K}^0)\big]=0,
\end{align}
\begin{align}
SumI_-^2[\Xi_b^0,\Delta^{++},\overline{K}^0,\overline{K}^0] &=2\big[\sqrt{3}\mathcal{A}(\Xi_b^0\to\Delta^0\overline{K}^0\overline{K}^0)+\mathcal{A}(\Xi_b^0\to\Delta^{++}K^-K^-) \notag \\ &-\sqrt{3}\big(\mathcal{A}(\Xi_b^0\to\Delta^+K^-\overline{K}^0)+\mathcal{A}(\Xi_b^0\to\Delta^+\overline{K}^0K^-)\big)\big]=0,
\end{align}
\begin{align}
SumI_-^2[\Xi_b^-,\Delta^+,\overline{K}^0,\overline{K}^0] &=2\big[-2\mathcal{A}(\Xi_b^0\to\Delta^0\overline{K}^0\overline{K}^0)+\mathcal{A}(\Xi_b^0\to\Delta^+K^-\overline{K}^0) \notag \\ &+\mathcal{A}(\Xi_b^0\to\Delta^+\overline{K}^0K^-)-2\mathcal{A}(\Xi_b^-\to\Delta^0K^-\overline{K}^0) \notag \\ &-2\mathcal{A}(\Xi_b^-\to\Delta^0\overline{K}^0K^-)+\mathcal{A}(\Xi_b^-\to\Delta^+K^-K^-) \notag \\ &+\sqrt{3}\mathcal{A}(\Xi_b^-\to\Delta^-\overline{K}^0\overline{K}^0)\big]=0,
\end{align}
\begin{align}
SumI_-^2[\Xi_b^-,\Delta^{++},K^-,\overline{K}^0] &=2\big[\mathcal{A}(\Xi_b^0\to\Delta^{++}K^-K^-)-\sqrt{3}\big(\mathcal{A}(\Xi_b^0\to\Delta^+K^-\overline{K}^0) \notag \\ &-\mathcal{A}(\Xi_b^-\to\Delta^0K^-\overline{K}^0)+\mathcal{A}(\Xi_b^-\to\Delta^+K^-K^-)\big)\big]=0,
\end{align}
\begin{align}
SumI_-^2[\Xi_b^-,\Delta^{++},\overline{K}^0,K^-] &=2\big[\mathcal{A}(\Xi_b^0\to\Delta^{++}K^-K^-)-\sqrt{3}\big(\mathcal{A}(\Xi_b^0\to\Delta^+\overline{K}^0K^-) \notag \\ &-\mathcal{A}(\Xi_b^-\to\Delta^0\overline{K}^0K^-)+\mathcal{A}(\Xi_b^-\to\Delta^+K^-K^-)\big)\big]=0,
\end{align}
\begin{align}
SumI_-^2[\Lambda_b^0,\Delta^+,\pi^+,\overline{K}^0] &=-2\big[2\sqrt{2}\mathcal{A}(\Lambda_b^0\to\Delta^0\pi^0\overline{K}^0)+2\mathcal{A}(\Lambda_b^0\to\Delta^0\pi^+K^-) \notag \\ &-\sqrt{2}\mathcal{A}(\Lambda_b^0\to\Delta^+\pi^0K^-)+\mathcal{A}(\Lambda_b^0\to\Delta^+\pi^-\overline{K}^0) \notag \\ &-\sqrt{3}\mathcal{A}(\Lambda_b^0\to\Delta^-\pi^+\overline{K}^0)\big]=0,
\end{align}
\begin{align}
SumI_-^2[\Lambda_b^0,\Delta^{++},\pi^0,\overline{K}^0] &=2\big[\sqrt{3}\mathcal{A}(\Lambda_b^0\to\Delta^0\pi^0\overline{K}^0)-\sqrt{3}\mathcal{A}(\Lambda_b^0\to\Delta^+\pi^0K^-) \notag \\ &-\sqrt{2}\mathcal{A}(\Lambda_b^0\to\Delta^{++}\pi^-K^-)+\sqrt{6}\mathcal{A}(\Lambda_b^0\to\Delta^+\pi^-\overline{K}^0)\big]=0,
\end{align}
\begin{align}
SumI_-^2[\Lambda_b^0,\Delta^{++},\pi^+,K^-] &=2\sqrt{3}\mathcal{A}(\Lambda_b^0\to\Delta^0\pi^+K^-)-2\big[\sqrt{6}\mathcal{A}(\Lambda_b^0\to\Delta^+\pi^0K^-) \notag \\ &+\mathcal{A}(\Lambda_b^0\to\Delta^{++}\pi^-K^-)\big]=0,
\end{align}
\begin{align}
SumI_-^3[\Lambda_b^0,\Sigma^+,\pi^+,\pi^+] &=6\sqrt{2}\big[-2\mathcal{A}(\Lambda_b^0\to\Sigma^0\pi^0\pi^0)+\mathcal{A}(\Lambda_b^0\to\Sigma^0\pi^-\pi^+) \notag \\ &+\mathcal{A}(\Lambda_b^0\to\Sigma^0\pi^+\pi^-)+\mathcal{A}(\Lambda_b^0\to\Sigma^+\pi^0\pi^-) \notag \\ &+\mathcal{A}(\Lambda_b^0\to\Sigma^-\pi^0\pi^+)+\mathcal{A}(\Lambda_b^0\to\Sigma^+\pi^-\pi^0) \notag \\ &+\mathcal{A}(\Lambda_b^0\to\Sigma^-\pi^+\pi^0)\big]=0,
\end{align}
\begin{align}
SumI_-^2[\Lambda_b^0,\Sigma^+,\pi^+,\eta_8] &=4\mathcal{A}(\Lambda_b^0\to\Sigma^0\pi^0\eta_8)-2\big(\mathcal{A}(\Lambda_b^0\to\Sigma^+\pi^-\eta_8) \notag \\ &+\mathcal{A}(\Lambda_b^0\to\Sigma^-\pi^+\eta_8)\big)=0,
\end{align}
\begin{align}
SumI_-^2[\Lambda_b^0,\Sigma^+,K^+,\overline{K}^0] &=-2\big[\sqrt{2}\mathcal{A}(\Lambda_b^0\to\Sigma^0K^0\overline{K}^0)-\sqrt{2}\mathcal{A}(\Lambda_b^0\to\Sigma^0K^+K^-) \notag \\ &+\mathcal{A}(\Lambda_b^0\to\Sigma^+K^0K^-)+\mathcal{A}(\Lambda_b^0\to\Sigma^-K^+\overline{K}^0)\big]=0,
\end{align}
\begin{align}
SumI_-^2[\Lambda_b^0,p,\pi^+,\overline{K}^0] &=-2\big[\sqrt{2}\mathcal{A}(\Lambda_b^0\to\ n\pi^0\overline{K}^0)+\mathcal{A}(\Lambda_b^0\to\ n\pi^+K^-) \notag \\ &-\sqrt{2}\mathcal{A}(\Lambda_b^0\to\ p\pi^0K^-)+\mathcal{A}(\Lambda_b^0\to\ p\pi^-\overline{K}^0)\big]=0,
\end{align}
\begin{align}
SumI_-^2[\Lambda_b^0,\Xi^0,\pi^+,K^+] &=-2\big[\sqrt{2}\mathcal{A}(\Lambda_b^0\to\Xi^0\pi^0K^0)+\mathcal{A}(\Lambda_b^0\to\Xi^0\pi^-K^+) \notag \\ &-\sqrt{2}\mathcal{A}(\Lambda_b^0\to\Xi^-\pi^0K^+)+\mathcal{A}(\Lambda_b^0\to\Xi^-\pi^+K^0)\big]=0,
\end{align}
\begin{align}
SumI_-^2[\Lambda_b^0,\Lambda^0,\pi^+,\pi^+] &=4\mathcal{A}(\Lambda_b^0\to\Lambda^0\pi^0\pi^0)-2\big[\mathcal{A}(\Lambda_b^0\to\Lambda^0\pi^-\pi^+) \notag \\ &+\mathcal{A}(\Lambda_b^0\to\Lambda^0\pi^+\pi^-)\big]=0,
\end{align}
\begin{align}
SumI_-^3[\Xi_b^-,\Sigma^+,\pi^+,\overline{K}^0] &=6\big[-2\mathcal{A}(\Xi_b^0\to\Sigma^0\pi^0\overline{K}^0)-\sqrt{2}\mathcal{A}(\Xi_b^0\to\Sigma^0\pi^+K^-) \notag \\ &-\sqrt{2}\mathcal{A}(\Xi_b^0\to\Sigma^+\pi^0K^-)+\mathcal{A}(\Xi_b^0\to\Sigma^-\pi^+\overline{K}^0) \notag \\ &+\mathcal{A}(\Xi_b^0\to\Sigma^+\pi^-\overline{K}^0)-2\mathcal{A}(\Xi_b^-\to\Sigma^0\pi^0K^-) \notag \\ &+\sqrt{2}\mathcal{A}(\Xi_b^-\to\Sigma^0\pi^-\overline{K}^0)+\sqrt{2}\mathcal{A}(\Xi_b^-\to\Sigma^-\pi^0\overline{K}^0) \notag \\ &+\mathcal{A}(\Xi_b^-\to\Sigma^-\pi^+K^-)+\mathcal{A}(\Xi_b^-\to\Sigma^+\pi^-K^-)\big]=0,
\end{align}
\begin{align}
SumI_-^2[\Xi_b^-,\Sigma^+,\overline{K}^0,\eta_8] &=2\big[\sqrt{2}\mathcal{A}(\Xi_b^0\to\Sigma^0\overline{K}^0\eta_8)+\mathcal{A}(\Xi_b^0\to\Sigma^+K^-\eta_8) \notag \\ &+\sqrt{2}\mathcal{A}(\Xi_b^-\to\Sigma^0K^-\eta_8)-\mathcal{A}(\Xi_b^-\to\Sigma^-\overline{K}^0\eta_8)\big]=0,
\end{align}
\begin{align}
SumI_-^2[\Xi_b^-,p,\overline{K}^0,\overline{K}^0] &=-2\big[\mathcal{A}(\Xi_b^0\to\ n\overline{K}^0\overline{K}^0)-\mathcal{A}(\Xi_b^0\to\ pK^-\overline{K}^0) \notag \\ &-\mathcal{A}(\Xi_b^0\to\ p\overline{K}^0K^-)+\mathcal{A}(\Xi_b^-\to\ nK^-\overline{K}^0) \notag \\ &+\mathcal{A}(\Xi_b^-\to\ n\overline{K}^0K^-)-\mathcal{A}(\Xi_b^-\to\ pK^-K^-)\big]=0,
\end{align}
\begin{align}
SumI_-^2[\Xi_b^-,\Xi^0,\pi^+,\eta_8] &=2\big[\sqrt{2}\mathcal{A}(\Xi_b^0\to\Xi^0\pi^0\eta_8)+\mathcal{A}(\Xi_b^0\to\Xi^-\pi^+\eta_8) \notag \\ &-\mathcal{A}(\Xi_b^-\to\Xi^0\pi^-\eta_8)+\sqrt{2}(\Xi_b^-\to\Xi^-\pi^0\eta_8)\big]=0,
\end{align}
\begin{align}
SumI_-^2[\Xi_b^-,\Xi^0,K^+,\overline{K}^0] &=-2\big[\mathcal{A}(\Xi_b^0\to\Xi^0K^0\overline{K}^0)-\mathcal{A}(\Xi_b^0\to\Xi^0K^+K^-) \notag \\ &-\mathcal{A}(\Xi_b^0\to\Xi^-K^+\overline{K}^0)+\mathcal{A}(\Xi_b^-\to\Xi^0K^0K^-) \notag \\ &+\mathcal{A}(\Xi_b^-\to\Xi^-K^0\overline{K}^0)-\mathcal{A}(\Xi_b^-\to\Xi^-K^+K^-)\big]=0,
\end{align}
\begin{align}
SumI_-^2[\Xi_b^-,\Lambda^0,\pi^+,\overline{K}^0] &=2\big[\sqrt{2}\mathcal{A}(\Xi_b^0\to\Lambda^0\pi^0\overline{K}^0)+\mathcal{A}(\Xi_b^0\to\Lambda^0\pi^+K^-) \notag \\ &+\sqrt{2}\mathcal{A}(\Xi_b^-\to\Lambda^0\pi^0K^-)-\mathcal{A}(\Xi_b^-\to\Lambda^0\pi^-\overline{K}^0)\big]=0,
\end{align}
\begin{align}
SumI_-^2[\Lambda_b^0,\Sigma^0,\pi^+,\pi^+] &=4\mathcal{A}(\Lambda_b^0\to\Sigma^0\pi^0\pi^0)-2\big[\mathcal{A}(\Lambda_b^0\to\Sigma^0\pi^-\pi^+) \notag \\ &+\mathcal{A}(\Lambda_b^0\to\Sigma^0\pi^+\pi^-)+2\big(\mathcal{A}(\Lambda_b^0\to\Sigma^-\pi^0\pi^+) \notag \\ &+\mathcal{A}(\Lambda_b^0\to\Sigma^-\pi^+\pi^0)\big)\big]=0,
\end{align}
\begin{align}
SumI_-^2[\Lambda_b^0,\Sigma^+,\pi^0,\pi^+] &=4\mathcal{A}(\Lambda_b^0\to\Sigma^0\pi^0\pi^0)-2\big[2\mathcal{A}(\Lambda_b^0\to\Sigma^0\pi^-\pi^+) \notag \\ &+\mathcal{A}(\Lambda_b^0\to\Sigma^+\pi^0\pi^-)+\mathcal{A}(\Lambda_b^0\to\Sigma^-\pi^0\pi^+) \notag \\ &+2\mathcal{A}(\Lambda_b^0\to\Sigma^+\pi^-\pi^0)\big]=0,
\end{align}
\begin{align}
SumI_-^2[\Xi_b^0,\Sigma^+,\pi^+,\overline{K}^0] &=2\big[2\mathcal{A}(\Xi_b^0\to\Sigma^0\pi^0\overline{K}^0)+\sqrt{2}\mathcal{A}(\Xi_b^0\to\Sigma^0\pi^+K^-) \notag \\ &+\sqrt{2}\mathcal{A}(\Xi_b^0\to\Sigma^+\pi^0K^-)-\mathcal{A}(\Xi_b^0\to\Sigma^-\pi^+\overline{K}^0) \notag \\ &-\mathcal{A}(\Xi_b^0\to\Sigma^+\pi^-\overline{K}^0)\big]=0,
\end{align}
\begin{align}
SumI_-^2[\Xi_b^-,\Sigma^0,\pi^+,\overline{K}^0] &=2\big[\sqrt{2}\mathcal{A}(\Xi_b^0\to\Sigma^0\pi^0\overline{K}^0)+\mathcal{A}(\Xi_b^0\to\Sigma^0\pi^+K^-) \notag \\ &-\sqrt{2}\mathcal{A}(\Xi_b^0\to\Sigma^-\pi^+\overline{K}^0)+\sqrt{2}\mathcal{A}(\Xi_b^-\to\Sigma^0\pi^0K^-) \notag \\ &-\mathcal{A}(\Xi_b^-\to\Sigma^0\pi^-\overline{K}^0)-2\mathcal{A}(\Xi_b^-\to\Sigma^-\pi^0\overline{K}^0) \notag \\ &-\sqrt{2}\mathcal{A}(\Xi_b^-\to\Sigma^-\pi^+K^-)\big]=0,
\end{align}
\begin{align}
SumI_-^2[\Xi_b^-,\Sigma^+,\pi^0,\overline{K}^0] &=2\big[\sqrt{2}\mathcal{A}(\Xi_b^0\to\Sigma^0\pi^0\overline{K}^0)+\mathcal{A}(\Xi_b^0\to\Sigma^+\pi^0K^-) \notag \\ &-\sqrt{2}\mathcal{A}(\Xi_b^0\to\Sigma^+\pi^-\overline{K}^0)+\sqrt{2}\mathcal{A}(\Xi_b^-\to\Sigma^0\pi^0K^-) \notag \\ &-2\mathcal{A}(\Xi_b^-\to\Sigma^0\pi^-\overline{K}^0)-\mathcal{A}(\Xi_b^-\to\Sigma^-\pi^0\overline{K}^0) \notag \\ &-\sqrt{2}\mathcal{A}(\Xi_b^-\to\Sigma^+\pi^-K^-)\big]=0,
\end{align}
\begin{align}
SumI_-^2[\Xi_b^-,\Sigma^+,\pi^+,K^-] &=2\big[\sqrt{2}\mathcal{A}(\Xi_b^0\to\Sigma^0\pi^+K^-)+\sqrt{2}\mathcal{A}(\Xi_b^0\to\Sigma^+\pi^0K^-) \notag \\ &+2\mathcal{A}(\Xi_b^-\to\Sigma^0\pi^0K^-)-\mathcal{A}(\Xi_b^-\to\Sigma^-\pi^+K^-) \notag \\ &-\mathcal{A}(\Xi_b^-\to\Sigma^+\pi^-K^-)\big]=0,
\end{align}
\begin{align}
SumI_-^2[\Xi_b^0,\Xi^0,\pi^+,\pi^+] &=4\mathcal{A}(\Xi_b^0\to\Xi^0\pi^0\pi^0)-2\big[\mathcal{A}(\Xi_b^0\to\Xi^0\pi^-\pi^+) \notag \\ &+\mathcal{A}(\Xi_b^0\to\Xi^0\pi^+\pi^-)-\sqrt{2}\big(\mathcal{A}(\Xi_b^0\to\Xi^-\pi^0\pi^+) \notag \\ &+\mathcal{A}(\Xi_b^0\to\Xi^-\pi^+\pi^0)\big)\big]=0,
\end{align}
\begin{align}
SumI_-^2[\Xi_b^-,\Xi^-,\pi^+,\pi^+] &=2\big[\sqrt{2}\mathcal{A}(\Xi_b^0\to\Xi^-\pi^0\pi^+)+\sqrt{2}\mathcal{A}(\Xi_b^0\to\Xi^-\pi^+\pi^0) \notag \\ &+2\mathcal{A}(\Xi_b^-\to\Xi^-\pi^0\pi^0)-\mathcal{A}(\Xi_b^-\to\Xi^-\pi^+\pi^-) \notag \\ &-\mathcal{A}(\Xi_b^-\to\Xi^-\pi^-\pi^+)\big]=0,
\end{align}
\begin{align}
SumI_-^2[\Xi_b^-,\Xi^0,\pi^0,\pi^+] &=2\big[\sqrt{2}\mathcal{A}(\Xi_b^0\to\Xi^0\pi^0\pi^0)-\sqrt{2}\mathcal{A}(\Xi_b^0\to\Xi^0\pi^-\pi^+) \notag \\ &+\mathcal{A}(\Xi_b^0\to\Xi^-\pi^0\pi^+)-\mathcal{A}(\Xi_b^-\to\Xi^0\pi^0\pi^-) \notag \\ &-2\mathcal{A}(\Xi_b^-\to\Xi^0\pi^-\pi^0)+\sqrt{2}\big(\mathcal{A}(\Xi_b^-\to\Xi^-\pi^0\pi^0) \notag \\ &-\mathcal{A}(\Xi_b^-\to\Xi^-\pi^-\pi^+)\big)\big]=0.
\end{align}

\subsection{$b\to u\overline c d/s$ modes}

\begin{align}
SumI_-^2[\Lambda_b^0,\Sigma^+,\overline{D}^0,K^+] &=-2\big[\sqrt{2}\mathcal{A}(\Lambda_b^0\to\Sigma^0D^-K^+)+\sqrt{2}\mathcal{A}(\Lambda_b^0\to\Sigma^0\overline{D}^0K^0) \notag \\ &-\mathcal{A}(\Lambda_b^0\to\Sigma^+D^-K^0)+\mathcal{A}(\Lambda_b^0\to\Sigma^-\overline{D}^0K^+)\big]=0,
\end{align}
\begin{align}
SumI_-^2[\Lambda_b^0,p,\overline{D}^0,\pi^+] &=2\mathcal{A}(\Lambda_b^0\to nD^-\pi^+)-2\big[\sqrt{2}\mathcal{A}(\Lambda_b^0\to n\overline{D}^0\pi^0) \notag \\ &+\sqrt{2}\mathcal{A}(\Lambda_b^0\to pD^-\pi^0)+\mathcal{A}(\Lambda_b^0\to p\overline{D}^0\pi^-)\big]=0,
\end{align}
\begin{align}
SumI_-^3[\Xi_b^-,\Sigma^+,\overline{D}^0,\pi^+] &=6\big[\sqrt{2}\mathcal{A}(\Xi_b^0\to\Sigma^0D^-\pi^+)-2\mathcal{A}(\Xi_b^0\to\Sigma^0\overline{D}^0\pi^0) \notag \\ &+\sqrt{2}\mathcal{A}(\Xi_b^0\to\Sigma^+D^-\pi^0)+\mathcal{A}(\Xi_b^0\to\Sigma^-\overline{D}^0\pi^+) \notag \\ &+\mathcal{A}(\Xi_b^0\to\Sigma^+\overline{D}^0\pi^-)+2\mathcal{A}(\Xi_b^-\to\Sigma^0D^-\pi^0) \notag \\ &+\sqrt{2}\mathcal{A}(\Xi_b^-\to\Sigma^0\overline{D}^0\pi^-)-\mathcal{A}(\Xi_b^-\to\Sigma^-D^-\pi^+) \notag \\ &-\mathcal{A}(\Xi_b^-\to\Sigma^+D^-\pi^-)+\sqrt{2}\mathcal{A}(\Xi_b^-\to\Sigma^-\overline{D}^0\pi^0)\big]=0,
\end{align}
\begin{align}
SumI_-^2[\Xi_b^-,\Sigma^+,\overline{D}^0,\eta_8] &=2\sqrt{2}\mathcal{A}(\Xi_b^0\to\Sigma^0\overline{D}^0\eta_8)-2\big[\mathcal{A}(\Xi_b^0\to\Sigma^+D^-\eta_8) \notag \\ &+\sqrt{2}\mathcal{A}(\Xi_b^-\to\Sigma^0D^-\eta_8)+\mathcal{A}(\Xi_b^-\to\Sigma^-\overline{D}^0\eta_8)\big]=0,
\end{align}
\begin{align}
SumI_-^2[\Xi_b^-,\Sigma^+,D_s^-,K^+] &=2\sqrt{2}\mathcal{A}(\Xi_b^0\to\Sigma^0D_s^-K^+)-2\big[\mathcal{A}(\Xi_b^0\to\Sigma^+D_s^-K^0) \notag \\ &+\sqrt{2}\mathcal{A}(\Xi_b^-\to\Sigma^0D_s^-K^0)+\mathcal{A}(\Xi_b^-\to\Sigma^-D_s^-K^+)\big]=0,
\end{align}
\begin{align}
SumI_-^2[\Xi_b^-,p,\overline{D}^0,\overline{K}^0] &=-2\big[\mathcal{A}(\Xi_b^0\to n\overline{D}^0\overline{K}^0)+\mathcal{A}(\Xi_b^0\to pD^-\overline{K}^0) \notag \\ &-\mathcal{A}(\Xi_b^0\to p\overline{D}^0K^-)-\mathcal{A}(\Xi_b^-\to nD^-\overline{K}^0) \notag \\ &+\mathcal{A}(\Xi_b^-\to n\overline{D}^0K^-)+\mathcal{A}(\Xi_b^-\to pD^-K^-)\big]=0,
\end{align}
\begin{align}
SumI_-^2[\Xi_b^-,p,D_s^-,\pi^+] &=-2\big[\mathcal{A}(\Xi_b^0\to nD_s^-\pi^+)-\sqrt{2}\mathcal{A}(\Xi_b^0\to pD_s^-\pi^0) \notag \\ &+\sqrt{2}\mathcal{A}(\Xi_b^-\to nD_s^-\pi^0)+\mathcal{A}(\Xi_b^-\to pD_s^-\pi^-)\big]=0,
\end{align}
\begin{align}
SumI_-^2[\Xi_b^-,\Xi^0,\overline{D}^0,K^+] &=-2\big[\mathcal{A}(\Xi_b^0\to\Xi^0D^-K^+)+\mathcal{A}(\Xi_b^0\to\Xi^0\overline{D}^0K^0) \notag \\ &-\mathcal{A}(\Xi_b^0\to\Xi^-\overline{D}^0K^+)-\mathcal{A}(\Xi_b^-\to\Xi^0D^-K^0) \notag \\ &+\mathcal{A}(\Xi_b^-\to\Xi^-D^-K^+)+\mathcal{A}(\Xi_b^-\to\Xi^-\overline{D}^0K^0)\big]=0,
\end{align}
\begin{align}
SumI_-^2[\Xi_b^-,\Lambda^0,\overline{D}^0,\pi^+] &=-2\big[\mathcal{A}(\Xi_b^0\to\Lambda^0D^-\pi^+)-\sqrt{2}\mathcal{A}(\Xi_b^0\to\Lambda^0\overline{D}^0\pi^0) \notag \\ &+\sqrt{2}\mathcal{A}(\Xi_b^-\to\Lambda^0D^-\pi^0)+\mathcal{A}(\Xi_b^-\to\Lambda^0\overline{D}^0\pi^-)\big]=0,
\end{align}
\begin{align}
SumI_-^2[\Lambda_b^0,\Sigma^+,\overline{D}^0,\pi^+] &=-2\big[\sqrt{2}\mathcal{A}(\Lambda_b^0\to\Sigma^0D^-\pi^+)-2\mathcal{A}(\Lambda_b^0\to\Sigma^0\overline{D}^0\pi^0) \notag \\ &+\sqrt{2}\mathcal{A}(\Lambda_b^0\to\Sigma^+D^-\pi^0)+\mathcal{A}(\Lambda_b^0\to\Sigma^-\overline{D}^0\pi^+) \notag \\ &+\mathcal{A}(\Lambda_b^0\to\Sigma^+\overline{D}^0\pi^-)\big]=0,
\end{align}
\begin{align}
SumI_-[\Lambda_b^0,\Sigma^+,\overline{D}^0,\eta_8] &=-\sqrt{2}\mathcal{A}(\Lambda_b^0\to\Sigma^0\overline{D}^0\eta_8)+\mathcal{A}(\Lambda_b^0\to\Sigma^+D^-\eta_8)=0,
\end{align}
\begin{align}
SumI_-[\Lambda_b^0,\Sigma^+,D_s^-,K^+] &=-\sqrt{2}\mathcal{A}(\Lambda_b^0\to\Sigma^0D_s^-K^+)+\mathcal{A}(\Lambda_b^0\to\Sigma^+D_s^-K^0)=0,
\end{align}
\begin{align}
SumI_-[\Lambda_b^0,p,\overline{D}^0,\overline{K}^0] &=\mathcal{A}(\Lambda_b^0\to n\overline{D}^0\overline{K}^0)+\mathcal{A}(\Lambda_b^0\to pD^-\overline{K}^0) \notag \\ &-\mathcal{A}(\Lambda_b^0\to p\overline{D}^0K^-)=0,
\end{align}
\begin{align}
SumI_-[\Lambda_b^0,p,D_s^-,\pi^+] &=\mathcal{A}(\Lambda_b^0\to nD_s^-\pi^+)-\sqrt{2}\mathcal{A}(\Lambda_b^0\to pD_s^-\pi^0)=0,
\end{align}
\begin{align}
SumI_-[\Lambda_b^0,\Xi^0,\overline{D}^0,K^+] &=\mathcal{A}(\Lambda_b^0\to\Xi^0D^-K^+)+\mathcal{A}(\Lambda_b^0\to\Xi^0\overline{D}^0K^0) \notag \\ &-\mathcal{A}(\Lambda_b^0\to\Xi^-\overline{D}^0K^+)=0,
\end{align}
\begin{align}
SumI_-[\Lambda_b^0,\Lambda^0,\overline{D}^0,\pi^+] &=\mathcal{A}(\Lambda_b^0\to\Lambda^0D^-\pi^+)-\sqrt{2}\mathcal{A}(\Lambda_b^0\to\Lambda^0\overline{D}^0\pi^0)=0,
\end{align}
\begin{align}
SumI_-^2[\Xi_b^-,\Sigma^+,\overline{D}^0,\overline{K}^0] &=2\sqrt{2}\mathcal{A}(\Xi_b^0\to\Sigma^0\overline{D}^0\overline{K}^0)-2\big[\mathcal{A}(\Xi_b^0\to\Sigma^+D^-\overline{K}^0) \notag \\ &-\mathcal{A}(\Xi_b^0\to\Sigma^+\overline{D}^0K^-)+\sqrt{2}\mathcal{A}(\Xi_b^-\to\Sigma^0D^-\overline{K}^0) \notag \\ &-\sqrt{2}\mathcal{A}(\Xi_b^-\to\Sigma^0\overline{D}^0K^-)+\mathcal{A}(\Xi_b^-\to\Sigma^+D^-K^-) \notag \\ &+\mathcal{A}(\Xi_b^-\to\Sigma^-\overline{D}^0\overline{K}^0)\big]=0,
\end{align}
\begin{align}
SumI_-^2[\Xi_b^-,\Sigma^+,D_s^-,\pi^+] &=2\big[\sqrt{2}\mathcal{A}(\Xi_b^0\to\Sigma^0D_s^-\pi^+)+\sqrt{2}\mathcal{A}(\Xi_b^0\to\Sigma^+D_s^-\pi^0) \notag \\ &+2\mathcal{A}(\Xi_b^-\to\Sigma^0D_s^-\pi^0)-\mathcal{A}(\Xi_b^-\to\Sigma^-D_s^-\pi^+) \notag \\ &-\mathcal{A}(\Xi_b^-\to\Sigma^+D_s^-\pi^-)\big]=0,
\end{align}
\begin{align}
SumI_-[\Xi_b^-,\Sigma^+,D_s^-,\eta_8] &=-\mathcal{A}(\Xi_b^0\to\Sigma^+D_s^-\eta_8)-\sqrt{2}\mathcal{A}(\Xi_b^-\to\Sigma^0D_s^-\eta_8)=0,
\end{align}
\begin{align}
SumI_-[\Xi_b^-,p,D_s^-,\overline{K}^0] &=-\mathcal{A}(\Xi_b^0\to pD_s^-\overline{K}^0)+\mathcal{A}(\Xi_b^-\to nD_s^-\overline{K}^0) \notag \\ &-\mathcal{A}(\Xi_b^-\to pD_s^-K^-)=0,
\end{align}
\begin{align}
SumI_-^2[\Xi_b^-,\Xi^0,\overline{D}^0,\pi^+] &=-2\big[\mathcal{A}(\Xi_b^0\to\Xi^0D^-\pi^+)-\sqrt{2}\mathcal{A}(\Xi_b^0\to\Xi^0\overline{D}^0\pi^0) \notag \\ &-\mathcal{A}(\Xi_b^0\to\Xi^-\overline{D}^0\pi^+)+\sqrt{2}\mathcal{A}(\Xi_b^-\to\Xi^0D^-\pi^0) \notag \\ &+\mathcal{A}(\Xi_b^-\to\Xi^0\overline{D}^0\pi^-)+\mathcal{A}(\Xi_b^-\to\Xi^-D^-\pi^+) \notag \\ &-\sqrt{2}\mathcal{A}(\Xi_b^-\to\Xi^-\overline{D}^0\pi^0)\big]=0,
\end{align}
\begin{align}
SumI_-[\Xi_b^-,\Xi^0,\overline{D}^0,\eta_8] &=-\mathcal{A}(\Xi_b^0\to\Xi^0\overline{D}^0\eta_8)+\mathcal{A}(\Xi_b^-\to\Xi^0D^-\eta_8) \notag \\ &-\mathcal{A}(\Xi_b^-\to\Xi^-\overline{D}^0\eta_8)=0,
\end{align}
\begin{align}
SumI_-[\Xi_b^-,\Xi^0,D_s^-,K^+] &=-\mathcal{A}(\Xi_b^0\to\Xi^0D_s^-K^+)+\mathcal{A}(\Xi_b^-\to\Xi^0D_s^-K^0) \notag \\ &-\mathcal{A}(\Xi_b^-\to\Xi^-D_s^-K^+)=0,
\end{align}
\begin{align}
SumI_-[\Xi_b^-,\Lambda^0,\overline{D}^0,\overline{K}^0] &=-\mathcal{A}(\Xi_b^0\to\Lambda^0\overline{D}^0\overline{K}^0)+\mathcal{A}(\Xi_b^-\to\Lambda^0D^-\overline{K}^0) \notag \\ &-\mathcal{A}(\Xi_b^-\to\Lambda^0\overline{D}^0K^-)=0,
\end{align}
\begin{align}
SumI_-[\Xi_b^-,\Lambda^0,D_s^-,\pi^+] &=-\mathcal{A}(\Xi_b^0\to\Lambda^0D_s^-\pi^+)-\sqrt{2}\mathcal{A}(\Xi_b^-\to\Lambda^0D_s^-\pi^0)=0,
\end{align}
\begin{align}
SumI_-^2[\Xi_b^0,\Sigma^+,\overline{D}^0,\pi^+] &=-2\big[\sqrt{2}\mathcal{A}(\Xi_b^0\to\Sigma^0D^-\pi^+)-2\mathcal{A}(\Xi_b^0\to\Sigma^0\overline{D}^0\pi^0) \notag \\ &+\sqrt{2}\mathcal{A}(\Xi_b^0\to\Sigma^+D^-\pi^0)+\mathcal{A}(\Xi_b^0\to\Sigma^-\overline{D}^0\pi^+) \notag\\ &+\mathcal{A}(\Xi_b^0\to\Sigma^+\overline{D}^0\pi^-)\big]=0,
\end{align}
\begin{align}
SumI_-^2[\Xi_b^-,\Sigma^0,\overline{D}^0,\pi^+] &=-2\big[\mathcal{A}(\Xi_b^0\to\Sigma^0D^-\pi^+)-\sqrt{2}\mathcal{A}(\Xi_b^0\to\Sigma^0\overline{D}^0\pi^0) \notag \\ &+\sqrt{2}\mathcal{A}(\Xi_b^0\to\Sigma^-\overline{D}^0\pi^+)+\sqrt{2}\mathcal{A}(\Xi_b^-\to\Sigma^0D^-\pi^0) \notag \\ &+\mathcal{A}(\Xi_b^-\to\Sigma^0\overline{D}^0\pi^-)-\sqrt{2}\mathcal{A}(\Xi_b^-\to\Sigma^-D^-\pi^+) \notag \\ &+2\mathcal{A}(\Xi_b^-\to\Sigma^-\overline{D}^0\pi^0)\big]=0,
\end{align}
\begin{align}
SumI_-^2[\Xi_b^-,\Sigma^+,D^-,\pi^+] &=2\big[\sqrt{2}\mathcal{A}(\Xi_b^0\to\Sigma^0D^-\pi^+)+\sqrt{2}\mathcal{A}(\Xi_b^0\to\Sigma^+D^-\pi^0) \notag \\ &+2\mathcal{A}(\Xi_b^-\to\Sigma^0D^-\pi^0)-\mathcal{A}(\Xi_b^-\to\Sigma^-D^-\pi^+) \notag \\ &-\mathcal{A}(\Xi_b^-\to\Sigma^+D^-\pi^-)\big]=0,
\end{align}
\begin{align}
SumI_-^2[\Xi_b^-,\Sigma^+,\overline{D}^0,\pi^0] &=2\sqrt{2}\mathcal{A}(\Xi_b^0\to\Sigma^0\overline{D}^0\pi^0)-2\big[\mathcal{A}(\Xi_b^0\to\Sigma^+D^-\pi^0) \notag \\ &+\sqrt{2}\mathcal{A}(\Xi_b^0\to\Sigma^+\overline{D}^0\pi^-)+\sqrt{2}\mathcal{A}(\Xi_b^-\to\Sigma^0D^-\pi^0) \notag \\ &+2\mathcal{A}(\Xi_b^-\to\Sigma^0\overline{D}^0\pi^-)-\sqrt{2}\mathcal{A}(\Xi_b^-\to\Sigma^+D^-\pi^-) \notag \\ &+\mathcal{A}(\Xi_b^-\to\Sigma^-\overline{D}^0\pi^0)\big]=0,
\end{align}
\begin{align}
SumI_-[\Lambda_b^0,\Sigma^0,\overline{D}^0,\pi^+] &=\mathcal{A}(\Lambda_b^0\to\Sigma^0D^-\pi^+)+\sqrt{2}\big[-\mathcal{A}(\Lambda_b^0\to\Sigma^0\overline{D}^0\pi^0) \notag \\ &+\mathcal{A}(\Lambda_b^0\to\Sigma^-\overline{D}^0\pi^+)\big]=0,
\end{align}
\begin{align}
SumI_-[\Lambda_b^0,\Sigma^+,D^-,\pi^+] &=-\sqrt{2}\big[\mathcal{A}(\Lambda_b^0\to\Sigma^0D^-\pi^+)+\mathcal{A}(\Lambda_b^0\to\Sigma^+D^-\pi^0)\big]=0,
\end{align}
\begin{align}
SumI_-[\Lambda_b^0,\Sigma^+,\overline{D}^0,\pi^0] &=-\sqrt{2}\mathcal{A}(\Lambda_b^0\to\Sigma^0\overline{D}^0\pi^0)+\mathcal{A}(\Lambda_b^0\to\Sigma^+D^-\pi^0) \notag \\ &+\sqrt{2}\mathcal{A}(\Lambda_b^0\to\Sigma^+\overline{D}^0\pi^-)=0,
\end{align}
\begin{align}
SumI_-[\Xi_b^0,\Sigma^+,\overline{D}^0,\overline{K}^0] &=-\sqrt{2}\mathcal{A}(\Xi_b^0\to\Sigma^0\overline{D}^0\overline{K}^0)+\mathcal{A}(\Xi_b^0\to\Sigma^+D^-\overline{K}^0) \notag \\ &-\mathcal{A}(\Xi_b^0\to\Sigma^+\overline{D}^0K^-)=0,
\end{align}
\begin{align}
SumI_-[\Xi_b^-,\Sigma^0,\overline{D}^0,\overline{K}^0] &=-\mathcal{A}(\Xi_b^0\to\Sigma^0\overline{D}^0\overline{K}^0)+\mathcal{A}(\Xi_b^-\to\Sigma^0D^-\overline{K}^0) \notag \\ &-\mathcal{A}(\Xi_b^-\to\Sigma^0\overline{D}^0K^-)+\sqrt{2}\mathcal{A}(\Xi_b^-\to\Sigma^-\overline{D}^0\overline{K}^0)=0,
\end{align}
\begin{align}
SumI_-[\Xi_b^-,\Sigma^+,D^-,\overline{K}^0] &=-\mathcal{A}(\Xi_b^0\to\Sigma^+D^-\overline{K}^0)-\sqrt{2}\mathcal{A}(\Xi_b^-\to\Sigma^0D^-\overline{K}^0) \notag \\ &-\mathcal{A}(\Xi_b^-\to\Sigma^+D^-K^-)=0,
\end{align}
\begin{align}
SumI_-[\Xi_b^-,\Sigma^+,\overline{D}^0,K^-] &=-\mathcal{A}(\Xi_b^0\to\Sigma^+\overline{D}^0K^-)-\sqrt{2}\mathcal{A}(\Xi_b^-\to\Sigma^0\overline{D}^0K^-) \notag \\ &+\mathcal{A}(\Xi_b^-\to\Sigma^+D^-K^-)=0,
\end{align}
\begin{align}
SumI_-[\Xi_b^0,\Sigma^+,D_s^-,\pi^+] &=-\sqrt{2}\big[\mathcal{A}(\Xi_b^0\to\Sigma^0D_s^-\pi^+)+\mathcal{A}(\Xi_b^0\to\Sigma^+D_s^-\pi^0)\big]=0,
\end{align}
\begin{align}
SumI_-[\Xi_b^-,\Sigma^0,D_s^-,\pi^+] &=-\mathcal{A}(\Xi_b^0\to\Sigma^0D_s^-\pi^+)+\sqrt{2}\big[-\mathcal{A}(\Xi_b^-\to\Sigma^0D_s^-\pi^0) \notag \\ &+\mathcal{A}(\Xi_b^-\to\Sigma^-D_s^-\pi^+)\big]=0,
\end{align}
\begin{align}
SumI_-[\Xi_b^-,\Sigma^+,D_s^-,\pi^0] &=-\mathcal{A}(\Xi_b^0\to\Sigma^+D_s^-\pi^0)+\sqrt{2}\big[-\mathcal{A}(\Xi_b^-\to\Sigma^0D_s^-\pi^0) \notag \\ &+\mathcal{A}(\Xi_b^-\to\Sigma^+D_s^-\pi^-)\big]=0,
\end{align}
\begin{align}
SumI_-[\Xi_b^0,\Xi^0,\overline{D}^0,\pi^+] &=\mathcal{A}(\Xi_b^0\to\Xi^0D^-\pi^+) -\sqrt{2}\mathcal{A}(\Xi_b^0\to\Xi^0\overline{D}^0\pi^0) \notag \\ &-\mathcal{A}(\Xi_b^0\to\Xi^-\overline{D}^0\pi^+)=0,
\end{align}
\begin{align}
SumI_-[\Xi_b^-,\Xi^-,\overline{D}^0,\pi^+] &=-\mathcal{A}(\Xi_b^0\to\Xi^-\overline{D}^0\pi^+)+\mathcal{A}(\Xi_b^-\to\Xi^-D^-\pi^+) \notag \\ &-\sqrt{2}\mathcal{A}(\Xi_b^-\to\Xi^-\overline{D}^0\pi^0)=0,
\end{align}
\begin{align}
SumI_-[\Xi_b^-,\Xi^0,\overline{D}^0,\pi^0] &=-\mathcal{A}(\Xi_b^0\to\Xi^0\overline{D}^0\pi^0)+\mathcal{A}(\Xi_b^-\to\Xi^0D^-\pi^0) \notag \\ &+\sqrt{2}\mathcal{A}(\Xi_b^-\to\Xi^0\overline{D}^0\pi^-)-\mathcal{A}(\Xi_b^-\to\Xi^-\overline{D}^0\pi^0)=0,
\end{align}
\begin{align}
SumI_-[\Xi_b^-,\Xi^0,D^-,\pi^+] &=-\mathcal{A}(\Xi_b^0\to\Xi^0D^-\pi^+)-\sqrt{2}\mathcal{A}(\Xi_b^-\to\Xi^0D^-\pi^0) \notag \\ &-\mathcal{A}(\Xi_b^-\to\Xi^-D^-\pi^+)=0,
\end{align}
\begin{align}
SumI_-^2[\Lambda_b^0,\Sigma^{*+},\overline{D}^0,K^+] &=2\big[\sqrt{2}\mathcal{A}(\Lambda_b^0\to\Sigma^{*0}D^-K^+)+\sqrt{2}\mathcal{A}(\Lambda_b^0\to\Sigma^{*0}\overline{D}^0K^0) \notag \\ &+\mathcal{A}(\Lambda_b^0\to\Sigma^{*+}D^-K^0)+\mathcal{A}(\Lambda_b^0\to\Sigma^{*-}\overline{D}^0K^+)\big]=0,
\end{align}
\begin{align}
SumI^2_-[\Lambda_b^0,\Delta^{++},\overline{D}^0,\eta_8] &=2\sqrt{3}\big[\mathcal{A}(\Lambda_b^0\to\Delta^0\overline{D}^0\eta_8)+\mathcal{A}(\Lambda_b^0\to\Delta^+D^-\eta_8)\big]=0,
\end{align}
\begin{align}
SumI_-^3[\Lambda_b^0,\Delta^{++},\overline{D}^0,\pi^+] &=-6\big[-\sqrt{3}\mathcal{A}(\Lambda_b^0\to\Delta^0D^-\pi^+)+\sqrt{6}\mathcal{A}(\Lambda_b^0\to\Delta^0\overline{D}^0\pi^0) \notag \\ &+\mathcal{A}(\Lambda_b^0\to\Delta^{++}D^-\pi^-)+\sqrt{6}\mathcal{A}(\Lambda_b^0\to\Delta^+D^-\pi^0) \notag \\ &-\mathcal{A}(\Lambda_b^0\to\Delta^-\overline{D}^0\pi^+)+\sqrt{3}\mathcal{A}(\Lambda_b^0\to\Delta^+\overline{D}^0\pi^-)\big]=0,
\end{align}
\begin{align}
SumI_-^2[\Lambda_b^0,\Delta^+,\overline{D}^0,\pi^+] &=4\mathcal{A}(\Lambda_b^0\to\Delta^0D^-\pi^+)-2\big[2\sqrt{2}\mathcal{A}(\Lambda_b^0\to\Delta^0\overline{D}^0\pi^0) \notag \\ &+\sqrt{2}\mathcal{A}(\Lambda_b^0\to\Delta^+D^-\pi^0)-\sqrt{3}\mathcal{A}(\Lambda_b^0\to\Delta^-\overline{D}^0\pi^+) \notag \\ &+\mathcal{A}(\Lambda_b^0\to\Delta^+\overline{D}^0\pi^-)\big]=0,
\end{align}
\begin{align}
SumI_-^2[\Lambda_b^0,\Delta^{++},D^-,\pi^+] &=2\sqrt{3}\mathcal{A}(\Lambda_b^0\to\Delta^0D^-\pi^+)-2\big(\mathcal{A}(\Lambda_b^0\to\Delta^{++}D^-\pi^-) \notag \\ &+\sqrt{6}\mathcal{A}(\Lambda_b^0\to\Delta^+D^-\pi^0)\big)=0,
\end{align}
\begin{align}
SumI_-^2[\Lambda_b^0,\Delta^{++},\overline{D}^0,\pi^0] &=2\big[\sqrt{3}\mathcal{A}(\Lambda_b^0\to\Delta^0\overline{D}^0\pi^0)+\sqrt{2}\mathcal{A}(\Lambda_b^0\to\Delta^{++}D^-\pi^-) \notag \\ &+\sqrt{3}\mathcal{A}(\Lambda_b^0\to\Delta^+D^-\pi^0)+\sqrt{6}\mathcal{A}(\Lambda_b^0\to\Delta^+\overline{D}^0\pi^-)\big]=0,
\end{align}
\begin{align}
SumI^2_-[\Lambda_b^0,\Delta^{++},D_s^-,K^+] &=2\sqrt{3}\big[\mathcal{A}(\Lambda_b^0\to\Delta^0D_s^-K^+)+\mathcal{A}(\Lambda_b^0\to\Delta^+D_s^-K^0)\big]=0,
\end{align}
\begin{align}
SumI_-[\Lambda_b^0,\Xi^{*0},\overline{D}^0,K^+] &=\mathcal{A}(\Lambda_b^0\to\Xi^{*0}D^-K^+)+\mathcal{A}(\Lambda_b^0\to\Xi^{*0}\overline{D}^0K^0) \notag \\ &+\mathcal{A}(\Lambda_b^0\to\Xi^{*-}\overline{D}^0K^+)=0,
\end{align}
\begin{align}
SumI_-^2[\Xi_b^-,\Xi^{*0},\overline{D}^0,K^+] &=-2\big[\mathcal{A}(\Xi_b^0\to\Xi^{*0}D^-K^+)+\mathcal{A}(\Xi_b^0\to\Xi^{*0}\overline{D}^0K^0) \notag \\ &+\mathcal{A}(\Xi_b^0\to\Xi^{*-}\overline{D}^0K^+)-\mathcal{A}(\Xi_b^-\to\Xi^{*0}D^-K^0) \notag \\ &-\mathcal{A}(\Xi_b^-\to\Xi^{*-}D^-K^+)-\mathcal{A}(\Xi_b^-\to\Xi^{*-}\overline{D}^0K^0)\big]=0,
\end{align}
\begin{align}
SumI_-^3[\Xi_b^-,\Sigma^{*+},\overline{D}^0,\pi^+] &=-6\big[\sqrt{2}\mathcal{A}(\Xi_b^0\to\Sigma^{*0}D^-\pi^+)-2\mathcal{A}(\Xi_b^0\to\Sigma^{*0}\overline{D}^0\pi^0) \notag \\ &-\sqrt{2}\mathcal{A}(\Xi_b^0\to\Sigma^{*+}D^-\pi^0)+\mathcal{A}(\Xi_b^0\to\Sigma^{*-}\overline{D}^0\pi^+) \notag \\ &-\mathcal{A}(\Xi_b^0\to\Sigma^{*+}\overline{D}^0\pi^-)+2\mathcal{A}(\Xi_b^-\to\Sigma^{*0}D^-\pi^0) \notag \\ &+\sqrt{2}\mathcal{A}(\Xi_b^-\to\Sigma^{*0}\overline{D}^0\pi^-)-\mathcal{A}(\Xi_b^-\to\Sigma^{*-}D^-\pi^+) \notag \\ &+\mathcal{A}(\Xi_b^-\to\Sigma^{*+}D^-\pi^-)+\sqrt{2}\mathcal{A}(\Xi_b^-\to\Sigma^{*-}\overline{D}^0\pi^0)\big]=0,
\end{align}
\begin{align}
SumI_-^2[\Xi_b^-,\Sigma^{*+},\overline{D}^0,\eta_8] &=2\big[-\sqrt{2}\mathcal{A}(\Xi_b^0\to\Sigma^{*0}\overline{D}^0\eta_8)-\mathcal{A}(\Xi_b^0\to\Sigma^{*+}D^-\eta_8) \notag \\ &+\sqrt{2}\mathcal{A}(\Xi_b^-\to\Sigma^{*0}D^-\eta_8)+\mathcal{A}(\Xi_b^-\to\Sigma^{*-}\overline{D}^0\eta_8)\big]=0,
\end{align}
\begin{align}
SumI_-^2[\Xi_b^-,\Sigma^{*+},D_s^-,K^+] &=2\big[-\sqrt{2}\mathcal{A}(\Xi_b^0\to\Sigma^{*0}D_s^-K^+)-\mathcal{A}(\Xi_b^0\to\Sigma^{*+}D_s^-K^0) \notag \\ &+\sqrt{2}\mathcal{A}(\Xi_b^-\to\Sigma^{*0}D_s^-K^0)+\mathcal{A}(\Xi_b^-\to\Sigma^{*-}D_s^-K^+)\big]=0,
\end{align}
\begin{align}
SumI^3_-[\Xi_b^-,\Delta^{++},\overline{D}^0,\overline{K}^0] &=6\big[-\sqrt{3}\mathcal{A}(\Xi_b^0\to\Delta^0\overline{D}^0\overline{K}^0)+\mathcal{A}(\Xi_b^0\to\Delta^{++}D^-K^-) \notag \\ &-\sqrt{3}\mathcal{A}(\Xi_b^0\to\Delta^+D^-\overline{K}^0)+\sqrt{3}\mathcal{A}(\Xi_b^0\to\Delta^+\overline{D}^0K^-) \notag \\ &+\sqrt{3}\mathcal{A}(\Xi_b^-\to\Delta^0D^-\overline{K}^0)-\sqrt{3}\mathcal{A}(\Xi_b^-\to\Delta^0\overline{D}^0K^-) \notag \\ &-\sqrt{3}\mathcal{A}(\Xi_b^-\to\Delta^+D^-K^-)+\mathcal{A}(\Xi_b^-\to\Delta^-\overline{D}^0\overline{K}^0)\big]=0,
\end{align}
\begin{align}
SumI^3_-[\Xi_b^-,\Delta^{++},D_s^-,\pi^+] &=6\big[-\sqrt{3}\mathcal{A}(\Xi_b^0\to\Delta^0D_s^-\pi^+)+\mathcal{A}(\Xi_b^0\to\Delta^{++}D_s^-\pi^-) \notag \\ &+\sqrt{6}\mathcal{A}(\Xi_b^0\to\Delta^+D_s^-\pi^0)-\sqrt{6}\mathcal{A}(\Xi_b^-\to\Delta^0D_s^-\pi^0) \notag \\ &+\mathcal{A}(\Xi_b^-\to\Delta^-D_s^-\pi^+)-\sqrt{3}\mathcal{A}(\Xi_b^-\to\Delta^+D_s^-\pi^-)\big]=0,
\end{align}
\begin{align}
SumI^2_-[\Xi_b^-,\Delta^{++},D_s^-,\eta_8] &=2\sqrt{3}\big[-\mathcal{A}(\Xi_b^0\to\Delta^+D_s^-\eta_8)+\mathcal{A}(\Xi_b^-\to\Delta^0D_s^-\eta_8)\big]=0,
\end{align}
\begin{align}
SumI_-[\Lambda_b^0,\Xi^{*0},\overline{D}^0,K^+] &=\mathcal{A}(\Lambda_b^0\to\Xi^{*0}D^-K^+)+\mathcal{A}(\Lambda_b^0\to\Xi^{*0}\overline{D}^0K^0) \notag \\ &+\mathcal{A}(\Lambda_b^0\to\Xi^{*-}\overline{D}^0K^+)=0,
\end{align}
\begin{align}
SumI_-^2[\Lambda_b^0,\Sigma^{*+},\overline{D}^0,\pi^+] &=2\big[\sqrt{2}\mathcal{A}(\Lambda_b^0\to\Sigma^{*0}D^-\pi^+)-2\mathcal{A}(\Lambda_b^0\to\Sigma^{*0}\overline{D}^0\pi^0) \notag \\ &-\sqrt{2}\mathcal{A}(\Lambda_b^0\to\Sigma^{*+}D^-\pi^0)+\mathcal{A}(\Lambda_b^0\to\Sigma^{*-}\overline{D}^0\pi^+) \notag \\ &-\mathcal{A}(\Lambda_b^0\to\Sigma^{*+}\overline{D}^0\pi^-)\big]=0,
\end{align}
\begin{align}
SumI_-[\Lambda_b^0,\Sigma^{*+},\overline{D}^0,\eta_8] &=\sqrt{2}\mathcal{A}(\Lambda_b^0\to\Sigma^{*0}\overline{D}^0\eta_8)+\mathcal{A}(\Lambda_b^0\to\Sigma^{*+}D^-\eta_8)=0,
\end{align}
\begin{align}
SumI_-[\Lambda_b^0,\Sigma^{*+},D_s^-,K^+] &=\sqrt{2}\mathcal{A}(\Lambda_b^0\to\Sigma^{*0}D_s^-K^+)+\mathcal{A}(\Lambda_b^0\to\Sigma^{*+}D_s^-K^0)=0,
\end{align}
\begin{align}
SumI_-^2[\Lambda_b^0,\Delta^{++},\overline{D}^0,\overline{K}^0] &=2\big[\sqrt{3}\mathcal{A}(\Lambda_b^0\to\Delta^0\overline{D}^0\overline{K}^0)-\mathcal{A}(\Lambda_b^0\to\Delta^{++}D^-K^-) \notag \\ &+\sqrt{3}\big(\mathcal{A}(\Lambda_b^0\to\Delta^+D^-\overline{K}^0)-\mathcal{A}(\Lambda_b^0\to\Delta^+\overline{D}^0K^-)\big)\big]=0,
\end{align}
\begin{align}
SumI_-^2[\Lambda_b^0,\Delta^{++},D_s^-,\pi^+] &=2\sqrt{3}\mathcal{A}(\Lambda_b^0\to\Delta^0D_s^-\pi^+)-2\big[\mathcal{A}(\Lambda_b^0\to\Delta^{++}D_s^-\pi^-) \notag \\ &+\sqrt{6}\mathcal{A}(\Lambda_b^0\to\Delta^+D_s^-\pi^0)\big]=0,
\end{align}
\begin{align}
SumI_-[\Lambda_b^0,\Delta^{++},D_s^-,\eta_8] &=\sqrt{3}\mathcal{A}(\Lambda_b^0\to\Delta^+D_s^-\eta_8)=0,
\end{align}
\begin{align}
SumI_-[\Xi_b^-,\Omega^-,\overline{D}^0,K^+] &=-\mathcal{A}(\Xi_b^0\to\Omega^-\overline{D}^0K^+)+\mathcal{A}(\Xi_b^-\to\Omega^-D^-K^+) \notag \\ &+\mathcal{A}(\Xi_b^-\to\Omega^-K^0\overline{D}^0)=0,
\end{align}
\begin{align}
SumI_-^2[\Xi_b^-,\Xi^{*0},\overline{D}^0,\pi^+] &=-2\big[\mathcal{A}(\Xi_b^0\to\Xi^{*0}D^-\pi^+)-\sqrt{2}\mathcal{A}(\Xi_b^0\to\Xi^{*0}\overline{D}^0\pi^0) \notag \\ &+\mathcal{A}(\Xi_b^0\to\Xi^{*-}\overline{D}^0\pi^+)+\sqrt{2}\mathcal{A}(\Xi_b^-\to\Xi^{*0}D^-\pi^0) \notag \\ &+\mathcal{A}(\Xi_b^-\to\Xi^{*0}\overline{D}^0\pi^-)-\mathcal{A}(\Xi_b^-\to\Xi^{*-}D^-\pi^+) \notag \\ &+\sqrt{2}\mathcal{A}(\Xi_b^-\to\Xi^{*-}\overline{D}^0\pi^0)\big]=0,
\end{align}
\begin{align}
SumI_-[\Xi_b^-,\Xi^{*0},\overline{D}^0,\eta_8] &=-\mathcal{A}(\Xi_b^0\to\Xi^{*0}\overline{D}^0\eta_8)+\mathcal{A}(\Xi_b^-\to\Xi^{*0}D^-\eta_8) \notag \\ &+\mathcal{A}(\Xi_b^-\to\Xi^{*-}\overline{D}^0\eta_8)=0,
\end{align}
\begin{align}
SumI_-[\Xi_b^-,\Xi^{*0},D_s^-,K^+] &=-\mathcal{A}(\Xi_b^0\to\Xi^{*0}D_s^-K^+)+\mathcal{A}(\Xi_b^-\to\Xi^{*0}D_s^-K^0) \notag \\ &+\mathcal{A}(\Xi_b^-\to\Xi^{*-}D_s^-K^+)=0,
\end{align}
\begin{align}
SumI_-^2[\Xi_b^-,\Sigma^{*+},\overline{D}^0,\overline{K}^0] &=-2\big[\sqrt{2}\mathcal{A}(\Xi_b^0\to\Sigma^{*0}\overline{D}^0\overline{K}^0)+\mathcal{A}(\Xi_b^0\to\Sigma^{*+}D^-\overline{K}^0) \notag \\ &-\mathcal{A}(\Xi_b^0\to\Sigma^{*+}\overline{D}^0K^-)-\sqrt{2}\mathcal{A}(\Xi_b^-\to\Sigma^{*0}D^-\overline{K}^0) \notag \\ &+\sqrt{2}\mathcal{A}(\Xi_b^-\to\Sigma^{*0}\overline{D}^0K^-)+\mathcal{A}(\Xi_b^-\to\Sigma^{*+}D^-K^-) \notag \\ &-\mathcal{A}(\Xi_b^-\to\Sigma^{*-}\overline{D}^0\overline{K}^0)\big]=0,
\end{align}
\begin{align}
SumI_-^2[\Xi_b^-,\Sigma^{*+},D_s^-,\pi^+] &=-2\big[\sqrt{2}\mathcal{A}(\Xi_b^0\to\Sigma^{*0}D_s^-\pi^+)-\sqrt{2}\mathcal{A}(\Xi_b^0\to\Sigma^{*+}D_s^-\pi^0) \notag \\ &+2\mathcal{A}(\Xi_b^-\to\Sigma^{*0}D_s^-\pi^0)-\mathcal{A}(\Xi_b^-\to\Sigma^{*-}D_s^-\pi^+) \notag \\ &+\mathcal{A}(\Xi_b^-\to\Sigma^{*+}D_s^-\pi^-)\big]=0,
\end{align}
\begin{align}
SumI_-[\Xi_b^-,\Sigma^{*+},D_s^-,\eta_8] &=-\mathcal{A}(\Xi_b^0\to\Sigma^{*+}D_s^-\eta_8)+\sqrt{2}\mathcal{A}(\Xi_b^-\to\Sigma^{*0}D_s^-\eta_8)=0,
\end{align}
\begin{align}
SumI_-^2[\Xi_b^-,\Delta^{++},D_s^-,\overline{K}^0] &=2\big[\mathcal{A}(\Xi_b^0\to\Delta^{++}D_s^-K^-)-\sqrt{3}\big(\mathcal{A}(\Xi_b^0\to\Delta^+D_s^-\overline{K}^0) \notag \\ &-\mathcal{A}(\Xi_b^-\to\Delta^0D_s^-\overline{K}^0)+\mathcal{A}(\Xi_b^-\to\Delta^+D_s^-K^-)\big)\big]=0,
\end{align}
\begin{align}
SumI_-^2[\Xi_b^0,\Sigma^{*+},\overline{D}^0,\pi^+] &=2\big[\sqrt{2}\mathcal{A}(\Xi_b^0\to\Sigma^{*0}D^-\pi^+)-2\mathcal{A}(\Xi_b^0\to\Sigma^{*0}\overline{D}^0\pi^0) \notag \\ &-\sqrt{2}\mathcal{A}(\Xi_b^0\to\Sigma^{*+}D^-\pi^0)+\mathcal{A}(\Xi_b^0\to\Sigma^{*-}\overline{D}^0\pi^+) \notag \\ &-\mathcal{A}(\Xi_b^0\to\Sigma^{*+}\overline{D}^0\pi^-)\big]=0,
\end{align}
\begin{align}
SumI_-^2[\Xi_b^-,\Sigma^{*0},\overline{D}^0,\pi^+] &=-2\big[\mathcal{A}(\Xi_b^0\to\Sigma^{*0}D^-\pi^+)-\sqrt{2}\mathcal{A}(\Xi_b^0\to\Sigma^{*0}\overline{D}^0\pi^0) \notag \\ &+\sqrt{2}\mathcal{A}(\Xi_b^0\to\Sigma^{*-}\overline{D}^0\pi^+)+\sqrt{2}\mathcal{A}(\Xi_b^-\to\Sigma^{*0}D^-\pi^0) \notag \\ &+\mathcal{A}(\Xi_b^-\to\Sigma^{*0}\overline{D}^0\pi^-)-\sqrt{2}\mathcal{A}(\Xi_b^-\to\Sigma^{*-}D^-\pi^+) \notag \\ &+2\mathcal{A}(\Xi_b^-\to\Sigma^{*-}\overline{D}^0\pi^0)\big]=0,
\end{align}
\begin{align}
SumI_-^2[\Xi_b^-,\Sigma^{*+},D^-,\pi^+] &=-2\big[\sqrt{2}\mathcal{A}(\Xi_b^0\to\Sigma^{*0}D^-\pi^+)-\sqrt{2}\mathcal{A}(\Xi_b^0\to\Sigma^{*+}D^-\pi^0) \notag \\ &+2\mathcal{A}(\Xi_b^-\to\Sigma^{*0}D^-\pi^0)-\mathcal{A}(\Xi_b^-\to\Sigma^{*-}D^-\pi^+) \notag \\ &+\mathcal{A}(\Xi_b^-\to\Sigma^{*+}D^-\pi^-)\big]=0,
\end{align}
\begin{align}
SumI_-^2[\Xi_b^-,\Sigma^{*+},\overline{D}^0,\pi^0] &=2\big[-\sqrt{2}\mathcal{A}(\Xi_b^0\to\Sigma^{*0}\overline{D}^0\pi^0)-\mathcal{A}(\Xi_b^0\to\Sigma^{*+}D^-\pi^0) \notag \\ &-\sqrt{2}\mathcal{A}(\Xi_b^0\to\Sigma^{*+}\overline{D}^0\pi^-)+\sqrt{2}\mathcal{A}(\Xi_b^-\to\Sigma^{*0}D^-\pi^0) \notag \\ &+2\mathcal{A}(\Xi_b^-\to\Sigma^{*0}\overline{D}^0\pi^-)+\sqrt{2}\mathcal{A}(\Xi_b^-\to\Sigma^{*+}D^-\pi^-) \notag \\ &+\mathcal{A}(\Xi_b^-\to\Sigma^{*-}\overline{D}^0\pi^0)\big]=0,
\end{align}
\begin{align}
SumI_-^2[\Xi_b^0,\Delta^{++},\overline{D}^0,\overline{K}^0] &=2\big[\sqrt{3}\mathcal{A}(\Xi_b^0\to\Delta^0\overline{D}^0\overline{K}^0)-\mathcal{A}(\Xi_b^0\to\Delta^{++}D^-K^-) \notag \\ &+\sqrt{3}\big(\mathcal{A}(\Xi_b^0\to\Delta^+D^-\overline{K}^0)-\mathcal{A}(\Xi_b^0\to\Delta^+\overline{D}^0K^-)\big)\big]=0,
\end{align}
\begin{align}
SumI_-^2[\Xi_b^-,\Delta^+,\overline{D}^0,\overline{K}^0] &=-2\big[2\mathcal{A}(\Xi_b^0\to\Delta^0\overline{D}^0\overline{K}^0)+\mathcal{A}(\Xi_b^0\to\Delta^+D^-\overline{K}^0) \notag \\ &-\mathcal{A}(\Xi_b^0\to\Delta^+\overline{D}^0K^-)-2\mathcal{A}(\Xi_b^-\to\Delta^0D^-\overline{K}^0) \notag \\ &+2\mathcal{A}(\Xi_b^-\to\Delta^0\overline{D}^0K^-)+\mathcal{A}(\Xi_b^-\to\Delta^+D^-K^-) \notag \\ &+\sqrt{3}\mathcal{A}(\Xi_b^-\to\Delta^-\overline{D}^0\overline{K}^0)\big]=0,
\end{align}
\begin{align}
SumI_-^2[\Xi_b^-,\Delta^{++},D^-,\overline{K}^0] &=2\big[\mathcal{A}(\Xi_b^0\to\Delta^{++}D^-K^-)-\sqrt{3}\big(\mathcal{A}(\Xi_b^0\to\Delta^+D^-\overline{K}^0) \notag \\ &-\mathcal{A}(\Xi_b^-\to\Delta^0D^-\overline{K}^0)+\mathcal{A}(\Xi_b^-\to\Delta^+D^-K^-)\big)\big]=0,
\end{align}
\begin{align}
SumI_-^2[\Xi_b^-,\Delta^{++},\overline{D}^0,K^-] &=-2\big[\mathcal{A}(\Xi_b^0\to\Delta^{++}D^-K^-)+\sqrt{3}\big(\mathcal{A}(\Xi_b^0\to\Delta^+\overline{D}^0K^-) \notag \\ &-\mathcal{A}(\Xi_b^-\to\Delta^0\overline{D}^0K^-)-\mathcal{A}(\Xi_b^-\to\Delta^+D^-K^-)\big)\big]=0,
\end{align}
\begin{align}
SumI_-^2[\Xi_b^0,\Delta^{++},D_s^-,\pi^+] &=2\sqrt{3}\mathcal{A}(\Xi_b^0\to\Delta^0D_s^-\pi^+)-2\big[\mathcal{A}(\Xi_b^0\to\Delta^{++}D_s^-\pi^-) \notag \\ &+\sqrt{6}\mathcal{A}(\Xi_b^0\to\Delta^+D_s^-\pi^0)\big]=0,
\end{align}
\begin{align}
SumI_-^2[\Xi_b^-,\Delta^+,D_s^-,\pi^+] &=-2\big[2\mathcal{A}(\Xi_b^0\to\Delta^0D_s^-\pi^+)-\sqrt{2}\mathcal{A}(\Xi_b^0\to\Delta^+D_s^-\pi^0) \notag \\ &+2\sqrt{2}\mathcal{A}(\Xi_b^-\to\Delta^0D_s^-\pi^0)-\sqrt{3}\mathcal{A}(\Xi_b^-\to\Delta^-D_s^-\pi^+) \notag \\ &+\mathcal{A}(\Xi_b^-\to\Delta^+D_s^-\pi^-)\big]=0,
\end{align}
\begin{align}
SumI_-^2[\Xi_b^-,\Delta^{++},D_s^-,\pi^0] &=-2\sqrt{2}\mathcal{A}(\Xi_b^0\to\Delta^{++}D_s^-\pi^-)+2\sqrt{3}\big[-\mathcal{A}(\Xi_b^0\to\Delta^+D_s^-\pi^0) \notag \\ &+\mathcal{A}(\Xi_b^-\to\Delta^0D_s^-\pi^0)+\sqrt{2}\mathcal{A}(\Xi_b^-\to\Delta^+D_s^-\pi^-)\big]=0,
\end{align}
\begin{align}
SumI_-[\Lambda_b^0,\Sigma^{*0},\overline{D}^0,\pi^+] &=\mathcal{A}(\Lambda_b^0\to\Sigma^{*0}D^-\pi^+)+\sqrt{2}\big[-\mathcal{A}(\Lambda_b^0\to\Sigma^{*0}\overline{D}^0\pi^0) \notag \\ &+\mathcal{A}(\Lambda_b^0\to\Sigma^{*-}\overline{D}^0\pi^+)\big]=0,
\end{align}
\begin{align}
SumI_-[\Lambda_b^0,\Sigma^{*+},D^-,\pi^+] &=\sqrt{2}\big[\mathcal{A}(\Lambda_b^0\to\Sigma^{*0}D^-\pi^+)-\mathcal{A}(\Lambda_b^0\to\Sigma^{*+}D^-\pi^0)\big]=0,
\end{align}
\begin{align}
SumI_-[\Lambda_b^0,\Sigma^{*+},\overline{D}^0,\pi^0] &=\sqrt{2}\mathcal{A}(\Lambda_b^0\to\Sigma^{*0}\overline{D}^0\pi^0)+\mathcal{A}(\Lambda_b^0\to\Sigma^{*+}D^-\pi^0) \notag \\ &+\sqrt{2}\mathcal{A}(\Lambda_b^0\to\Sigma^{*+}\overline{D}^0\pi^-)=0,
\end{align}
\begin{align}
SumI_-[\Lambda_b^0,\Delta^+,\overline{D}^0,\overline{K}^0] &=2\mathcal{A}(\Lambda_b^0\to\Delta^0\overline{D}^0\overline{K}^0)+\mathcal{A}(\Lambda_b^0\to\Delta^+D^-\overline{K}^0) \notag \\ &-\mathcal{A}(\Lambda_b^0\to\Delta^+\overline{D}^0K^-)=0,
\end{align}
\begin{align}
SumI_-[\Lambda_b^0,\Delta^{++},D^-,\overline{K}^0] &=-\mathcal{A}(\Lambda_b^0\to\Delta^{++}D^-K^-)+\sqrt{3}\mathcal{A}(\Lambda_b^0\to\Delta^+D^-\overline{K}^0)=0,
\end{align}
\begin{align}
SumI_-[\Lambda_b^0,\Delta^{++},\overline{D}^0,K^-] &=\mathcal{A}(\Lambda_b^0\to\Delta^{++}D^-K^-)+\sqrt{3}\mathcal{A}(\Lambda_b^0\to\Delta^+\overline{D}^0K^-)=0,
\end{align}
\begin{align}
SumI_-[\Lambda_b^0,\Delta^+,D_s^-,\pi^+] &=2\mathcal{A}(\Lambda_b^0\to\Delta^0D_s^-\pi^+)-\sqrt{2}\mathcal{A}(\Lambda_b^0\to\Delta^+D_s^-\pi^0)=0,
\end{align}
\begin{align}
SumI_-[\Lambda_b^0,\Delta^{++},D_s^-,\pi^0] &=\sqrt{2}\mathcal{A}(\Lambda_b^0\to\Delta^{++}D_s^-\pi^-)+\sqrt{3}\mathcal{A}(\Lambda_b^0\to\Delta^+D_s^-\pi^0)=0,
\end{align}
\begin{align}
SumI_-[\Xi_b^0,\Xi^{*0},\overline{D}^0,\pi^+] &=\mathcal{A}(\Xi_b^0\to\Xi^{*0}D^-\pi^+)-\sqrt{2}\mathcal{A}(\Xi_b^0\to\Xi^{*0}\overline{D}^0\pi^0) \notag \\ &+\mathcal{A}(\Xi_b^0\to\Xi^{*-}\overline{D}^0\pi^+)=0,
\end{align}
\begin{align}
SumI_-[\Xi_b^-,\Xi^{*-},\overline{D}^0,\pi^+] &=-\mathcal{A}(\Xi_b^0\to\Xi^{*-}\overline{D}^0\pi^+)+\mathcal{A}(\Xi_b^-\to\Xi^{*-}D^-\pi^+) \notag \\ &-\sqrt{2}\mathcal{A}(\Xi_b^-\to\Xi^{*-}\overline{D}^0\pi^0)=0,
\end{align}
\begin{align}
SumI_-[\Xi_b^-,\Xi^{*0},D^-,\pi^+] &=-\mathcal{A}(\Xi_b^0\to\Xi^{*0}D^-\pi^+)-\sqrt{2}\mathcal{A}(\Xi_b^-\to\Xi^{*0}D^-\pi^0) \notag \\ &+\mathcal{A}(\Xi_b^-\to\Xi^{*-}D^-\pi^+)=0,
\end{align}
\begin{align}
SumI_-[\Xi_b^-,\Xi^{*0},\overline{D}^0,\pi^0] &=-\mathcal{A}(\Xi_b^0\to\Xi^{*0}\overline{D}^0\pi^0)+\mathcal{A}(\Xi_b^-\to\Xi^{*0}D^-\pi^0) \notag \\ &+\sqrt{2}\mathcal{A}(\Xi_b^-\to\Xi^{*0}\overline{D}^0\pi^-)+\mathcal{A}(\Xi_b^-\to\Xi^{*-}\overline{D}^0\pi^0)=0,
\end{align}
\begin{align}
SumI_-[\Xi_b^0,\Sigma^{*+},\overline{D}^0,\overline{K}^0] &=\sqrt{2}\mathcal{A}(\Xi_b^0\to\Sigma^{*0}\overline{D}^0\overline{K}^0)+\mathcal{A}(\Xi_b^0\to\Sigma^{*+}D^-\overline{K}^0) \notag \\ &-\mathcal{A}(\Xi_b^0\to\Sigma^{*+}\overline{D}^0K^-)=0,
\end{align}
\begin{align}
SumI_-[\Xi_b^-,\Sigma^{*0},\overline{D}^0,\overline{K}^0] &=-\mathcal{A}(\Xi_b^0\to\Sigma^{*0}\overline{D}^0\overline{K}^0)+\mathcal{A}(\Xi_b^-\to\Sigma^{*0}D^-\overline{K}^0) \notag \\ &-\mathcal{A}(\Xi_b^-\to\Sigma^{*0}\overline{D}^0K^-)+\sqrt{2}\mathcal{A}(\Xi_b^-\to\Sigma^{*-}\overline{D}^0\overline{K}^0)=0,
\end{align}
\begin{align}
SumI_-[\Xi_b^-,\Sigma^{*+},D^-,\overline{K}^0] &=-\mathcal{A}(\Xi_b^0\to\Sigma^{*+}D^-\overline{K}^0)+\sqrt{2}\mathcal{A}(\Xi_b^-\to\Sigma^{*0}D^-\overline{K}^0) \notag \\ &-\mathcal{A}(\Xi_b^-\to\Sigma^{*+}D^-K^-)=0,
\end{align}
\begin{align}
SumI_-[\Xi_b^-,\Sigma^{*+},\overline{D}^0,K^-] &=-\mathcal{A}(\Xi_b^0\to\Sigma^{*+}\overline{D}^0K^-)+\sqrt{2}\mathcal{A}(\Xi_b^-\to\Sigma^{*0}\overline{D}^0K^-) \notag \\ &+\mathcal{A}(\Xi_b^-\to\Sigma^{*+}D^-K^-)=0,
\end{align}
\begin{align}
SumI_-[\Xi_b^0,\Sigma^{*+},D_s^-,\pi^+] &=\sqrt{2}\big[\mathcal{A}(\Xi_b^0\to\Sigma^{*0}D_s^-\pi^+)-\mathcal{A}(\Xi_b^0\to\Sigma^{*+}D_s^-\pi^0\big]=0,
\end{align}
\begin{align}
SumI_-[\Xi_b^-,\Sigma^{*0},D_s^-,\pi^+] &=-\mathcal{A}(\Xi_b^0\to\Sigma^{*0}D_s^-\pi^+)+\sqrt{2}\big[-\mathcal{A}(\Xi_b^-\to\Sigma^{*0}D_s^-\pi^0) \notag \\ &+\mathcal{A}(\Xi_b^-\to\Sigma^{*-}D_s^-\pi^+)\big]=0,
\end{align}
\begin{align}
SumI_-[\Xi_b^-,\Sigma^{*+},D_s^-,\pi^0] &=-\mathcal{A}(\Xi_b^0\to\Sigma^{*+}D_s^-\pi^0)+\sqrt{2}\big[\mathcal{A}(\Xi_b^-\to\Sigma^{*0}D_s^-\pi^0) \notag \\ &+\mathcal{A}(\Xi_b^-\to\Sigma^{*+}D_s^-\pi^-)=0,
\end{align}
\begin{align}
SumI_-[\Xi_b^0,\Delta^{++},D_s^-,\overline{K}^0] &=-\mathcal{A}(\Xi_b^0\to\Delta^{++}D_s^-K^-)+\sqrt{3}\mathcal{A}(\Xi_b^0\to\Delta^+D_s^-\overline{K}^0)=0,
\end{align}
\begin{align}
SumI_-[\Xi_b^-,\Delta^+,D_s^-,\overline{K}^0] &=-\mathcal{A}(\Xi_b^0\to\Delta^+D_s^-\overline{K}^0)+2\mathcal{A}(\Xi_b^-\to\Delta^0D_s^-\overline{K}^0) \notag \\ &-\mathcal{A}(\Xi_b^-\to\Delta^+D_s^-K^-)=0,
\end{align}
\begin{align}
SumI_-[\Xi_b^-,\Delta^{++},D_s^-,K^-] &=-\mathcal{A}(\Xi_b^0\to\Delta^{++}D_s^-K^-)
+\sqrt{3}\mathcal{A}(\Xi_b^-\to\Delta^+D_s^-K^-)=0.
\end{align}

\end{appendix}

%%%%%%%%%%%%%%%%%%%%%%%%%%%%%%%%%%%%%%%%%%%%%%%%%%%%%%%%%%%%%%%%%%%%%%%%%%%%%%%

%\include{reference}

\end{document}